\documentclass[aps,prd,superscriptaddress,nofootinbib,amsmath,amsfonts,preprintnumbers,groupedaddress,showpacs,10pt,english]
{revtex4-1}
\usepackage{amsmath}
\usepackage{amssymb}
\usepackage[mathlines]{lineno}% Enable numbering of text and display math
\usepackage{array,multirow,graphicx}
\usepackage{soul}
\usepackage{dcolumn}
\usepackage{newlfont}
\usepackage{bm}
\usepackage[colorlinks,citecolor=blue,urlcolor=blue,linkcolor=blue]{hyperref}
\usepackage[figtopcap]{subfigure}
\usepackage{color}
\usepackage{verbatim}
\renewcommand{\k}{\kappa}

\usepackage{scalerel}
\usepackage{tikz}
\usetikzlibrary{svg.path}
\definecolor{orcidlogocol}{HTML}{A6CE39}
\tikzset{
 orcidlogo/.pic={
 \fill[orcidlogocol] svg{M256,128c0,70.7-57.3,128-128,128C57.3,256,0,198.7,0,128C0,57.3,57.3,0,128,0C198.7,0,256,57.3,256,128z};
 \fill[white] svg{M86.3,186.2H70.9V79.1h15.4v48.4V186.2z}
 svg{M108.9,79.1h41.6c39.6,0,57,28.3,57,53.6c0,27.5-21.5,53.6-56.8,53.6h-41.8V79.1z M124.3,172.4h24.5c34.9,0,42.9-26.5,42.9-39.7c0-21.5-13.7-39.7-43.7-39.7h-23.7V172.4z}
 svg{M88.7,56.8c0,5.5-4.5,10.1-10.1,10.1c-5.6,0-10.1-4.6-10.1-10.1c0-5.6,4.5-10.1,10.1-10.1C84.2,46.7,88.7,51.3,88.7,56.8z};}}
\newcommand\orcid[1]{\href{https://orcid.org/#1}{\mbox{\scalerel*{
\begin{tikzpicture}[yscale=-1,transform shape]
\pic{orcidlogo};
\end{tikzpicture}
}{|}}}}

\graphicspath{{./}{Figs/}}
\begin{document}
\date{\today}

\title{Stable soliton dark matter wormhole in non-minimally coupled $f({\cal Q},{\cal T})$ gravity}
\author{G.G.L. Nashed \orcid{0000-0001-5544-1119}}
\email{nashed@bue.edu.eg}
\author{Waleed El Hanafy \orcid{0000-0002-0097-6412}}
\email{waleed.elhanafy@bue.edu.eg}
\affiliation{Centre for Theoretical Physics, The British University in Egypt, P.O. Box 43, El Sherouk City, Cairo 11837, Egypt}
%%%%%%%%%%%%%%%%%%%%%%%%%%%%%%%%%%%%%  DATE  %%%%%%%%%%%%%%%%%%%%%%%%%%%%%%%%%%%%
\date{\today}
%%%%%%%%%%%%%%%%%%%%%%%%%%%%%  Abstract  %%%%%%%%%%%%%%%%%%%%%%%%%%%%%
\begin{abstract}
We show that non-minimal coupling between matter and geometry can indeed help in constructing stable, traversable, wormholes (WHs) without requiring exotic matter under certain conditions. In models like $f({\cal Q},{\cal T})={\cal Q}+\beta {\cal T}$ gravity, where ${\cal Q}$ is the non-metricity scalar, and ${\cal T}$ is the trace of the energy-momentum tensor, the coupling between matter and geometry introduces additional degrees of freedom in terms of the parameter $\beta$. These can mimic the effects of exotic matter or even replace it entirely under specific parameter choice. The analysis involves deriving WH shape functions based on two dark matter (DM) density profiles: a solitonic core at the center of DM halos, and the outer halo follows the universal Navarro-Frenk-White (NFW) density profile of cold DM (CDM). The WH solutions derived in these models satisfy important geometric conditions like: Flaring-out condition (necessary for traversability) and asymptotic flatness condition. For large positive coupling parameter, the null energy condition (NEC) can be satisfied at the WH throat, meaning exotic matter is not needed, while the WH is no longer Lorentzian and the flaring-out condition is broken. However, for large negative coupling parameter, the NEC can be satisfied, allowing for healthy WHs without exotic matter, provided the coupling strength remains within certain bounds. In the latter case, the NEC is broken only effectively. We investigate the stability of the obtained WH solutions by virtue of a modified version of Tolman–Oppenheimer–Volkoff (TOV) equation, which includes a new force due to matter-geometry non-minimal, showing that these WHs can be dynamically stable.
\end{abstract}
%%%%%%%%%%%%%%%%%%%%  Title  %%%%%%%%%%%%%%%%%%%%%%%
\maketitle
%\tableofcontents
%%%%%%%%%%%%%%%%%%%%  Keywords  %%%%%%%%%%%%%%%%%%%%%%%
%\textbf{Keywords: $f({\cal Q},{\cal T})$ gravity, wormhole, energy conditions}
%%%%%%%%%%%%%%%%%%%%%%%%%%%%%%%%%%%%%%%%%%%%%%%%%%%%%%%%%%%%%%%%%%%%%%%%
%\newpage
%%%%%%%%%%%%%%%%%%%%%%%%%%%%%%%%%%%%%%%%%%%%%%%%%%%%%%%%%%%%%%%%%%%%%%%%%%%%%%%%%%%%%%%%%%%%%%%%
\section{Introduction}\label{SEC:I}
%%%%%%%%%%%%%%%%%%%%%%%%%%%%%%%%%%%%%%%%%%%%%%%%%%%%%%%%%%%%%%%%%%%%%%%%%%%%%%%%%%%%%%%%%%%%%%%%
Current cutting-edge studies in astrophysics and cosmology are propelling us towards a fresh phase of revelations concerning the  Universe.
Partnerships like the collaboration of LIGO \cite{Gourgoulhon:2019iyu}, EHT \cite{Chael_2016, EventHorizonTelescope:2019dse}, INTEGRAL \cite{Winkler:2003nn}, Virgo \cite{GRAVITY:2020gka}, ATHENA \cite{https://doi.org/10.1002/asna.201713323}, IXPE \cite{Soffitta:2013hla}, Swift \cite{SWIFT:2005ngz}, as well as CHIME \cite{CHIMEFRB:2018mlh} are conducting unprecedented experiments on gravity and astrophysical objects.   Additionally, upcoming investigations of  BINGO \cite{Abdalla:2021nyj}, SKA \cite{Hall2005TheSK} and LISA \cite{eLISA:2013xep} could significantly limit the scope of gravity theories, narrowing down the plethora of theoretical ideas circulating in academic literature.

The WH is a fascinating and unusual solution to Einstein's equations in general relativity \cite{Visser:1995cc}. The WH serves as a topological connection between points in separate universes or within the same universe. Although it could be considered as a theoretical concept, numerous studies have been conducted on the possibility of discovering WHs. In the study carried out by Bueno et al. (2018)  \cite{Bueno:2017hyj} an intriguing method was suggested to locate detectable WHs by utilizing gravitational wave data to analyze echoes of the gravitational wave signal near black hole (BH) horizons to improve the sentence flow. The rings produced after a merger are associated with the emission of gravitational waves during the post-merger phase of binary mergers. Additionally, Paul et al. \cite{Paul:2019trt} examined unique features of accretion disk images to differentiate a WH shape from a BH, putting forward an alternative fascinating opportunity for observation. Furthermore, extensive research has been conducted on the gravitational lensing caused by a WH and the deflection of light  \cite{Ovgun:2018prw,Ovgun:2018fnk,Ovgun:2020yuv}.

Another sought-after aspect regarding WHs is the quest for traversable options,  that is, WHs that are sufficiently large for a person to pass through and withstand the powerful tidal forces. Morris and Thorne's groundbreaking paper \cite{Morris:1988cz} first introduced this type of WHs, which remains a popular topic, evident in the impressive research presented in \cite{Maldacena:2020sxe}. In their study \cite{Maldacena:2020sxe}  authors introduced a novel dark sector in the Randall-Sundrum model \cite{Randall:1999vf} with a $U(1)$ gauge field which only interacts with particles in the standard model through the force of gravity, allowing for the potential existence of traversable WHs. WH has been investigated within modified gravity, it has been shown that traversable WHs can be constructed within in $f(\mathcal{G})$ modified gravity, where $\mathcal{G}$ denotes the Gauss-Bonnet term, for anisotropic, isotropic and barotropic fluids cases \cite{Ilyas:2023rde}.

Up to now, general relativity remains the prevailing explanation supported by many for its gravitational force influence through   observations and experiments. Nevertheless, certain events that are governed by general relativity fails to fully explain, like the accelerated cosmic expansion, the gravitation at galactic scales, as well as the search for a quantum theory to describe gravity. Alternative explanations for these issues have been suggested in the form of different gravity theories. These theories make alterations or expansions to general relativity in different manners. An example of extending general relativity is $f({\cal R})$ gravitational theory. Modifying the Einstein-Hilbert action, this theory is changed by substituting a function of the scalar curvature for the Ricci scalar \cite{Buchdahl:1970ldb,Starobinsky:1979ty,DeFelice:2010aj,Sotiriou:2008rp,Nojiri:2017ncd,Nojiri:2010wj,Nojiri:2006ri}. This alteration results in modifications to the equations governing the gravitational field, potentially impacting how gravity acts on various scales. This theory explains the speeding up of the Universe's expansion, supports limits on early inflation, and removes the necessity of DM to account for galaxy rotation, star movement, and galaxy shape \cite{Capozziello:2012ie}.

In the year 2011, Harko and colleagues \cite{Harko:2011kv} presented a modified version of $f({\cal R})$ known as $f({\cal R},{\cal T})$ gravitational theory.  Such a  theory adds an additional factor to Einstein's equations based on the energy-momentum tensor trace.  This contribution would explain the typical display of significant quantum phenomena related to features of conformal anomaly within $f({\cal R},{\cal T})$ modified gravity.  The theory has undergone testing using various realistic and conceptual methods as evident in \cite{Sardar:2023iha, Pappas:2022gtt, Bose:2022xoc,Moraes:2016gpe}.

In addition to prior general relativity extensions, we  emphasize the $f({\cal Q})$ gravitational theory proposed by Jimenez and colleagues \cite{BeltranJimenez:2017tkd}. This theory posits that  gravitational interaction is caused by the non-metricity scalar ${\cal Q}$. In recent years, various observational data have been used to test the $f({\cal Q})$ gravity theory, as demonstrated in the research conducted by Lazkoz et al. \cite{Lazkoz:2019sjl}. The researchers used information from the growth rate,  Gamma-Ray Bursts, Type Ia Supernovae, Cosmic Microwave Background, Quasars, Baryon Acoustic Oscillations data and in order to constrain $f({\cal Q})$ gravity.  Furthermore, the impact of viscosity on cosmic acceleration has been examined within the framework of $f({\cal Q})$ gravity \cite{Solanki:2021qni,Solanki:2022ofw}. Furthermore, Mandal et al. confirmed the appropriateness of $f({\cal Q})$ in the domain of cosmology using the energy conditions \cite{Mandal:2020lyq}. In their study \cite{Mandal:2020lyq}, the researchers presented the embedding procedure, allowing for the incorporation of complex aspects from the absence of metricity in the energy requirement.  A recent research showed that $f({\cal Q})$ gravity has been effectively utilized in both Casimir WHs and the GUP-corrected Casimir WH \cite{Hassan:2022hcb,Hassan:2022ibc}.

Xu et al. \cite{Xu:2019sbp} expanded the $f({\cal Q})$ gravity to include $f({\cal Q},{\cal T})$ gravity, with  ${\cal T}$ representing the energy-momentum tensor trace. This trace is in charge of including additional quantum realm's contributions to traditional gravity. Arora et al. recently demonstrated the feasibility of the theory of  gravitation of  $f({\cal Q},{\cal T})$ in the domain of cosmology in relation to energy requirements  \cite{Arora:2020met}, while Tayde et al. \cite{Tayde:2023pjh} employed the matching condition to exam the stability of a slender shell encasing the WH and the potential it generates.

{ The flaring-out condition plays a vital role in WH physics, which necessarily violates the NEC of the matter fluid introducing an exotic matter where the GR theory is applied \cite{Rueda:2025rlj}. One way to possibly avoid the NEC violation while keeping the flaring-out condition is to modify the gravity.  For example $f(\mathcal{R})$ modified gravity theories, and also $f(T)$ teleparallel gravity which replaces the Ricci scalar by teleparallel torsion scalar $T$, allow WH solutions satisfying energy conditions but this could require specific forms of $f(\mathcal{R})$ or $f(T)$ \cite{Lobo:2009ip,Pavlovic:2014gba,Bhattacharya:2015oma,Boehmer:2012uyw}. Gauss-Bonnet gravity allows normal matter WHs in higher dimensions, whereas exotic matter is still required in 4D \cite{Bhawal:1992sz,Mehdizadeh:2015jra,Mehdizadeh:2021kgv}. Brans-Dicke theory allows WH solution where the additional scalar field coupled to gravity mimics exotic matter, but it requires the scalar field to be carefully tuned \cite{Nandi:1997mx}. Another approach to modify the GR gravity is by considering possible non-minimal coupling between matter and curvature \cite{MontelongoGarcia:2010xd,Moraes:2017mir}. Several WH solutions have been investigated within this framework using different approaches like $f(\mathcal{R},L_m)$, $f(\mathcal{R},\mathcal{T})$ and $f(\mathcal{R},L_m,\mathcal{T})$ where $L_m$ denotes the lagrangian matter density, c.f. \cite{Maurya:2024jos,Banerjee:2020uyi,Tayde:2025ncz}.}

{Static spherically symmetric WH configurations have been investigated within the context of $f(\mathcal{Q})$ gravity, assuming power-law and inverse power-law models \cite{Banerjee:2021mqk}. It has been shown that the NEC is violated if the matter density is positive. The present study aims to derive new WH solutions within the framework of $f(\mathcal{Q},\mathcal{T})$ gravity, where the coupling between matter and the non-metricity scalar introduces additional degrees of freedom. The analysis investigates how this coupling can sustain WH geometries that satisfy the flaring-out condition and asymptotic flatness without the need for exotic matter.}

During our study, we thoroughly analyze  a specific $f({\cal Q},{\cal T})\equiv{\cal Q}+\beta{\cal T}$ theory, where the WH models are computed by employing fuzzy dark matter which includes soliton quantum wave DM at the core of the halo while the outer regions follow NFW parameterization of the CDM. The ideas covered in this research are delineated in the subsequent sections as follows:
In Sec. \ref{SEC:II}, we provided an overview of the $f({\cal Q},{\cal T})$ gravity theory.
In Sec. \ref{SEC:III}, we discuss Morris-Thorne spacetime which describes WH models and the corresponding traversability conditions. We apply the field equations of $f({\cal Q},{\cal T})$ theory to the Morris-Thorne spacetime and their consequence on the energy conditions, with a particular focus on the linear $f({\cal Q},{\cal T})\equiv{\cal Q}+\beta{\cal T}$ theory.
In Sec. \ref{SEC:IV}, we present the matter, assumed to fill the WH, in this study. We discuss two density profiles corresponds to Soliton+NFW, where solitons quantum wave DM at the core of dwarf galaxies and the outer regions follow the NFW density profile which characterizes the CDM.
In Sec. \ref{SEC:V}, we derive the WH shape functions correspond to the two density profiles assumed in the present study. We investigate possible constraints on the coupling parameter $\beta$ from traversability conditions of WH geometry.
In Sec. \ref{SEC:VI}, we investigate additional possible constraints on the coupling parameter $\beta$ from the modified energy conditions corresponds to the linear $f({\cal Q},{\cal T})$ gravity.
{In Sec. \ref{SEC:VII}, we study the stability of the obtained WH solutions via a modified version of the TOV equation of hydrostatic equilibrium.}
In Sec. \ref{SEC:VIII}, we discuss the observational signatures which distinguish WHs from BHs. Also, we derive the basic equations which describe the photon trajectories around the WHs presented in this study.
The final section is devoted to summarize the results presented in this study and future work.
%%%%%%%%%%%%%%%%%%%%%%%%%%%%%%%%%%%%%%%%%%%%%%%%%%%%%%%%%%%%%%%%%%%%%%%%%%%%%%%%%%%%%%%%%%%%%%%%
\section{Bases of $f({\cal Q},{\cal T})$ gravitational theory}\label{SEC:II}
%%%%%%%%%%%%%%%%%%%%%%%%%%%%%%%%%%%%%%%%%%%%%%%%%%%%%%%%%%%%%%%%%%%%%%%%%%%%%%%%%%%%%%%%%%%%%%%%
In this section, we provide the main characteristics of a general $f({\cal Q},{\cal T})$ theory. We write the action \cite{Xu:2019sbp}
\begin{equation}\label{eq:action}
\mathcal{S}=\int\frac{1}{2\kappa^2}\,f({\cal Q},{\cal T})\sqrt{-g}\,d^4x+\int \mathcal{L}_m\,\sqrt{-g}\,d^4x,
\end{equation}
where $\kappa^2=8\pi G/c^4$, with $G$ being the gravitational constant and $c$ is the speed of light,  $g$ denotes the determinant of the metric tensor $g_{\mu\nu}$ and $\mathcal{L}_m$ is the matter Lagrangian density. The scalar $\mathcal{Q}$ is related to the non-metricity tensor,
\begin{equation}\label{eq:nonmetricity}
{\mathcal Q}_{\lambda\mu\nu}=\bigtriangledown_{\lambda} g_{\mu\nu},
\end{equation}
by the following relation \cite{BeltranJimenez:2017tkd}
\begin{equation}\label{eq:nonmetricity_scalar}
{\cal Q} = -{\cal Q}_{\alpha\mu\nu}\,P^{\alpha\mu\nu}\equiv  -g^{\mu\nu}\left(L^\beta_{\,\,\,\alpha\mu}\,L^\alpha_{\,\,\,\nu\beta}-L^\beta_{\,\,\,\alpha\beta}\,L^\alpha_{\,\,\,\mu\nu}\right)\,.
\end{equation}
The tensor $P$ is commonly referred to as the superpotential tensor
\begin{equation}\label{eq:superpotential}
P{^\alpha}{_{\mu\nu}}=\frac{1}{4}\left[2{\cal Q}{_{(\mu}}{^\alpha}{_{\nu)}}-{\cal Q}^\alpha\;_{\mu\nu}+{\cal Q}^\alpha g_{\mu\nu}-\tilde{{\cal Q}}^\alpha g_{\mu\nu}-\delta^\alpha_{(\mu}{\cal Q}_{\nu)}\right],
\end{equation}
and the tensor $L$ is known as disformation tensor
\begin{equation}\label{eq:disformation}
L{^\beta}{_{\mu\nu}}=\frac{1}{2}{\cal Q}{^\beta}{_{\mu\nu}}-{\cal Q}{{_{(\mu}}{^\beta}{_{\nu)}}}.
\end{equation}
In the above expressions, we used
\begin{equation}\label{eq:nonmetricity_vector}
{\cal Q}_{\alpha}={\cal Q}{{_\alpha}{^\mu}{_\mu}}, \qquad \tilde{{\cal Q}}_\alpha={\cal Q}{^\mu}{_{\alpha\mu}}.
\end{equation}
The variation of the action \eqref{eq:action} with respect to the metric tensor $g_{\mu\nu}$ gives the following field equations of $f({\cal Q},{\cal T})$ modified gravity as \cite{Xu:2019sbp,Arora:2020met}:
\begin{equation}\label{eq:fieldeqns1}
\frac{-2}{\sqrt{-g}}\bigtriangledown_\alpha\left(\sqrt{-g}\,f_{\cal Q}\,P^\alpha\;_{\mu\nu}\right)-\frac{1}{2}g_{\mu\nu}f \\
+f_{\cal T} \left({\cal T}_{\mu\nu} +\Theta_{\mu\nu}\right)\\
-f_{\cal Q}\left(P_{\mu\alpha\beta}\,{\cal Q}_\nu\;^{\alpha\beta}-2\,{\cal Q}^
{\alpha\beta}\,\,_{\mu}\,P_{\alpha\beta\nu}\right)=\kappa^2 {\cal T}_{\mu\nu}\,.
\end{equation}
Here $f_{\cal Q}$ refers to $f_{\cal Q}=\frac{\partial f}{\partial {\cal Q}}$ and $f_{\cal T}=\frac{\partial f}{\partial {\cal T}}$.  Additionally, the tensors $\Theta_{\mu\nu}$ and ${\cal T}_{\mu\nu}$ are defined as:
\begin{equation}\label{eq:stress_tensor1}
\Theta_{\mu\nu}=g^{\alpha\beta}\frac{\delta {\cal T}_{\alpha\beta}}{\delta g^{\mu\nu}}\,, \qquad \qquad
{\cal T}_{\mu\nu}=-\frac{2}{\sqrt{-g}}\frac{\delta\left(\sqrt{-g}\,\mathcal{L}_m\right)}{\delta g^{\mu\nu}}\,.
\end{equation}
Remarkably $f({\cal Q},{\cal T})$ gravity theory has attracted interest in the last few years from several aspects. In cosmological applications, it has been shown that the theory can account for late accelerated expansion without the need for dark energy \cite{Myrzakulov:2024esv}. Viability of $f({\cal Q},{\cal T})$ gravity has been tested with cosmological observations \cite{Hazarika:2024alm}. On the other hand, in the realm of astrophysics, stability of compact stellar models within $f({\cal Q},{\cal T})$ gravity have been investigated, where a compact star heavier than the GR predictions can be obtained \cite{Kausar:2025uqm}. Additionally, the impact of anisotropic matter and electromagnetism on the compactness and sound speed limits, within stars, has been studied \cite{Pradhan:2024rpp,Das:2024ytl}.
%%%%%%%%%%%%%%%%%%%%%%%%%%%%%%%%%%%%%%%%%%%%%%%%%%%%%%%%%%%%%%%%%%%%%%%%%%%%%%%%%%%%%%%%%%%%%%%%
\section{Study of wormholes in $f({\cal Q},{\cal T})$ gravitational theory}\label{SEC:III}
%%%%%%%%%%%%%%%%%%%%%%%%%%%%%%%%%%%%%%%%%%%%%%%%%%%%%%%%%%%%%%%%%%%%%%%%%%%%%%%%%%%%%%%%%%%%%%%%
In this section, we discuss two unknown functions which characterize WH solutions, those are the redshift and the shape functions.  Additionally, we discuss the constraints which govern those functions in order to obtain a traversable WH. Next, we derive the corresponding field equations of a general $f(\cal{Q}, \cal{T})$ form in presence of anisotropic fluid. We show possible implications of $f(\cal{Q}, \cal{T})$ modifications on the energy conditions which play essential roles in WH solutions. Finally, we discuss a particular theory with linear behavior, i.e. $f(\cal{Q}, \cal{T})=\cal{Q}+\beta \cal{T}$, which represents a simple case of non-minimal coupling between matter and geometry.
%%%%%%%%%%%%%%%%%%%%%%%%%%%%%%%%%%%%%%%%%%%%%%%%%%%%%%%%%%%%%%
\subsection{Traversable wormhole spacetime configuration}\label{Sec:travWH}
To discuss the general aspects of WH models in $f({\cal Q},{\cal T})$ theory, we will examine a spherically symmetric WH metric.  This analysis follows the influential researches presented in \cite{Visser:1995cc,Morris:1988cz}, where the line element is considered as the Morris-Thorne WH metric
\begin{equation}\label{eq:MTWHmetric}
ds^2=e^{2\xi(r)}dt^2-\left(1-\frac{h(r)}{r}\right)^{-1}dr^2-r^2\,d\theta^2-r^2\,\sin^2\theta\,d\Phi^2\,.
\end{equation}
Here $h(r)$ and $\xi(r)$ represent the shape and redshift functions, respectively. We believe that the models of WH in $f({\cal Q},{\cal T})$ can be consistent within Birkhoff's theorem because the  line element \eqref{eq:MTWHmetric} in presented in  polar form $(t,r,\theta,\Phi)$. The hypothesis about the applicability of Birkhoff's theorem is supported  by the study presented in \cite{Meng:2011ne} as well as the recent review by Bahamond et al. \cite{Bahamonde:2021gfp}.
The relevance of the Birkhoff theorem in the realm of teleparallel gravity was discussed in \cite{Meng:2011ne}. Short time ago, Bahamonde et al. assessed the relevance of Birkhoff's theorem in an extended version of teleparallel gravity, including influences from a boundary term and scalar fields.  In their study \cite{Bahamonde:2021gfp}, they demonstrated that Birkhoff's theorem is limited solely if the scalar fields are based on the variables $t$ or $r$ coordinate. In Eq. (5.49) presented in \cite{Bahamonde:2021gfp}, the authors  demonstrated a mapping correlation between the  non-metricity and torsion  scalars.  This correlation, combined with the considerations of Birkhoff's theorem in the frame of teleparallel theory, backs up the hypothesis about the applicability of this theorem for spherically symmetric  solutions within  $f({\cal Q},{\cal T})$ theory.

When looking for a traversable WH model, specific conditions on the redshift and shape function must be fulfilled
\cite{Visser:1995cc,Morris:1988cz}:
\begin{itemize}
  \item[(i)] In general the shape function $h(r)<r$, whereas $h(r_0)=r_0$ at a minimal radius (the WH throat), known as the throat condition.
  \item[(ii)] The function $h(r)$ must satisfy the flaring out condition $\frac{h-rh'}{2h^2}>0$, where $h'(r_0)<1$ at the WH throat.
  \item[(iii)] The WH must be asymptotically flat condition, i.e. the ratio $\frac{h(r)}{r}\to 0$ as $r\to \infty$.
  \item[(iv)] The redshift function $\xi(r)$ must have a finite value at all points.
\end{itemize}
%%%%%%%%%%%%%%%%%%%%%%%%%%%%%%%%%%%%%%%%%%%%%%%%%%%%%%%%%%%%
\subsection{The field equations}\label{Sec:fieldeqns}
In this study, we consider that the matter of WH solutions is defined by a non-uniform energy-momentum tensor supplied by:
\begin{equation}\label{eq:aniso_fluid}
T_{\mu}^{\nu}=\left(\sigma c^2+p_\theta \right)U_{\mu}\,U^{\nu}-p_\theta\,\delta_{\mu}^{\nu}+\left(p_r-p_\theta\right)V_{\mu}\,V^{\nu}\,.
\end{equation}
The vectors $U_{\mu}$ and $V_{\mu}$ denote the 4-velocity vector and space-like unit vector, where $U_{\mu}U^{\nu}=-V_{\mu}V^{\nu}=1$. In addition, the energy density, $\sigma$, the radial pressure, $p_r$, and the tangential pressure, $p_\theta$, all are functions of the radial coordinate $r$. The trace of the stress-energy tensor \eqref{eq:aniso_fluid} is given as follows $T=\sigma c^2-p_r-2p_\theta$. We note that Letelier introduced the anisotropic energy-momentum tensor as a means to explore a theoretical framework in plasma physics that deals with two distinct fluids \cite{Letelier:1980mxb}. Additionally, it  has been used in a variety of scenarios to mimic magnetized neutron stars \cite{Deb:2021ftm}.

Now we are ready to replicate the methods used in  \cite{Moraes:2016akv} to discover solutions with  static WH in $f({\cal Q},{\cal T})$. In this study we choose $\mathcal{L}_m=-P$, where the average pressure $P\equiv \frac{p_r+2\,p_\theta}{3}$. This Lagrangian reads Eq. \eqref{eq:stress_tensor1} as follows
\begin{equation}
    \Theta_{\mu\nu}=-g_{\mu\nu}\,P-2\,T_{\mu\nu}.
\end{equation}
Furthermore, for the spacetime metric \eqref{eq:MTWHmetric}, the scalar ${\cal Q}$ is explicitly expressed as \cite{Tayde:2022lxd}
\begin{equation}\label{eq:Qscalar}
{\cal Q}=-\frac{h}{r^2}\left[\frac{rh^{'}-h}{r(r-h)}+2\xi^{'}\right].
\end{equation}
By virtue of Eqs. \eqref{eq:aniso_fluid} and \eqref{eq:Qscalar}, we write the field  equations \eqref{eq:fieldeqns1} as follows \cite{Tayde:2022vbn}
\begin{align}
&\frac{2 (r-h)}{(2 r-h) f_{\cal Q}}\left[\sigma c^2-\frac{(r-h)}{\kappa^2  r^3} \left(\frac{h r f_{\cal QQ} {\cal Q}'}{r-h}+h f_{\cal Q} \left(\frac{ r \xi '+1}{r-h}-\frac{2 r-h}{2 (r-h)^2}\right)+\frac{f r^3}{2 (r-h)}\right)+\frac{f_{\cal T} (P+\sigma )}{\kappa^2 }\right]=\frac{h'}{\kappa^2  r^2}, \label{eq:fieldeq1}\\
&\frac{2 h}{f r^3}\left[p_r +\frac{(r-h)}{2\kappa^2 r^3} \left(f_{\cal Q} \left(\frac{h \left(\frac{r h'-h}{r-h}+2 r \xi '+2\right)}{r-h}-4 r h '\right)+\frac{2 h r f_{\cal{QQ}} {\cal Q}'}{r-h}\right)+\frac{fr^3 (r-h)\xi'}{\kappa^2 h  r^2}-\frac{f_{\cal T} \left(P-p_r\right)}{\kappa^2 }\right]\nonumber\\
&=\frac{1}{\kappa^2}\left[2\left(1-\frac{h}{r}\right)\frac{\xi '}{r}-\frac{h}{r^3}\right],  \label{eq:fieldeq2}\\
&\frac{1}{f_{\cal Q} \left(\frac{r}{r-h}+r \xi '\right)}\left[p_\theta +\frac{(r-h)}{4\kappa^2 r^2}\left(f_{\cal Q} \left(\frac{4 (2 h-r) \xi '}{r-h}-4 r \left(\xi'\right)^2-4 r \xi''\right)+\frac{2 f r^2}{r-b}-4 r f_{\cal{QQ}} Q' \xi'\right) -\frac{f_{\cal T} \left(P-p_\theta\right)}{\kappa^2 }\right.\nonumber\\
&\left.
+\frac{(r-h)}{\kappa^2 r}\left(\xi '' +{\xi '}^2-\frac{(rh'-h)\xi '}{2r(r-h)}+\frac{\xi '}{r}\right)f_{\cal Q} \left(\frac{r}{r-h}+r \xi '\right)\right]=\frac{1}{\kappa^2}\left(1-\frac{h}{r}\right)\left[\xi '' +{\xi '}^2-\frac{(rh'-h)\xi '}{2r(r-h)}-\frac{rh'-h}{2r^2 (r-h)}+\frac{\xi '}{r}\right]\,.\label{eq:fieldeq3}
\end{align}
The above equations are the non-vanishing components of the equations of motion of $f({\cal Q},{\cal T})$ theory \cite{Tayde:2022lxd}.  {Those allow to write the density and pressures of the matter fluid as given in Appendix \ref{appA}, namely Eqs. \eqref{eq:density1}--\eqref{eq:tpress1}}.
%%%%%%%%%%%%%%%%%%%%%%%%%%%%%%%%%%%%%%%%%%%5
\subsection{The energy conditions of wormhole with constant redshift in $f(\mathcal{Q}, \mathcal{T})$ gravity}\label{Sec:EC1}

{ In the GR theory, the focusing theorem and Raychaudhuri equation set direct constraints on the matter fluid known as the energy conditions. However, in modified gravity these conditions should be extended to the effective fluid as will be discussed in some details in Sec. \ref{SEC:VI}. In this sense, we derive the effective density and effective pressures for a general $f(\mathcal{Q}, \mathcal{T})$ theory in Appendix \ref{appA}}. For the particular case of a constant redshift function $\xi(r) = \xi_0$, the effective density and pressures, %{ are  presented in Appendix \ref{appA},
namely Eqs. \eqref{eq:eff_dens2}, \eqref{eq:eff_rpress2} and \eqref{eq:eff_tpress2}, reduce to
\begin{equation}\label{eq:eff_dens3}
\tilde{\sigma} c^2=\frac{2 (r-h) }{(2 r-h) f_{\cal Q}}\left[\sigma c^2-\frac{\left(1-\frac{h}{r}\right) \left\{\frac{h r f_{\cal{QQ}} {\cal Q}'}{r-h}+\frac{h f_{\cal Q}}{r-h}-\frac{h (2 r-h) f_{\cal Q}}{2 (r-h)^2}+\frac{f r^3}{2 (r-h)}\right\}}{\kappa^2  r^2}+\frac{f_{\cal T} (P+\sigma c^2)}{\kappa^2 }\right]\,,
\end{equation}
\begin{equation}\label{eq:eff_rpress3}
\tilde{p_r}=\frac{2 h }{f r^3}\left[p_r- \frac{f_{\cal T} \left(P-p_r\right)}{\kappa^2 }+\frac{\left(1-\frac{h}{r}\right) \left\{\frac{h f_{\cal Q} \left(\frac{r h'-h}{r-b}+2\right)}{r-h}+\frac{2 h r f_{\cal{QQ}} {\cal Q}'}{r-h}\right\}}{2\kappa^2  r^2}\right]\,,
\end{equation}
\begin{equation}\label{eq:eff_tpress3}
\tilde{p_\theta}=\frac{(r-h)}{r f_{\cal Q}}\left[p_\theta-\frac{f_{\cal T} \left(P-p_\theta\right)}{\kappa^2 }+ \frac{f r \left(1-\frac{h}{r}\right)}{2\kappa^2  (r-h)}\right]\,.
\end{equation}

A couple of lines will be devoted to standard energy conditions, which are based on the Raychaudhuri equations. These equations allow one to explain how gravity affects the convergence and divergence of time-like, spacelike, and light-like curves. The Raychaudhuri equations establish limitations on the density and pressures of a WH, described by Arora et al. \cite{Arora:2020met} as:\\
$\bullet$ The weak energy condition (WEC) is satisfied whenever $\tilde{\sigma}\geq 0$ and $\tilde{\sigma} c^2+\tilde{p_j}\geq0$ $\forall j$ with $j=r$ or $\theta$.\\
$\bullet$ If the sum of the energy density and pressure in all directions is non-negative, then the Null energy condition (NEC) holds.\\
$\bullet$ If $\tilde{\sigma}\geq0$ and $\tilde{\sigma} c^2-|\tilde{p_j}|\geq0$ $\forall j$ with $j=r$ or $\theta$, then the Dominant energy condition (DEC) is satisfied.\\
$\bullet$ Strong energy condition (SEC) is satisfied whenever $\tilde{\sigma} c^2+\tilde{p_r}\geq0$, $\tilde{\sigma} c^2+\tilde{p_\theta}\geq0$ and $\tilde{\sigma} c^2+\tilde{p_r}+\tilde{p_\theta}\geq0$. { In Appendix \ref{appB}, we explicitly derive the above conditions ensuring the verification of these conditions for the system of Eqs.~\eqref{eq:eff_dens3}--\eqref{eq:eff_tpress3}.}
%%%%%%%%%%%%%%%%%%%%%%%%%%%%%%%%%%%%%%%%%%%%%%%%%%%%%%%%%%%%
\subsection{A particular $f(\mathcal{Q}, \mathcal{T})$ theory}\label{Sec:linearfQT}
To determine if traversable WH solutions are feasible, we will analyze the function $f({\cal Q},{\cal T})$ in the following way \cite{Xu:2019sbp}:
\begin{equation}\label{eq:linearfQT}
f({\cal Q},{\cal T})={\cal Q}+\beta\,{\cal T}.
\end{equation}
where $\beta$ is a dimensional parameter with a dimension [N$^{-1}$] similar to $\kappa^2$. Therefore, in the following we use the transformation $\beta\to \beta \kappa^2$ where $\beta$ is a dimensionless parameter. By utilizing the previously mentioned  formula of $f({\cal Q},{\cal T})$, assuming a constant redshift function $\xi(r)=\xi_0$, in Eqs.~\eqref{eq:density1}-\eqref{eq:rpress1}, it yields
 \begin{align}
 &\sigma =\frac{ (3+8\beta ) h'}{3 (1+2\beta)(1+4\beta)\kappa^2 c^2 r^2}\,,\label{eq:lineardensity}\\
 &p_r=-\frac{3(1+4\beta)h-4\beta r h'}{3(1+2\beta)(1+4\beta)\kappa^2 r^3}\label{eq:linearrpress}\,,\\
 &p_\theta=\frac{3(1+4\beta)h-(3+4\beta)rh'}{6(1+2\beta)(1+4\beta)\kappa^2 r^3}\label{eq:lineartpress}\,.
 \end{align}
For linear $f(\cal{Q}, \cal{T})$, we write the effective density and pressures in terms on the matter density and pressures as follows:
\begin{align}
 &\tilde{\sigma} c^2 = \sigma c^2+\beta(3\sigma c^2-p_r/3-2p_\theta/3),\label{eq:lineareffdensity}\\
&\tilde{p}_r=p_r-\beta(\sigma c^2-7 p_r/3-2 p_\theta/3),\label{eq:lineareffrpress}\\
&\tilde{p}_\theta=p_\theta-\beta(\sigma c^2-p_r/3-8 p_\theta/3)\label{eq:linearefftpress}.
\end{align}
Clearly the GR equations are recovered by setting $\beta=0$. The above equations enable us to, alternatively, write the matter density and pressure in terms of the effective ones. These relations are important for determining the validity of the energy conditions within any modified gravity in principle. This will be discussed in details in Sec. \ref{SEC:VI}.
%%%%%%%%%%%%%%%%%%%%%%%%%%%%%%%%%%%%%%%%%%%%%%%%%%%%%%%%%%%%%%%%%%%%%%%%%%%%%%%%%%%%%%%%%%%%%%%%
\section{Dark matter density profiles}\label{SEC:IV}
%%%%%%%%%%%%%%%%%%%%%%%%%%%%%%%%%%%%%%%%%%%%%%%%%%%%%%%%%%%%%%%%%%%%%%%%%%%%%%%%%%%%%%%%%%%%%%%%
It has been shown that DM at the galactic halo could be consistent with WHs structure \cite{Rahaman:2013xoa, Sarkar:2024klp} as well as at the central region of the halo \cite{Rahaman:2014pba}. In the present study, we use Solition+NFW DM density profiles to derive the corresponding shape functions, where soliton core model is applied at the central region of the DM halo and outer halo follows the NFW density profile which characterizes the CDM. This scenario is motivated by the famous core-cusp problem where N-body CDM simulations predict a cuspy density profile as $\sigma(r)\propto 1/r$ at small radii, while rotation curves of dwarf galaxies, where DM is dominating, point out a flat density profile at their cores. Another puzzle is known as the missing satellites problem, where CDM simulations predict more low-mass galaxies in the local group than the already observed. In this sense, the central masses are too low compared to the most massive (sub)halos predicted in $\Lambda$CDM.

Recent simulations of solition+NFW DM scenario show that the galactic halos surround a dense core of dwarf spheroidal galaxies with a transition between the soliton core--characterized by a flat density profile--and the CDM halo at a radius of $\simeq 1.0$ kpc \cite{Pozo:2020ukk}; and also dwarf galaxies \cite{Banares-Hernandez:2023axy}. Therefore, the soliton+NFW model provides a good candidate to solve small radii problems related to $\Lambda$CDM scenario. We use the numerical values of the model parameters as obtained for the dwarf galaxy NGS 2366, using fuzzy DM (soliton+NFW) simulation, by the recent analysis \cite{Banares-Hernandez:2023axy}, based on the rotation curves of the LITTLE THINGS in 3D catalog \cite{2017MNRAS.466.4159I}. In the following, we are going to briefly discuss the two density profiles mentioned above.
%%%%%%%%%%%%%%%%%%%%%%%%%%%%%%%%%%%%%%%%%%%%%%%%%%%%%%%%
\subsection{Soliton quantum wave dark matter}\label{Sec:sqwDM}

Axions and other ultralight bosons, with masses $m_b \sim 10^{-23}-10^{-21}$ eV, a well-known contender to address the aforementioned issues. At cosmological scales, these particles are consistent with the CDM. However, these particles behave as self-gravitating DM waves and populate the galactic halos with significant occupation numbers at distances comparable to their de Broglie wavelength, which can be in the kpc scale. The production of a flat core ``soliton" at the center of galaxies with a rather marked transition to a less dense outer region that follows a CDM-like distribution is one of the repercussions of this, as it appears to have a pressure-like impact on macroscopic scales.

Assuming the simple case of ultralight bosons, when self-interaction is ignored, then the boson mass is the only free parameter. If the corresponding de Broglie wavelength exceeds the mean free path set by the density of dark matter, these bosons can satisfy the ground state condition for a Bose-Einstein condensate described by the coupled Schr\"{o}dinger-Poisson equation. This can be written in comoving coordinates as
\begin{align*}
    &\left[i \frac{\partial}{\partial \tau}+\frac{1}{2}\nabla^2-a V\right]\psi=0,\\
    &\nabla^2 V=4\pi (|\psi|^2-1),
\end{align*}
where $\psi$ is the wave function, $V$ is the gravitation potential and $a$ is the cosmological scale factor. The fitting formula for the density profile of the solitonic core in a $\psi$DM halo is obtained from cosmological simulations \cite{Schive:2014dra,Schive:2014hza}:
\begin{equation} \label{eq:solitondensprof}
    \sigma_\text{sol}(r) = \frac{\sigma_c}{\left[1+\alpha \left(r/r_c\right)^2 \right]^{8}}\,,
\end{equation}
here $\sigma_c$ and $r_c$ stand for the central density and size of the soliton core \cite{Herrera-Martin:2017cux}. The leading study thoroughly examines the distribution of the matter that was previously mentioned~\cite{Schive:2014hza}. The precise calculation for the half-density radius will be a specific radius that is determined as a constant $\alpha = \sqrt[8]{2}-1 \sim 0.09051$ \cite{Schive:2014dra,Schive:2014hza}. Furthermore, the value of $\sigma_c$ in Eq.~\eqref{eq:solitondensprof} is specified as \cite{Herrera-Martin:2017cux}:
\begin{equation} \label{e2}
    \sigma_c = 2.4\times 10^{12} \left(\frac{m_b}{10^{-22}\text{eV}}\right)^{-2}  \left(\frac{r_c}{\text{pc}}\right)^{-4}\frac{M_\odot}{\text{pc}^{3}}.
\end{equation}
We use the numerical values of the model parameters, $r_c$ and $\rho_c$, using fuzzy DM simulation \cite{Banares-Hernandez:2023axy}, based on the rotation curves of the LITTLE THINGS in 3D catalog \cite{2017MNRAS.466.4159I}. For the dwarf galaxy NGC 2366, the core radius $r_c=3$ kpc and the central density $\sigma_c = 15\times 10^{-3}~\text{M}_{\odot}/\text{pc}^3$ \cite{Banares-Hernandez:2023axy}.
%%%%%%%%%%%%%%%%%%%%%%%%%%%%%%%%%%%%%%%%%%%%%%%%%%%%%%%%%%
\subsection{Cold dark matter halo}\label{Sec:NFWDM}
We assume that the CDM distribution in the halo obeys NFW profile, which is extensively used in $N$-body simulation, therefore we write the NFW density profile \cite{Navarro:1996gj}:
\begin{equation}\label{eq:NFWdensprof}
\sigma_\text{NFW}(r) = \frac{\sigma_s}{(r/r_s)(1+r/r_s)^2}\,.
\end{equation}
In the above, $r_s$ stands for the characteristic scale radius and $\sigma_s$ is the corresponding scale density. The NFW halo density clearly goes as $\sigma_\text{NFW} \propto 1/r$ at small radii, while at large radii it goes as $\sigma_\text{NFW} \propto 1/r^{3}$. The characteristic scale density is related to the concentration parameter $c=r_\text{vir}/r_s$, with $r_\text{vir}$ denotes the virial radius\footnote{The virial radius is defined as the radius where the average density falls to a critical threshold $\zeta(z)\sigma_{m0}$.}, and the cosmological matter density parameter $\Omega_{m}(z)$ as follows
\begin{align*}
    \sigma_s&=\frac{\zeta(z)\sigma_{m0}}{3}\frac{c^3}{\log(1+c)-\frac{c}{1+c}},\\
    \zeta(z)&=\frac{18\pi^2+82(\Omega_m(z)-1)-39(\Omega_m(z)-1)^2}{\Omega_m(z)},
\end{align*}
where $z$ denotes the redshift, and $\sigma_{m0}$ is the current cosmological matter density, related to the critical density $\sigma_{c0}=3H_0^2/8G$ with Hubble constant $H_0$, by the following relation $\sigma_{m0}=\Omega_{m0}\sigma_{c0}$. Following the calculations as given in \cite{Banares-Hernandez:2023axy}, we set $h=H_0/100=0.678$ and $\Omega_{m0}=0.308$ as measured by Planck 2015. For the dwarf galaxy NGC 2366, we use a virial radius $r_\text{vir}=5.5$ kpc as estimated for dwarf galaxies, where the concentration parameter for the NFW+baryons model is given by $c=0.80^{+0.64}_{-0.35}$ \cite{Banares-Hernandez:2023axy}, then we calculate the scale radius ${r_s} = 1.447$ kpc and the scale density $\sigma_s=3.11 \times 10^{-3}~\text{M}_{\odot}/\text{pc}^3$.
%%%%%%%%%%%%%%%%%%%%%%%%%%%%%%%%%%%%%%%%%%%%%%%%%%%%%%%%%%%%%%%%%%%%%%%%%%%%%%%%%%%%%%%%%%%%%%%%
\section{Examining dark matter wormholes}\label{SEC:V}
%%%%%%%%%%%%%%%%%%%%%%%%%%%%%%%%%%%%%%%%%%%%%%%%%%%%%%%%%%%%%%%%%%%%%%%%%%%%%%%%%%%%%%%%%%%%%%%%
In this section, we utilize the dark matter density profiles discussed above, namely Eqs. \eqref{eq:solitondensprof} and \eqref{eq:NFWdensprof}, checking their viability to construct WHs. This can be done by deriving the corresponding shape function, within linear $f(\mathcal{Q},\mathcal{T})$ theory, by virtue of Eq. \eqref{eq:lineardensity}. Then, we examine possible constraints on the coupling strength between matter and geometry, $\beta$, in order to satisfy traversable WH conditions discussed in Sec. \ref{Sec:travWH}.

%%%%%%%%%%%%%%%%%%%%%%%%%%%%%%%%%%%%%%%%%%%%
\subsection{Model I}\label{sec:WHmodel1}
%%%%%%%%%%%%%%%%%%%%%%%%%%%%%%%%%%%%%%%%%%%%
The exact shape function of a solitonic DM can be calculated by plugging the density profile \eqref{eq:solitondensprof} into Eq. \eqref{eq:lineardensity}:
\begin{equation}\label{eq:soliton_h}
    h(r)=r_0+A\, \left[\arctan(\sqrt{\alpha}r/r_c)-\arctan(\sqrt{\alpha}r_0/r_c) \right]+B\, \frac{r-r_0}{(r_c^2+\alpha r^2)^7}\mathcal{F}(r).
\end{equation}
The coefficients $A$ and $B$ are given as
\begin{eqnarray*}
    A&=&\frac {99 \sigma_s c^2 \kappa^2 (1+2\beta)(1+4\beta) r_c^3}{2048\alpha\sqrt{\alpha}(3+8\beta)},\\
    B&=&-\frac {99 \sigma_s c^2 \kappa^2 (1+2\beta)(1+4\beta) r_c^4}{2048\alpha(3+8\beta)(r_c^2+\alpha r_0^2)^7},
\end{eqnarray*}
where $\beta\neq -3/8$, { the explicit form of the complementary function $\mathcal{F}(r)$ is given in Appendix \ref{appC}.}

\begin{figure}
\centering
\subfigure[~The parameter space $\{r_0,\delta\}$]{\label{fig:space1}\includegraphics[width=.25\textwidth]{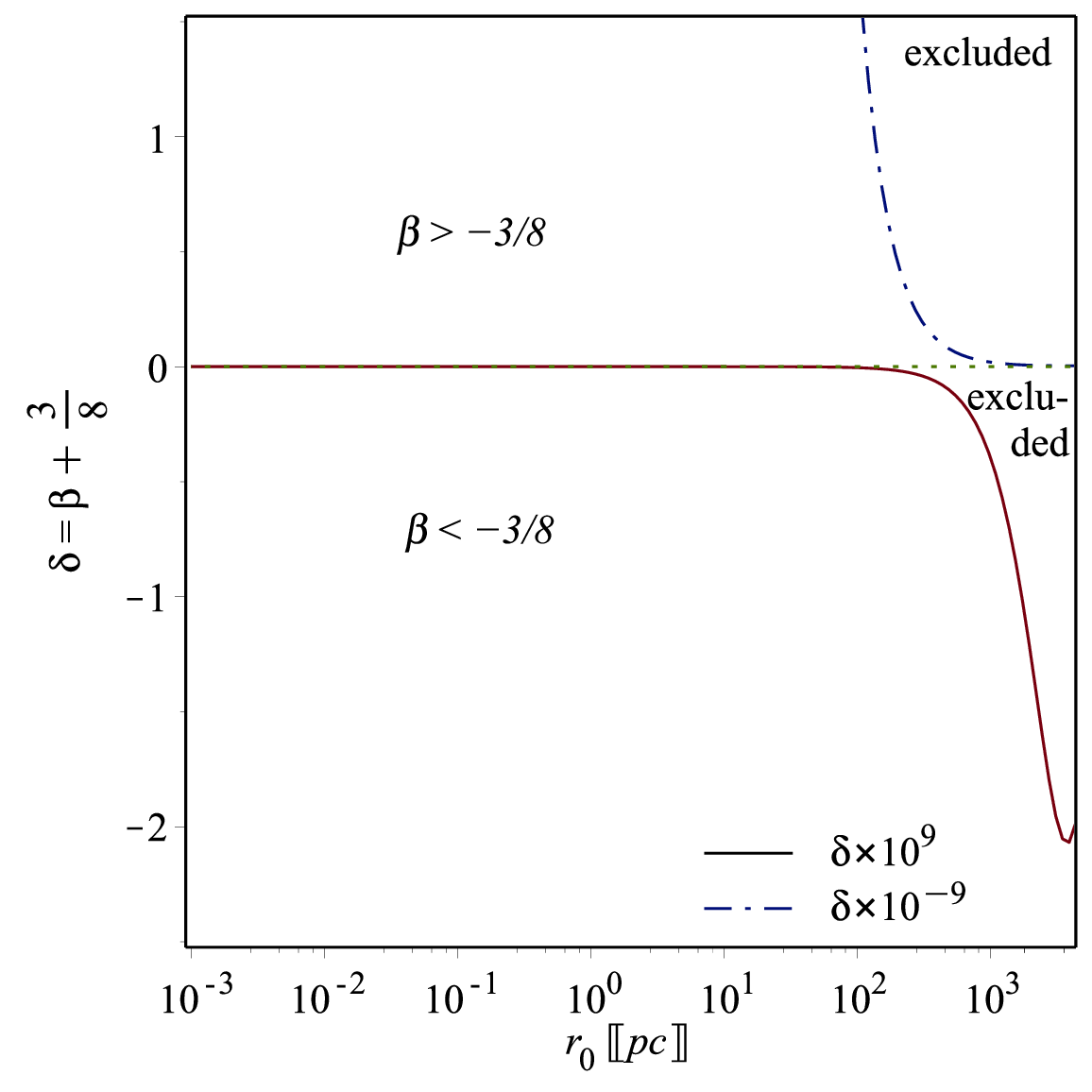}}\hspace{1cm}
\subfigure[~Flaring-out condition at throat]{\label{fig:flaring1}\includegraphics[width=.35\textwidth]{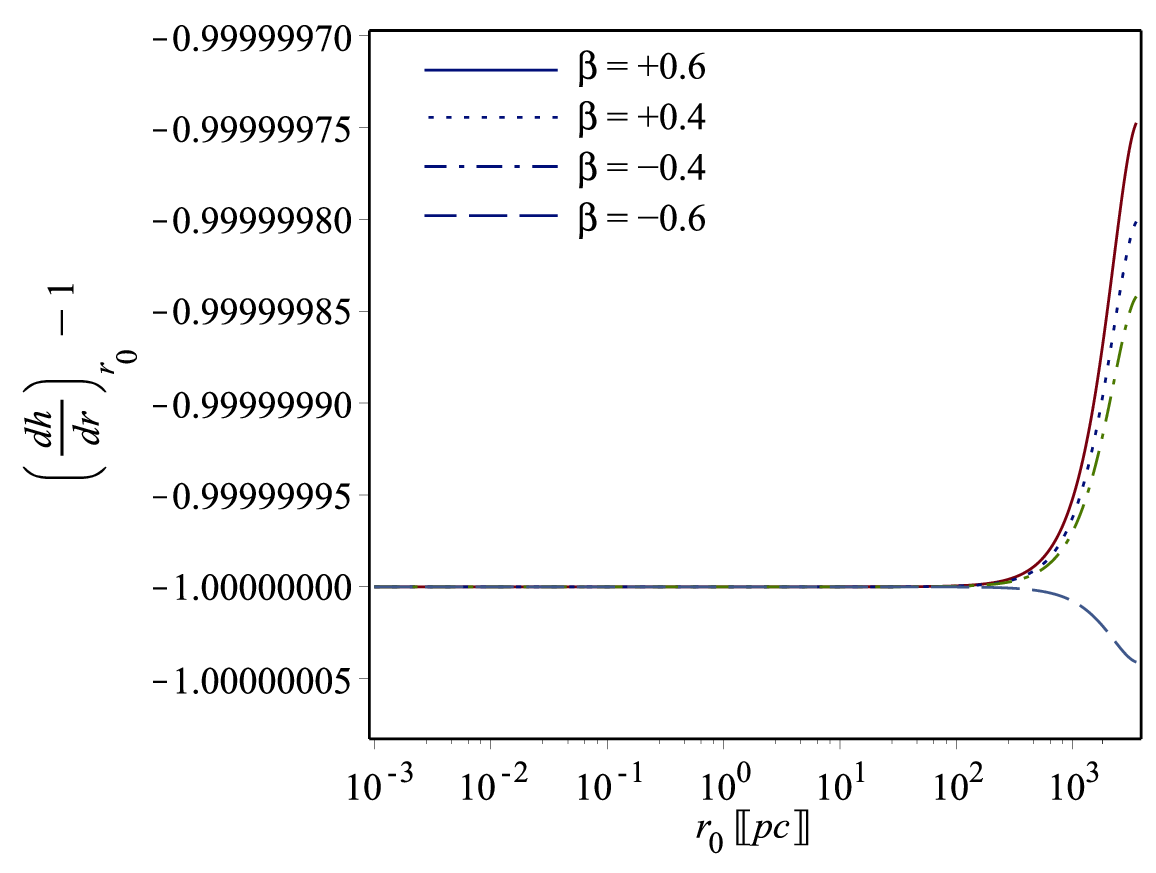}}
\caption{Model I: \subref{fig:space1} The parameter space $\{r_0,\delta\}$ of the soliton DM core WH model of galaxy NGC 2366 ($\sigma_c=15 \times 10^{-3}~M_\odot$/pc$^{3}$ and $r_c=3$ kpc \cite{Banares-Hernandez:2023axy}). We use the faring out constraint on the shape function at the WH throat $h'(r_0)<1$ to evaluate critical values of the non-minimal coupling parameter $\beta$, namely \eqref{eq:model1_beta1} and \eqref{eq:model1_beta2}, at arbitrary throat size $0<r_0<r_c$ where $r_c$ is the soliton core radius of galaxy NGC 2366. Given that $\beta\neq -3/8$, we show both cases when $\beta<-3/8$ ($\delta<0$) and $\beta>-3/8$ ($\delta>0$). The positive (negative) $\delta$ curve is multiplied by a factor $10^{-9}$ ($10^{9}$) to fit the curve into the scale of the graph. \subref{fig:flaring1} The flaring-out condition at the WH throat is satisfied for different selected values of the parameter $\beta$ where the WH throat is arbitrarily chosen $0<r_0<r_c$, it shows that $h'(r_0)<1$ is fulfilled.}
\label{Fig:Model1_beta}
\end{figure}

Obviously, at the WH throat, one obtains $h(r_0)=r_0$ as required by the throat condition of a traversable WH. Additionally, we apply the flaring-out condition at the the WH throat, $h'(r_0)<1$, which gives
\begin{equation}\label{eq:flaring_throat}
    \frac{3\kappa^2 \sigma_c c^2 r_0^2 r_c^{16}(1+2\beta)(1+4\beta)}{(3+8\beta)(r_c^2+\alpha r_0^2)^8}<1.
\end{equation}
The above inequality sets the following constraints
\begin{eqnarray}
    &\beta&<-\frac{3}{8}+\frac{4(r_c^2+\alpha r_0^2)^{8}-\sqrt{16(r_c^2+\alpha r_0^2)^{16}+9\kappa^4 \sigma_c^2 c^4 r_c^{32} r_0^{4}}}{24\k^2 \sigma_c c^2 r_c^{16} r_0^{2}},~\text{or} \label{eq:model1_beta1}\\
    -3/8<&\beta&<-\frac{3}{8}+\frac{4(r_c^2+\alpha r_0^2)^{8}+\sqrt{16(r_c^2+\alpha r_0^2)^{16}+9\kappa^4 \sigma_c^2 c^4 r_c^{32} r_0^{4}}}{24\k^2 \sigma_c c^2 r_c{16} r_0^{2}}.\label{eq:model1_beta2}
\end{eqnarray}
The obtained intervals show the dependence of the parameter $\beta$ on the WH size $r_0$. For example, \textit{assuming} $r_0=1$ pc, the flaring out condition is satisfied when $\beta<-3/8-8.494 \times 10^{-16}$ or $-3/8<\beta< -3/8+1.840 \times 10^{13}$. However, for arbitrary values of $0<r_0<r_c$ one can set similar constraints on the non-minimal coupling parameter $\beta$ satisfying $h'(r_0)<1$. We show the parameter space $\{r_0,\delta\}$ graphically as seen in Fig. \ref{Fig:Model1_beta}\subref{fig:space1} by introducing the deviation factor, $\delta\equiv \beta+3/8$, which measures the deviation from the restricted value $\beta=-3/8$. For $\delta>0$, the non-minimal coupling parameter becomes large as $r_0\to 0$ and smaller as $r\to r_c$. For $\delta<0$ the deviation factor is tiny at $0<r_0<r_c$, where $\delta\sim -2\times 10^{-9}$ as shown by the figure. We note that the excluded values in the parameter space $\{r_0,\delta\}$ must be considered when choosing the WH throat size $r_0$ to ensure its consistency with the flaring-out condition, see Fig. \ref{Fig:Model1_beta}\subref{fig:flaring1}. The figure shows that the flaring-out condition at the throat, $(dh/dr)_{r_0}<1$, is satisfied for different selected values of the parameter $\beta$.

\begin{figure}
\centering
\subfigure[~Shape function]{\label{fig:shapefn1}\includegraphics[width=.22\textwidth]{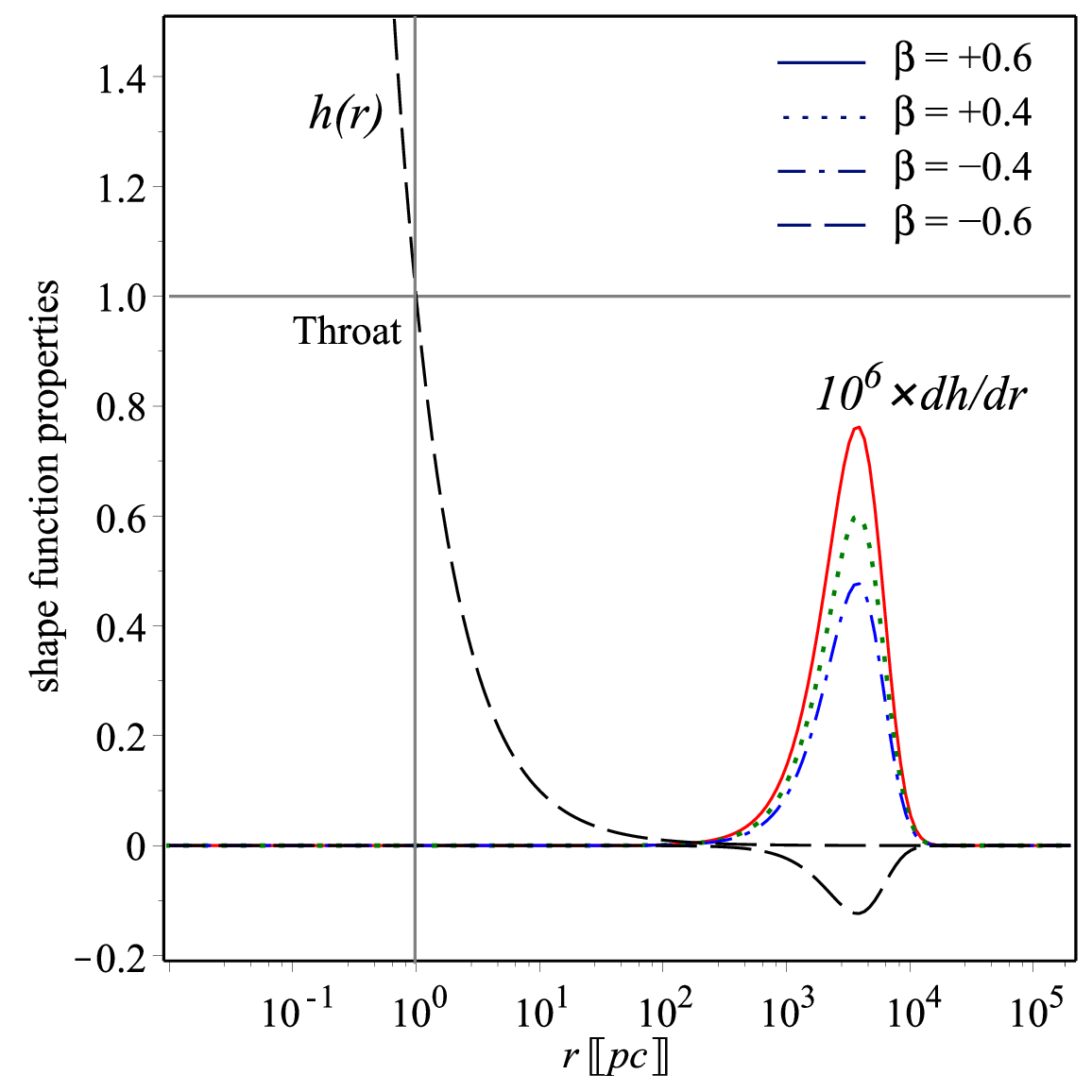}}\hspace{0.5cm}
\subfigure[~2D embedding]{\label{fig:2demedding1}\includegraphics[width=.22\textwidth]{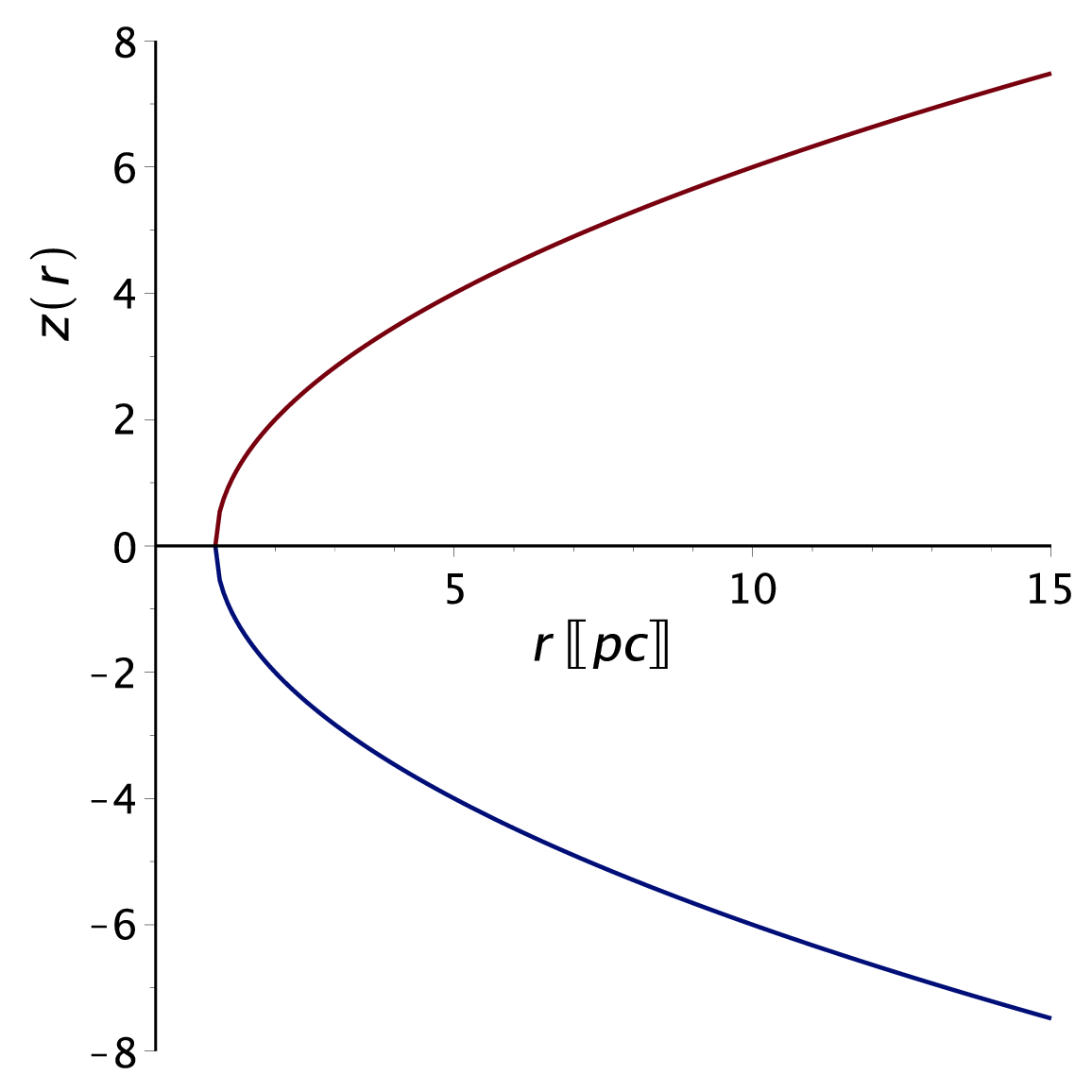}}\hspace{0.5cm}
\subfigure[~3D embedding]{\label{fig:3demedding1}\includegraphics[width=.22\textwidth]{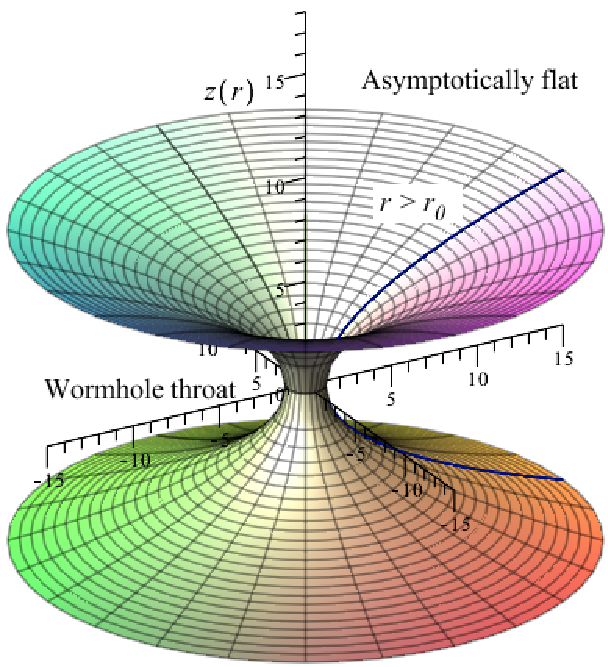}}
\caption{Model I: WH embedding diagrams: \subref{fig:shapefn1} The plots show that $h(r)<r$ and $h'(r)<1$ at $r>r_0$, while $h/r\to 0$ and $h'\to 0$ as $r\to \infty$. \subref{fig:2demedding1} The embedding surface integral, Eq. \eqref{eq:surface_int}, shows that $z'(r_0)\to \infty$, $z(r>r_0)$ is finite and $z(r)\to \infty$ as $r\to \infty$. \subref{fig:3demedding1} Since $z'(r_0)\to \infty$, then $z(r)$ is vertical in the 3D embedding diagram. We set $r_0=1$ pc, for the galaxy NGC 2366, where $\sigma_c=15 \times 10^{-3}~M_\odot$/pc$^{3}$ and $r_c=3$ kpc \cite{Banares-Hernandez:2023axy}.}
\label{Fig:emedding1}
\end{figure}

In Fig. \ref{Fig:emedding1}\subref{fig:shapefn1}, for $r_0=1$ pc, we graphically show that the shape function \eqref{eq:soliton_h} preserves the Lorentz signature, since $h(r)/r<1$ at $r>r_0$ and it fulfills the asymptotic flatness condition $h(r)/r\to 0$ as $r\to \infty$. The figure shows also that the conditions related to the first derivative of the shape functions are fulfilled, i.e. $dh/dr<1$ at $r>r_0$ and asymptotically $dh/dr\to 0$ as $r\to \infty$. Notably, the selected values of the parameter $\beta$ are consistent with the outcome of the flaring-out condition at the WH throat \eqref{eq:flaring_throat}.

For the static spherically symmetric spacetime solution we have just obtained, we take the time slice $t=constant$ on the equatorial plane $\theta=\pi/2$. Therefore the line element reads
\begin{equation}
    ds^2=\frac{dr^2}{1-h(r)/r}+r^2 d\Phi^2.
\end{equation}
To visualize the WH, we embed the above surface in $R^3$ Euclidean space cylindrical coordinate ($r,\Phi,z$)
\begin{equation}
    ds^2=\left[1+\left(\frac{dz}{dr}\right)^2\right]dr^2+r^2 d \Phi^2.
\end{equation}
Thus, the embedding surface integral is given as
\begin{equation}\label{eq:surface_int}
    z(r)=\pm \int_{r_0}^{r}\frac{d\zeta}{\sqrt{\zeta/h(\zeta)-1}}.
\end{equation}
In Fig. \ref{Fig:emedding1}\subref{fig:2demedding1}, we represent the embedding surface integral (2D embedding) diagram \eqref{eq:surface_int} corresponds to the soliton shape function \eqref{eq:soliton_h}.  It can be noted that at the WH throat $$\lim_{r\to r_0}dz/dr\to \infty,$$ therefore $z(r)$ is vertical in the 3D embedding diagram in Fig. \ref{Fig:emedding1}\subref{fig:3demedding1}.
%%%%%%%%%%%%%%%%%%%%%%%%%%%%%%%%%%%%%%%%%%%%
\subsection{Model II}\label{sec:WHmodel2}
\begin{figure}
\centering
\subfigure[~The parameter space $\{r_0,\delta\}$]{\label{fig:space2}\includegraphics[width=.25\textwidth]{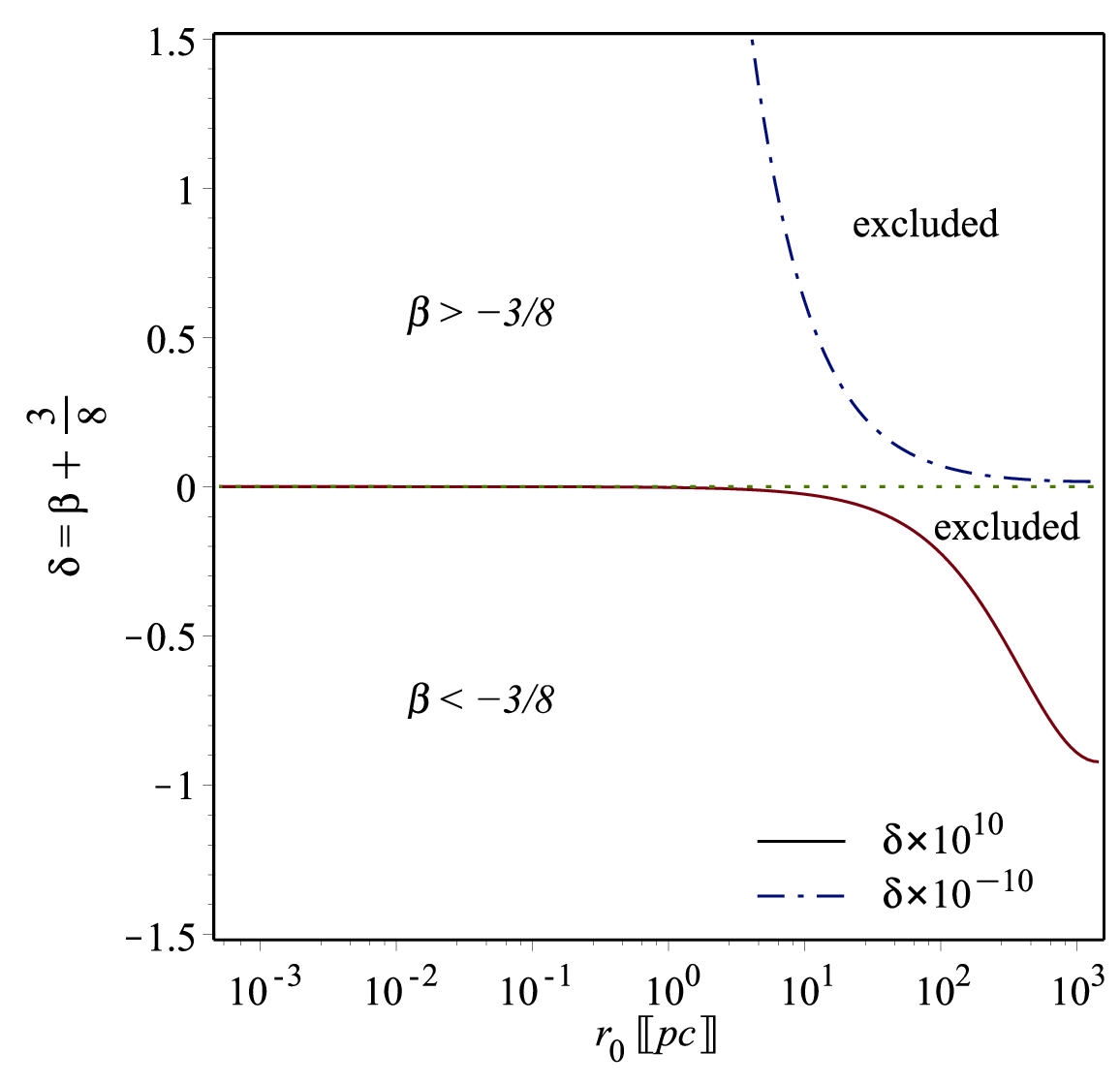}}\hspace{1cm}
\subfigure[~Flaring-out condition at throat]{\label{fig:flaring2}\includegraphics[width=.3\textwidth]{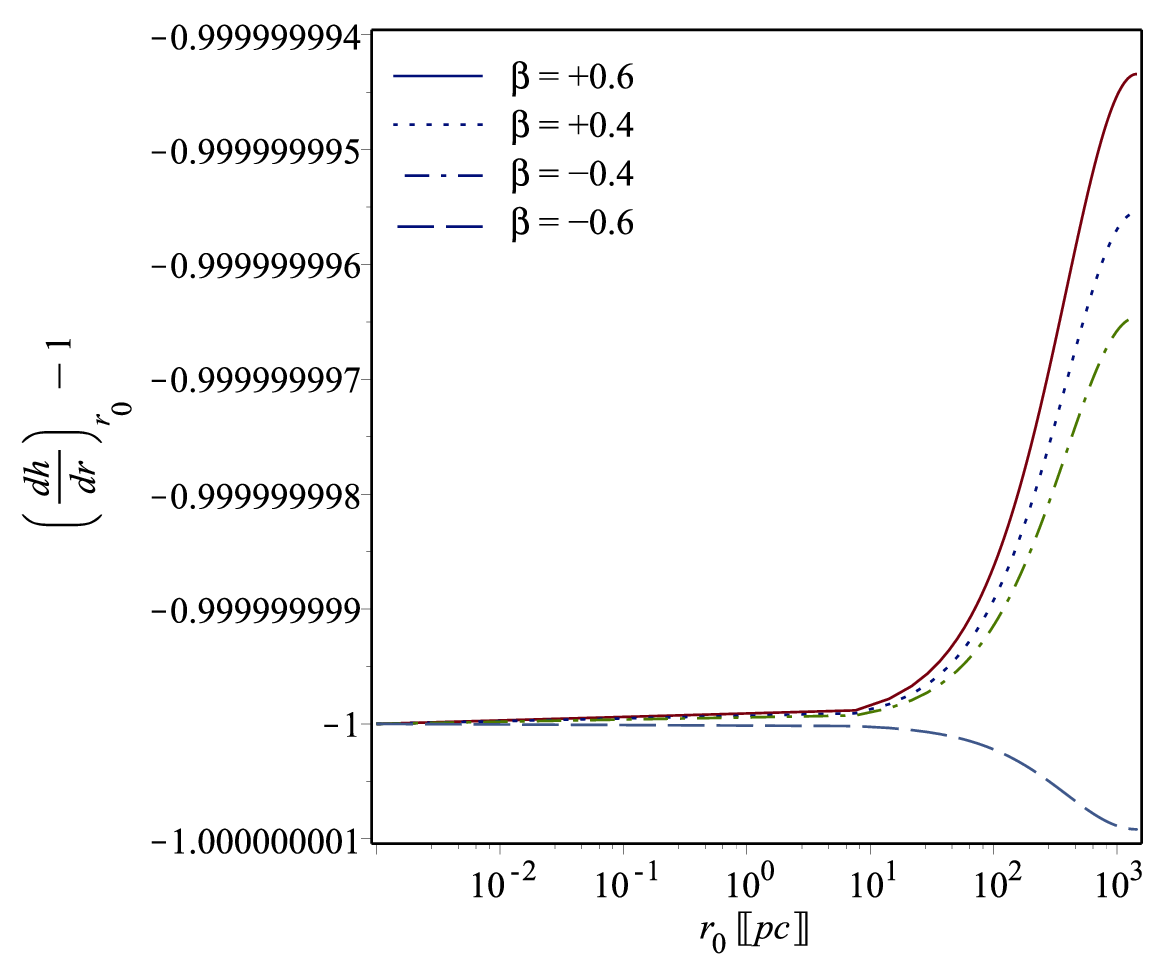}}
\caption{Model II: \subref{fig:space1} The parameter space $\{r_0,\delta\}$ of the NFW CDM WH model of galaxy NGC 2366 ($\sigma_s=3.11 \times 10^{-3}~\text{M}_{\odot}/\text{pc}^3$ and $r_s=1.447$ kpc). We use the faring out constraint on the shape function at the WH throat $h'(r_0)<1$ to evaluate critical values of the non-minimal coupling parameter $\beta$, namely \eqref{eq:model2_beta1} and \eqref{eq:model2_beta2}, at arbitrary throat size $0<r_0<r_c$ where $r_c$ is the soliton core radius of galaxy NGC 2366. Given that $\beta\neq -3/8$, we show both cases when $\beta<-3/8$ ($\delta<0$) and $\beta>-3/8$ ($\delta>0$). The positive (negative) $\delta$ curve is multiplied by a factor $10^{-10}$ ($10^{10}$) to fit the curve into the scale of the graph. \subref{fig:flaring2} The flaring-out condition at the WH throat is satisfied for different selected values of the parameter $\beta$ where the WH throat is arbitrarily chosen $0<r_0<r_s$, it shows that $h'(r_0)<1$ is fulfilled.}
\label{Fig:Model2_beta}
\end{figure}

%%%%%%%%%%%%%%%%%%%%%%%%%%%%%%%%%%%%%%%%%%%%
The exact shape function of CDM can be calculated by plugging the NFW density profile \eqref{eq:NFWdensprof}, into Eq. \eqref{eq:lineardensity}:
\begin{eqnarray}\label{eq:NFW_h}
h(r) = r_0+\tilde{A}\,\frac{r-r_0}{r+ r_s} + \tilde{B}\, \ln \left[\frac{r+r_s}{r_0+r_s}\right],
\end{eqnarray}
where the coefficients $\tilde{A}$ and $\tilde{B}$ are given as
\begin{eqnarray*}
    \tilde{A}&=&-\frac {3\sigma_s c^2 \kappa^2 (1+2\beta)(1+4\beta)r_s^4}{(3+8\beta)(r_0 + r_s)},\\
    \tilde{B}&=&-\tilde{A}\left(1+\frac{r_0}{r_s}\right)=\frac {3\sigma_s c^2 \kappa^2 (1+2\beta)(1+4\beta)r_s^3}{(3+8\beta)}.
\end{eqnarray*}
Obviously, at the WH throat, one obtains $h(r_0)=r_0$ as required by the throat condition of a traversable WH. Additionally, we apply the flaring-out condition at the the WH throat, $h'(r_0)<1$, which gives
\begin{equation}\label{eq:flaring_throat2}
    \frac{3\kappa^2 \sigma_s c^2 r_0 r_s^3 (1+2\beta)(1+4\beta)}{(3+8\beta)(r_s+r_0)^2}<1.
\end{equation}
The above inequality sets the following constraints
\begin{eqnarray}
    &\beta&<-\frac{3}{8}+\frac{4(r_s+ r_0)^{2}-\sqrt{16(r_s+r_0)^{4}+9\kappa^4 \sigma_s^2 c^4 r_s^{6} r_0^{2}}}{24\k^2 \sigma_s c^2 r_s^{3} r_0},~\text{or} \label{eq:model2_beta1}\\
    -3/8<&\beta&<-\frac{3}{8}+\frac{4(r_s+ r_0)^{2}+\sqrt{16(r_s+r_0)^{4}+9\kappa^4 \sigma_s^2 c^4 r_s^{6} r_0^{2}}}{24\k^2 \sigma_s c^2 r_s^{3} r_0}.\label{eq:model2_beta2}
\end{eqnarray}
The obtained intervals show the dependence of the parameter $\beta$ on the WH size $r_0$. For example, \textit{assuming} $r_0=1$ kpc, the flaring out condition is satisfied when $\beta<-3/8-2.545 \times 10^{-13}$ or $-3/8<\beta< -3/8+6.140 \times 10^{10}$. However, for arbitrary values of $0<r_0<r_c$ one can set similar constraints on the non-minimal coupling parameter $\beta$ satisfying $h'(r_0)<1$. We show the parameter space $\{r_0,\delta\}$ graphically as seen in Fig. \ref{Fig:Model2_beta}\subref{fig:space2} by introducing the deviation factor, $\delta\equiv \beta+3/8$, which measures the deviation from the restricted value $\beta=-3/8$. For $\delta>0$, the non-minimal coupling parameter becomes large as $r_0\to 0$ and smaller as $r\to r_s$. For $\delta<0$ the deviation factor is tiny at $0<r_0<r_s$, where $\delta\sim -1\times 10^{-10}$ as shown by the figure. We note that the excluded values in the parameter space $\{r_0,\delta\}$ must be considered when choosing the WH throat size $r_0$ to ensure its consistency with the flaring-out condition, see Fig. \ref{Fig:Model2_beta}\subref{fig:flaring2}. The figure shows that the flaring-out condition at the throat, $(dh/dr)_{r_0}<1$, is satisfied for different selected values of the parameter $\beta$.

\begin{figure}
\centering
\subfigure[~Shape function]{\label{fig:shapefn2}\includegraphics[width=.22\textwidth]{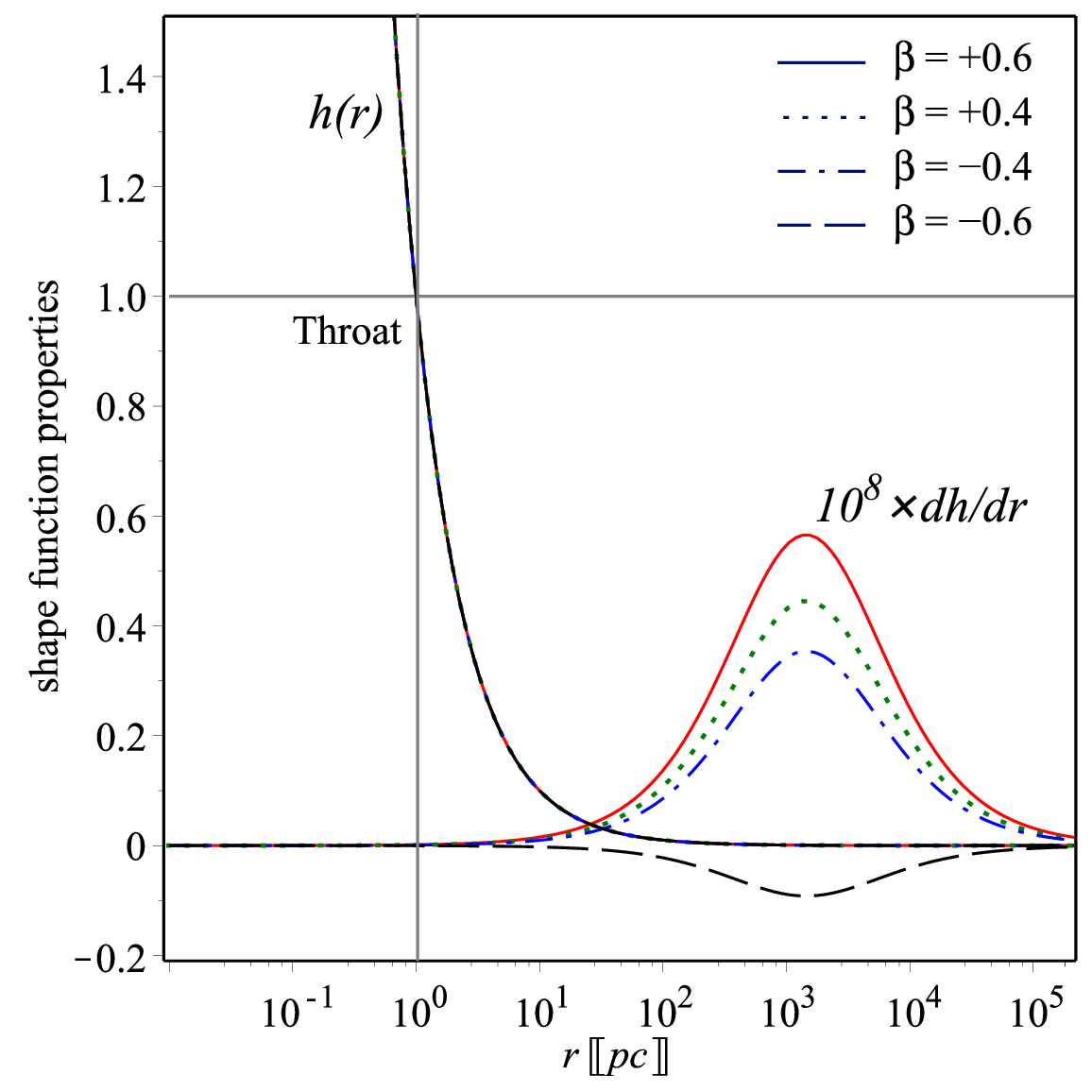}}\hspace{0.5cm}
\subfigure[~2D embedding]{\label{fig:2demedding2}\includegraphics[width=.22\textwidth]{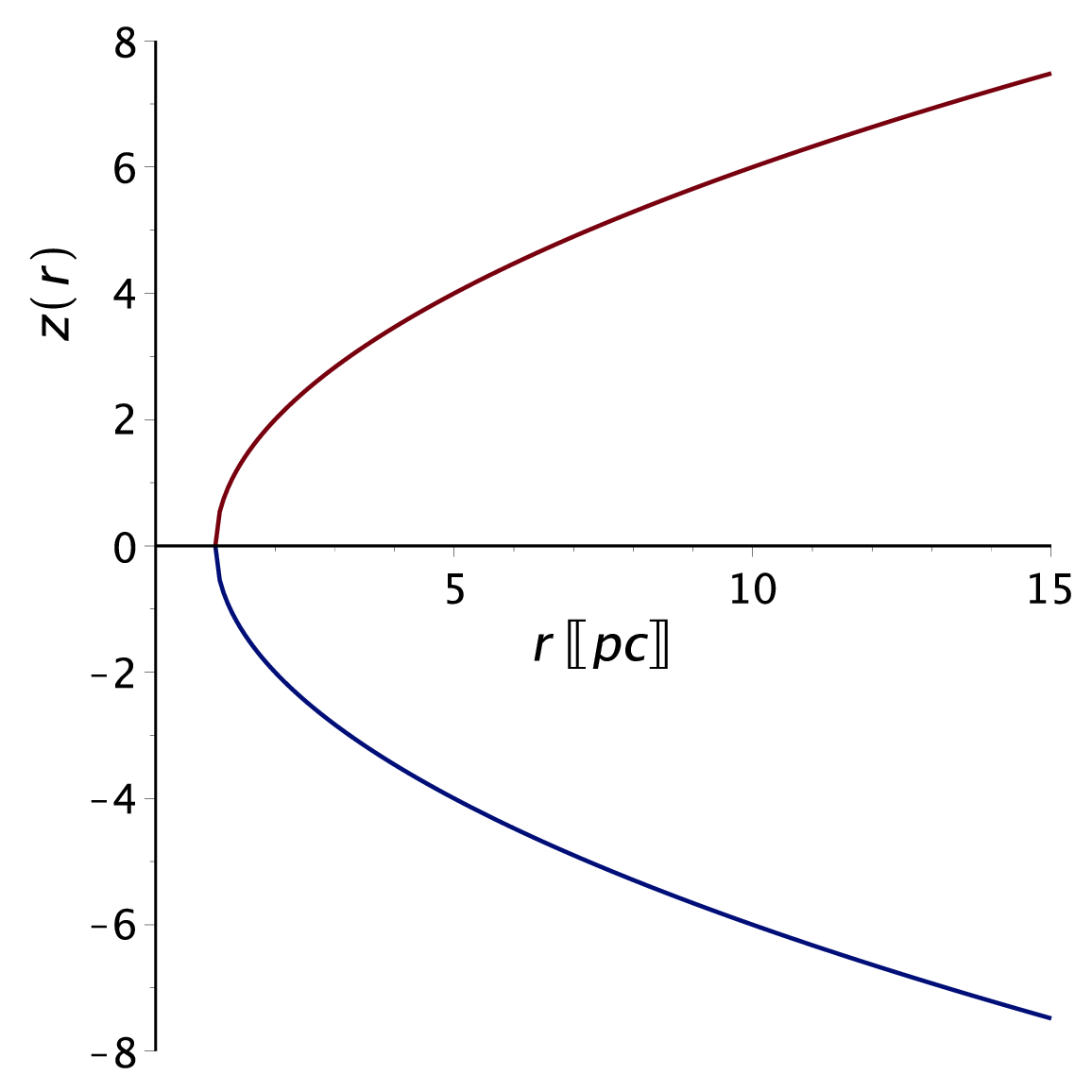}}\hspace{0.5cm}
\subfigure[~3D embedding]{\label{fig:3demedding2}\includegraphics[width=.22\textwidth]{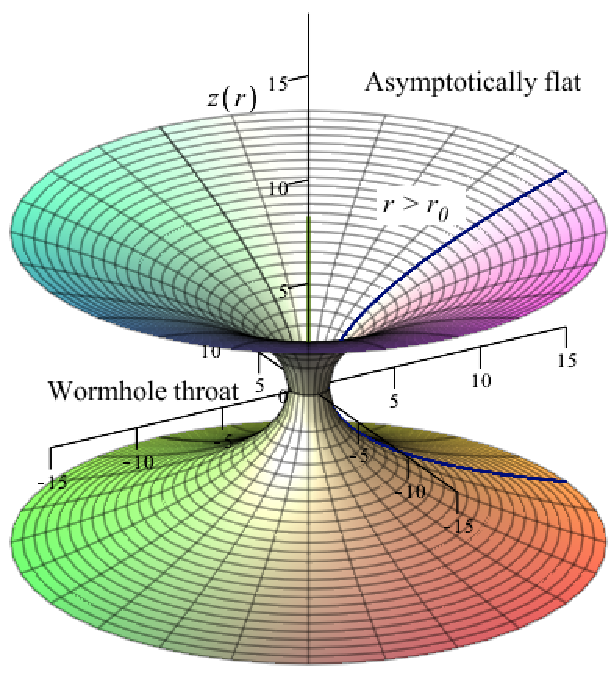}}
\caption{Model II: WH embedding diagrams: \subref{fig:shapefn2} The plots show that $h(r)<r$ and $h'(r)<1$ at $r>r_0$, while $h/r\to 0$ and $h'\to 0$ as $r\to \infty$. \subref{fig:2demedding2} The embedding surface integral, Eq. \eqref{eq:surface_int}, shows that $z'(r_0)\to \infty$, $z(r>r_0)$ is finite and $z(r)\to \infty$ as $r\to \infty$. \subref{fig:3demedding2} Since $z'(r_0)\to \infty$, then $z(r)$ is vertical in the 3D embedding diagram. We set $r_0=1$ pc, for the galaxy NGC 2366, $\sigma_s=3.11 \times 10^{-3}~\text{M}_{\odot}/\text{pc}^3$ and ${r_s} = 1.447$ kpc \cite{Banares-Hernandez:2023axy}.}
\label{Fig:emedding2}
\end{figure}

In Fig. \ref{Fig:emedding2}\subref{fig:shapefn2}, for $r_0=1$ kpc, we graphically show that the shape function \eqref{eq:NFW_h} preserves the Lorentz signature, since $h(r)/r<1$ at $r>r_0$ and it fulfills the asymptotic flatness condition $h(r)/r\to 0$ as $r\to \infty$. The figure shows also that the conditions related to the first derivative of the shape functions are fulfilled, i.e. $dh/dr<1$ at $r>r_0$ and asymptotically $dh/dr\to 0$ as $r\to \infty$. Notably, the selected values of the parameter $\beta$ are consistent with the outcome of the flaring-out condition at the WH throat \eqref{eq:flaring_throat2}.

Similar to Sec. \ref{sec:WHmodel1}, we visualize the WH solution \eqref{eq:NFW_h}, where the corresponding embedding surface integral \eqref{eq:surface_int} is as seen in Fig. \ref{Fig:emedding2}\subref{fig:2demedding2}. We note that at the WH throat $$\lim_{r\to r_0}dz/dr\to \infty,$$ therefore $z(r)$ is vertical in the 3D embedding diagram in Fig. \ref{Fig:emedding2}\subref{fig:3demedding2}.

In summary, we have shown that two DM models are consistent with the WH structure in presence of non-minimal coupling between matter and geometry. The first model assumes soliton quantum wave DM at the galactic core, where the density profile follows Eq. \eqref{eq:solitondensprof} and the corresponding WH shape function is obtained by Eq. \eqref{eq:soliton_h}. The second assumes CDM presence at the galactic halo, where the density profile follows the NFW ansatz \eqref{eq:NFWdensprof} and the corresponding WH shape function is obtained by Eq. \eqref{eq:NFW_h}. We investigated possible constraints on the non-minimal coupling parameter $\beta$ from the WH flaring-out condition. For the dwarf galaxy NGC 2366, recalling the values of the parameters of the soliton+NFW DM model as studied in \cite{Banares-Hernandez:2023axy}, we found the following constraints: In model I, we obtained the viable parameter space $\{\beta, r_0\}$ as given by \eqref{eq:model1_beta1} and \eqref{eq:model1_beta2}, and by setting $r_0=1$ pc, these give $\beta<-3/8-8.494\times 10^{-16}$ or $-3/8<\beta<1.840\times 10^{13}$. In model II, we obtained the viable parameter space $\{\beta, r_0\}$ as given by \eqref{eq:model2_beta1} and \eqref{eq:model2_beta2}, and by setting $r_0=1$ pc, these give $\beta<-3/8-2.545\times 10^{-13}$ or $-3/8<\beta<6.140\times 10^{10}$. In general, for the corresponding valid regions of the coupling parameter $\beta$, the WH models fulfill the following constraints:
\begin{itemize}
    \item [(i)] At the throat $r=r_0$; the shape function $h(r_0)=r_0$, $h'(r_0)<1$, $z(r_0)=0$ and $z'(r_0)\to \infty$.
    \item [(ii)] At finite radius $r>r_0$; the shape function $h(r)<r$, $h'(r)<1$, $z(r)$ is finite and $z'(r)$ is finite.
    \item [(iii)]  At infinite distance from the WH $r\to \infty$; the shape function satisfies $h(r)/r\to 0$, $h'(r)\to 0$, $z(r)\to \pm \infty$ and $z'(r)\to 0$.
\end{itemize}
%%%%%%%%%%%%%%%%%%%%%%%%%%%%%%%%%%%%%%%%%%%%%%%%%%%%%%%%%%%%%%%%%%%%%%%%%%%%%%%%%%%%%%%%%%%%%%%%
\section{Energy conditions of dark matter wormholes in $f(\cal{Q}, \cal{T})$ gravity}\label{SEC:VI}
%%%%%%%%%%%%%%%%%%%%%%%%%%%%%%%%%%%%%%%%%%%%%%%%%%%%%%%%%%%%%%%%%%%%%%%%%%%%%%%%%%%%%%%%%%%%%%%%
In the context of GR, the focusing theorem implies the positivity of the tidal tensor trace $\mathfrak{R}_{\alpha\beta} u^{\alpha} u^{\beta} \geq 0$ and $\mathfrak{R}_{\alpha\beta} \ell^{\alpha} \ell^{\beta} \geq 0$ in Raychaudhuri equation, where $u^{\alpha}$ is an arbitrary timelike vector and $\ell^{\alpha}$ is an arbitrary future directed null vector. This imposes four conditions on the energy-momentum tensor $\mathfrak{T}^{\alpha\beta}$, those are the energy conditions. These could be extended to modified gravity.  In the particular case of $f(\cal{Q}, \cal{T})$ gravity the energy conditions could be written in terms of the effective energy-momentum tensor $\widetilde{\mathfrak{T}}{^\alpha}{_\beta}=diag(-\tilde{\rho}c^2,\tilde{p}_r, \tilde{p}_t, \tilde{p}_t)$, since $\mathfrak{R}_{\alpha\beta}=\kappa\left(\widetilde{\mathfrak{T}}_{\alpha\beta}-\frac{1}{2} g_{\alpha\beta} \widetilde{\mathfrak{T}}\right)$. A physical model should satisfy the modified energy conditions as stated below:
\begin{itemize}
  \item[a.] Weak energy condition (WEC): $\tilde{\sigma}\geq 0$, $ \tilde{\sigma} c^2+ \tilde{p}_r > 0$, $\tilde{\sigma} c^2+\tilde{p}_\theta > 0$,
  \item[b.] Null energy condition (NEC): $\tilde{\sigma} c^2+ \tilde{p}_r \geq 0$, $\tilde{\sigma} c^2+  \tilde{p}_\theta \geq 0$,
  \item[c.] Strong energy condition (SEC): $\tilde{\sigma} c^2+\tilde{p}_r+2\tilde{p}_\theta \geq 0$, $\tilde{\sigma} c^2+\tilde{p}_r \geq 0$, $\tilde{\sigma} c^2+\tilde{p}_\theta \geq 0$,
  \item[d.] Dominant energy conditions (DEC): $\tilde{\sigma}\geq 0$, $\tilde{\sigma} c^2-\tilde{p}_r \geq 0$ and $\tilde{\sigma} c^2-\tilde{p}_\theta \geq 0$.
\end{itemize}
For linear $f(\cal{Q}, \cal{T})$, we have obtained the effective density and pressures, namely Eqs. \eqref{eq:lineareffdensity}--\eqref{eq:linearefftpress}, which allow us to write the corresponding energy conditions as follows:
\begin{align}
& \tilde{\sigma} c^2+ \tilde{p}_r= (1+2\beta)(\sigma c^2+p_r)\label{eq:linearEC1},\\
&\tilde{\sigma} c^2+ \tilde{p}_\theta=(1+2\beta)(\sigma c^2+p_\theta)\label{eq:linearEC2},\\
&\tilde{\sigma} c^2- \tilde{p}_r=\sigma c^2 - p_r+ 4\beta(\sigma c^2-p_r/3-2 p_\theta/3)\label{eq:linearEC3},\\
&\tilde{\sigma} c^2- \tilde{p}_\theta=\sigma c^2 - p_\theta+ 2\beta(2\sigma c^2-p_r/3-5 p_\theta/3)\label{eq:linearEC4},\\
&\tilde{\sigma} c^2+ \tilde{p}_r+  2\tilde{p}_\theta=\sigma c^2 + p_r+ 2 p_\theta+\frac{8}{3}\beta(p_r+2 p_\theta)\label{eq:linearEC5}.
\end{align}
{It is well known that, within the framework of GR, the flaring-out condition necessary for the existence of a traversable WH inevitably leads to a violation of the NEC, provided that the energy density is positive \cite{Rueda:2025rlj}.  Remarkably, our analysis shows that this incompatibility can be avoided in the context of $f(\cal{Q}, \cal{T})$ modified gravity.

From Eqs. \eqref{eq:eff_dens1} and \eqref{eq:eff_rpress1}, it follows that the effective NEC, i.e., $\tilde{\sigma} c^2+ \tilde{p}_r<0$, is required for satisfying the flaring-out condition just as in the GR case. However, unlike in GR, this does not imply that the NEC must be violated in the matter sector itself. Specifically, the physical matter NEC  $\sigma c^2+ p_r>0$, can still be fulfilled provided that the parameter $\beta<-1/2$. In other words, the presence of the $\beta$ parameter in the coupling function of  $f(\cal{Q}, \cal{T})$ gravity allows the flaring-out condition to coexist with standard energy conditions in the matter sector, thus avoiding the need for exotic matter. This case cannot be recovered in the GR limit $\beta=0$.

A similar conclusion applies to the condition \eqref{eq:linearEC2}, where the violation of the NEC appears only in the effective description, while the physical matter sector remains NEC-respecting. This distinction is a significant departure from GR and illustrates how modified gravity can relax traditional WH constraints. We also note that the condition  $\beta<-1/2$, which enables this reconciliation, is consistent with our previous findings derived from applying the flaring-out condition in the earlier section. { We note that the present results are similar to $f(\mathcal{R},L_m,\mathcal{T})$ gravity, where the NEC is still violated at the WH throat and exotic matter is still needed for positive couplings case. However, for negative coupling parameters, the NEC can be satisfied without exotic matter \cite{Loewer:2024lai}}. In what follows, we proceed to test the energy conditions for the specific WH models obtained}.

%%%%%%%%%%%%%%%%%%%%%%%%%%%%%%%%%%%%%%%%%%%%%%%%%%%%%%%%%%%%%%%%%%%%%%%
\subsection{Model I}\label{Sec:ECWHI}
\begin{figure}
\centering
\subfigure[~The WH fluid]{\label{fig:model1_fluid}\includegraphics[scale=.22]{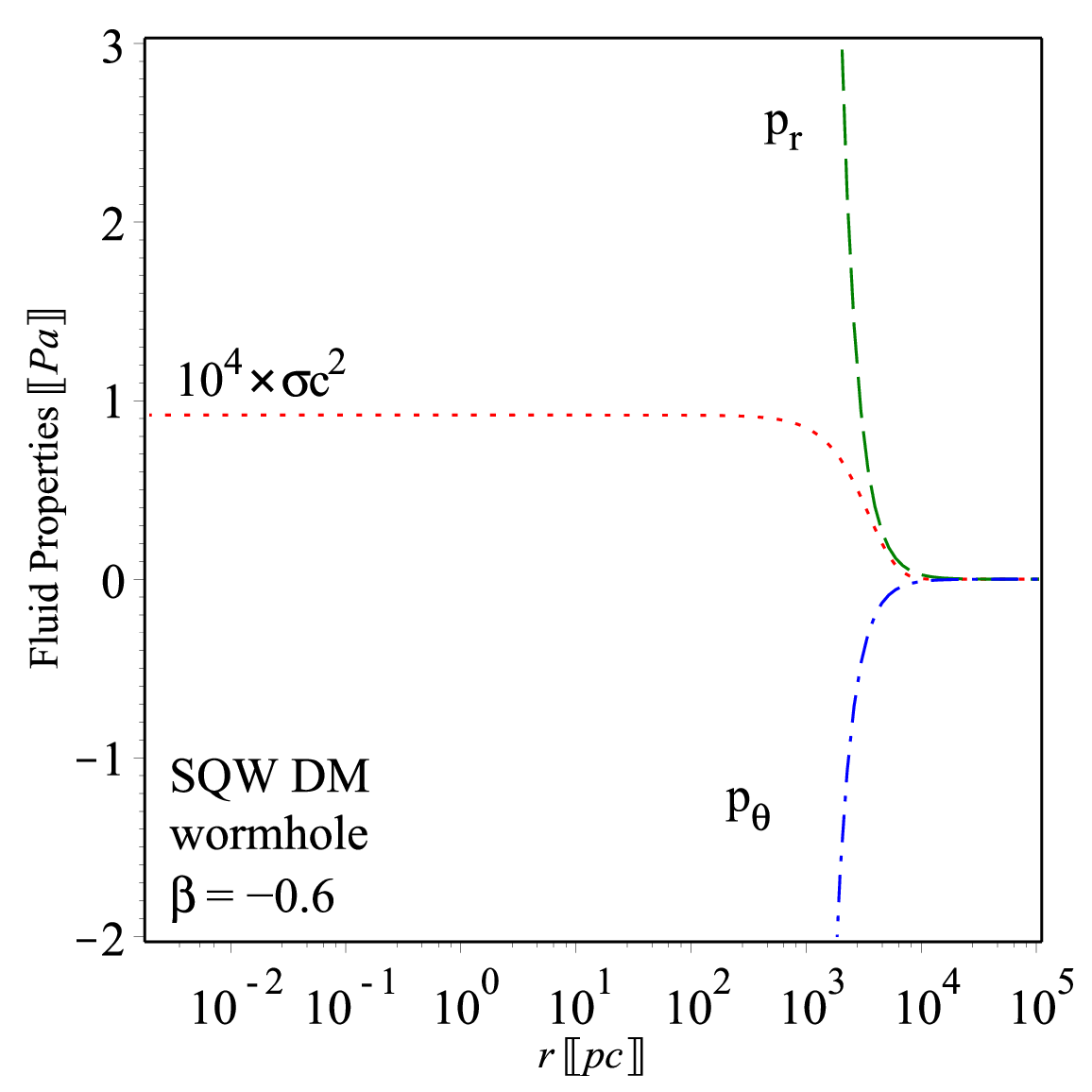}}\hspace{0.5cm}
\subfigure[~Matter energy conditions]{\label{fig:model1_MEC}\includegraphics[scale=.22]{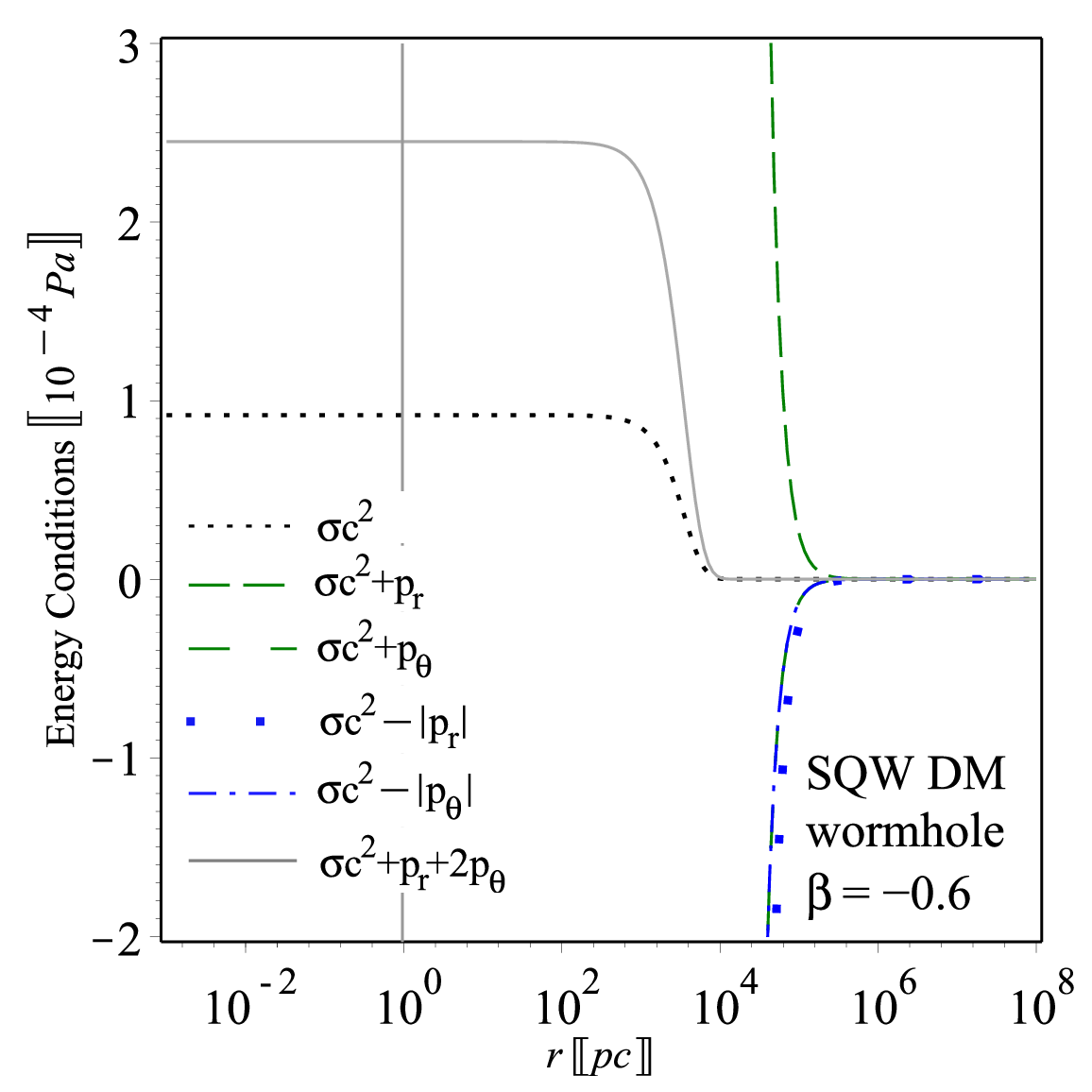}}\hspace{0.5cm}
\subfigure[~Effective energy conditions]{\label{fig:model1_EEC}\includegraphics[scale=.22]{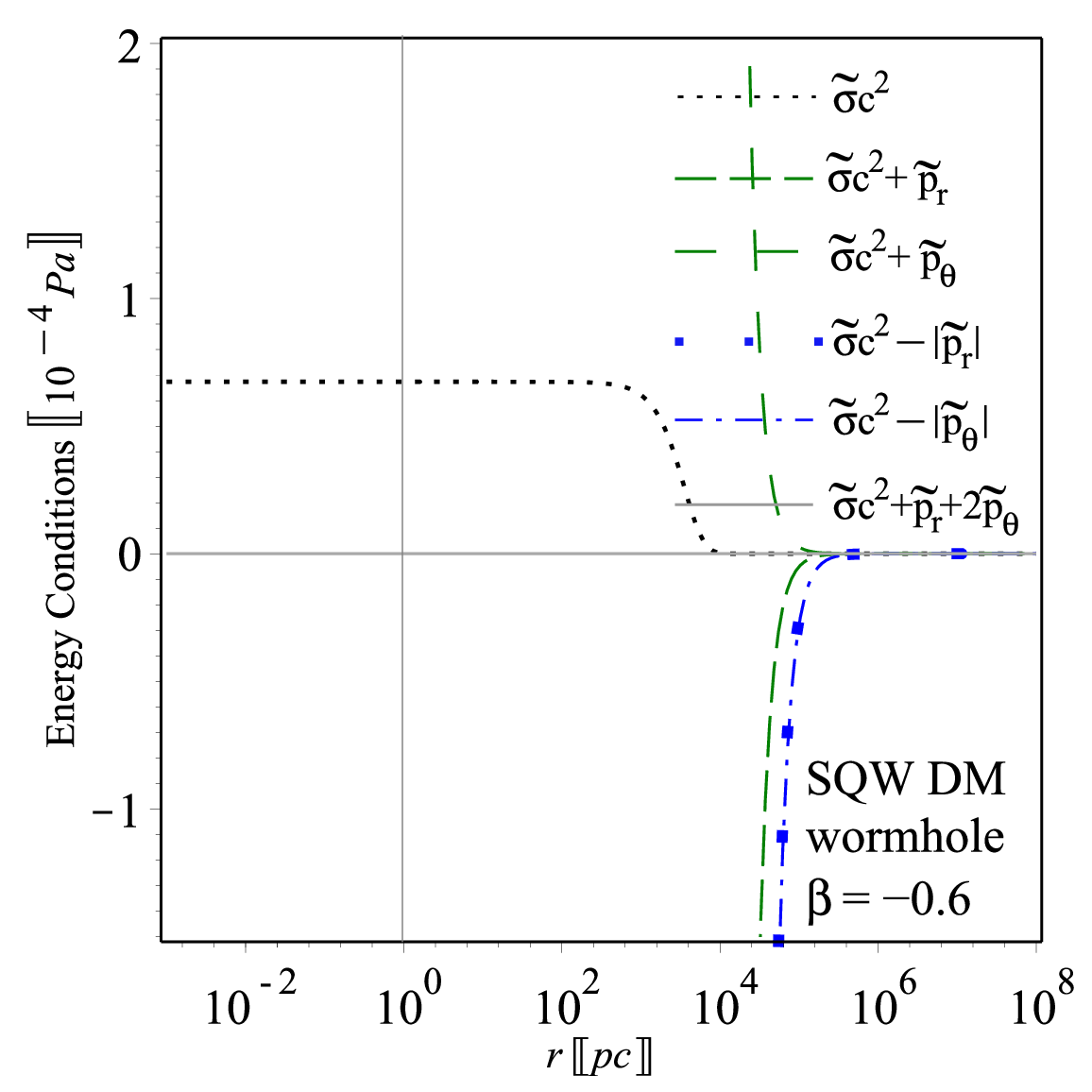}}
\caption[figtopcap]{Model I with $\beta=-0.6$: \subref{fig:model1_fluid} The matter fluid, the density has a flat profile at the core as suggested to solve the core-cusp problem with $p_r>0$ and $p_\theta<0$. \subref{fig:model1_MEC} The matter energy conditions, where the NEC $\sigma c^2+p_r>0$ is satisfied and $\sigma c^2+p_\theta<0$ at $r>r_0$. \subref{fig:model1_EEC} The effective energy conditions, where the NEC $\tilde{\sigma} c^2+\tilde{p}_r>0$ is broken, but $\tilde{\sigma} c^2+\tilde{p}_\theta>0$ is satisfied at $r>r_0$. The alternative behavior of the NEC in the matter and the effective sector is understood since $\beta=-0.6<-1/2$, see Eqs. \eqref{eq:linearEC1} and \eqref{eq:linearEC2}. Other energy conditions, namely SEC and DEC, are broken in both sectors.  We set $r_0=1$ pc, for the galaxy NGC 2366, where $\sigma_c=15 \times 10^{-3}~M_\odot$/pc$^{3}$ and $r_c=3$ kpc \cite{Banares-Hernandez:2023axy}.}
\label{Fig:Model1_EC}
\end{figure}

{ Plugging the soliton shape function \eqref{eq:soliton_h} into Eqs. \eqref{eq:linearrpress} and \eqref{eq:lineartpress}, we obtain the forms of $p_{r}$ and $p_{\theta}$ that we write them in Appendix \ref{appD}.} We select $\beta=-0.6$ which is consistent with the flaring-out condition and the NEC $\sigma c^2+p_r>0$ (i.e. $\beta<-1/2$). We plot the density profile of the soliton core model of galaxy NGC 2366, the radial and the tangential pressures as shown in Fig. \ref{Fig:Model1_EC}\subref{fig:model1_fluid}. The figure shows the flat density profile as suggested to solve the core-cusp problem, while $p_r>0$ and $p_\theta<0$ for $r>r_0$ where $r_0=1$ pc. The corresponding energy conditions on the matter fluid are presented in Fig. \ref{Fig:Model1_EC}\subref{fig:model1_MEC}, which clearly show that the NEC $\sigma c^2+p_r>0$ is fulfilled at $r>r_0$. Since $\beta<-1/2$, the NEC is broken effectively as $\tilde{\sigma} c^2+\tilde{p}_r<0$ as shown in Fig. \ref{Fig:Model1_EC}\subref{fig:model1_EEC}. On the other hand, one can find that $\sigma c^2+p_\theta<0$, where $\tilde{\sigma} c^2+\tilde{p}_\theta>0$. Other energy constraints on the matter and the effective sectors are also shown in Figs. \ref{Fig:Model1_EC}\subref{fig:model1_MEC} and \ref{Fig:Model1_EC}\subref{fig:model1_EEC}.

Substituting \eqref{eq:solitondensprof}, \eqref{s1pr} and \eqref{s1pt} into the energy conditions, one can accordingly set some constraints on the non-minimal coupling parameter $\beta$. By solving the energy conditions on the matter sector, we find the following constraints $\beta\leq -5.519\times 10^{13}$ or $\beta\geq 1.840\times 10^{13}$. Remarkably, for the same $r_0$, we have shown that the faring-out condition gives $\beta<-3/8-8.494\times 10^{-16}$ or $-3/8<\beta<1.840\times 10^{13}$. Therefore, the flaring-out condition excludes the interval $\beta\geq 1.840\times 10^{13}$. In the following we discuss both cases in more detail.

\begin{figure}
\centering
\subfigure[~The shape function]{\label{fig:model1_shape_pve}\includegraphics[scale=.21]{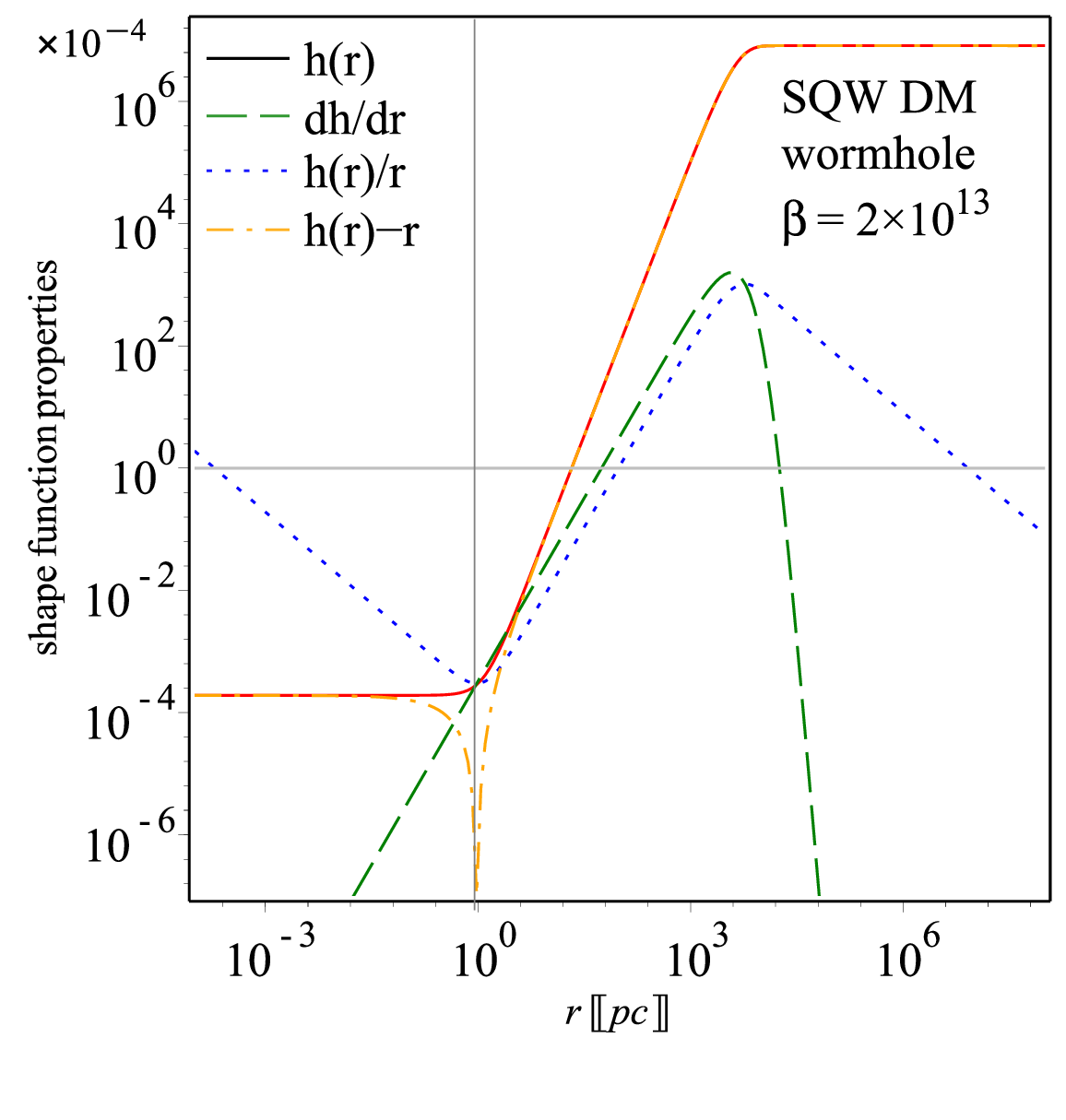}}
\subfigure[~The WH fluid]{\label{fig:model1_fluid_pve}\includegraphics[scale=.22]{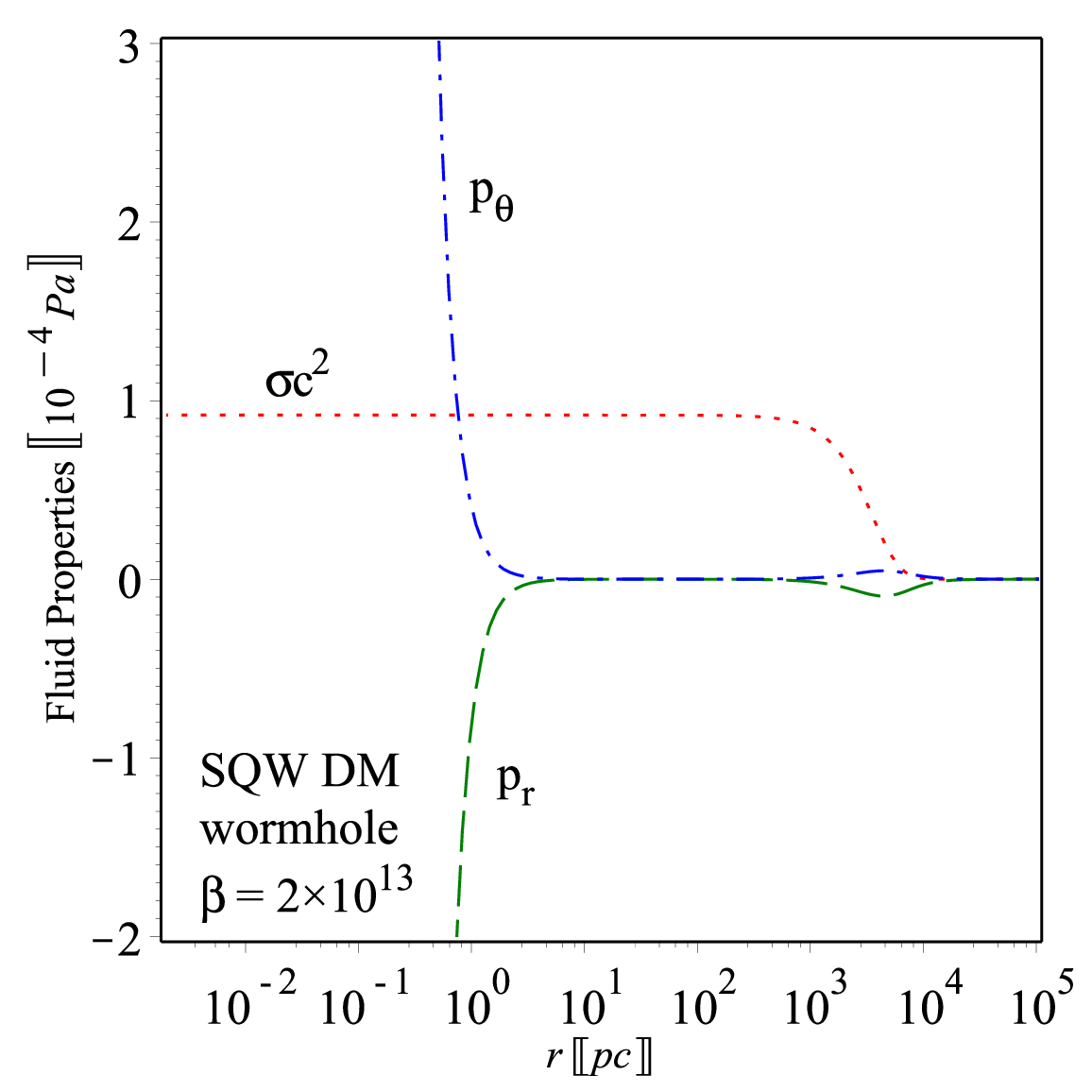}}
\subfigure[~Matter energy conditions]{\label{fig:model1_MEC_pve}\includegraphics[scale=.22]{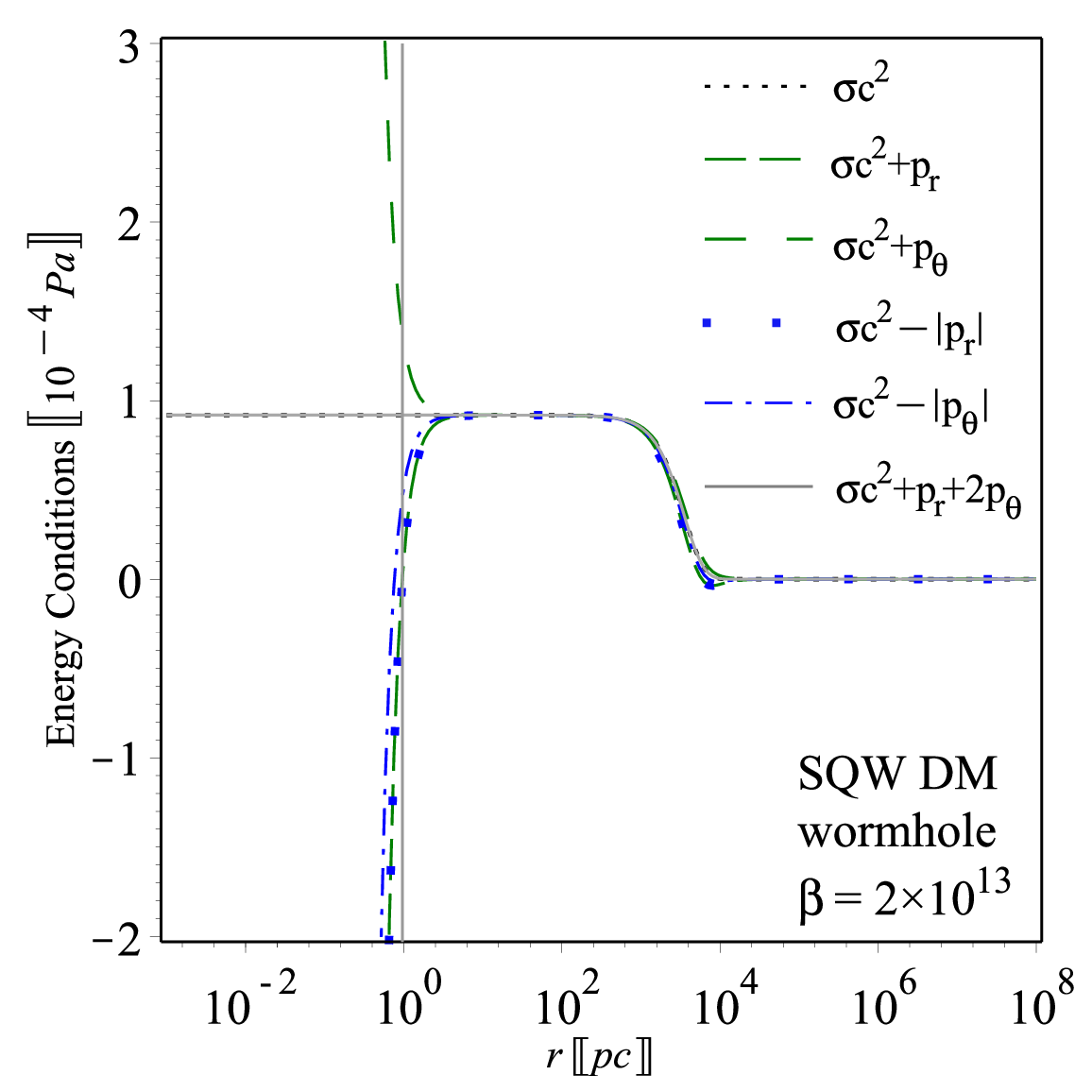}}
\subfigure[~Effective energy conditions]{\label{fig:model1_EEC_pve}\includegraphics[scale=.22]{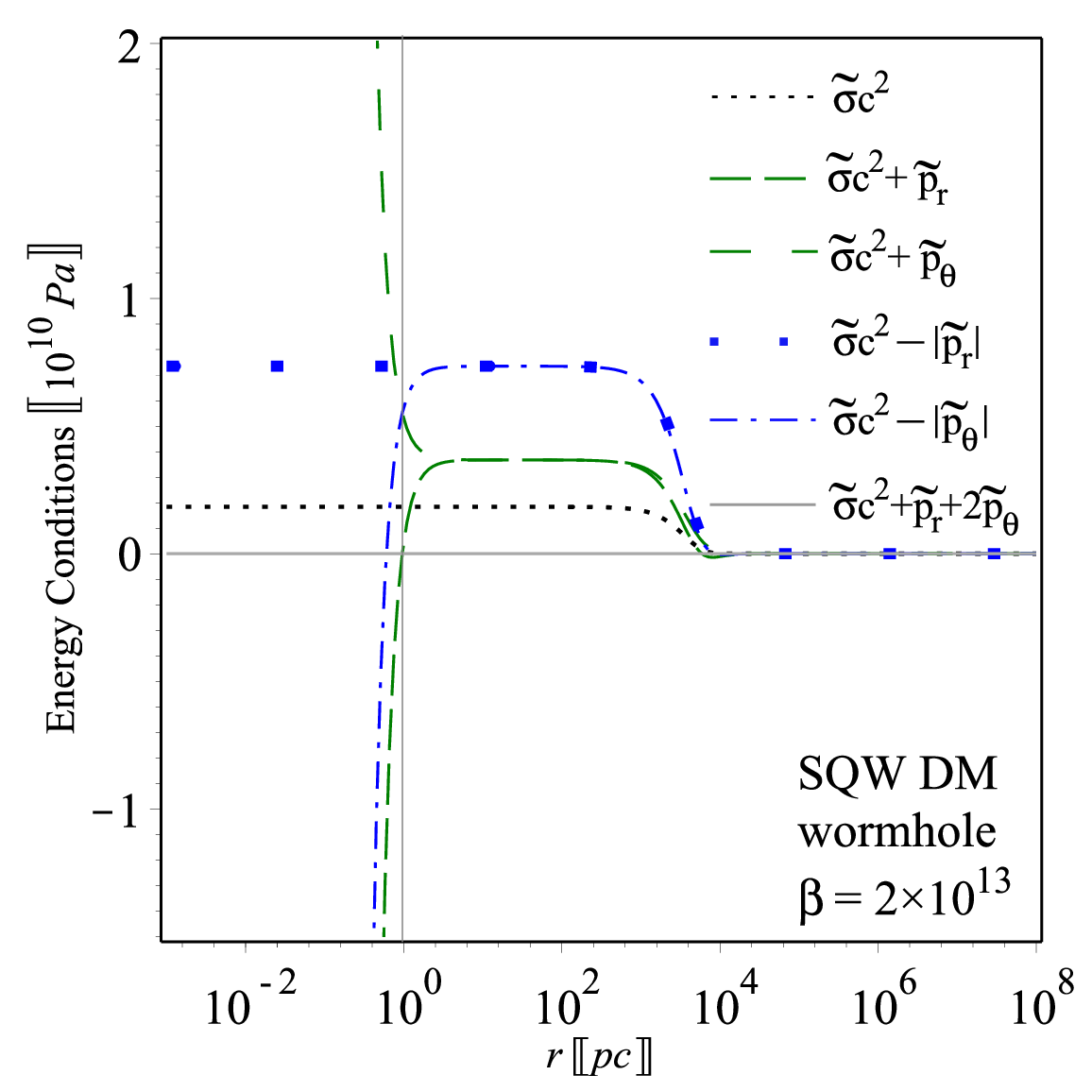}}
\caption[figtopcap]{Model I with large positive coupling $\beta\geq 1.840\times 10^{13}$: \subref{fig:model1_shape_pve} The WH is not Lorentzian as $h(r)>r$ at $r>r_0$, additionally the flaring-out condition is broken. \subref{fig:model1_fluid_pve} The matter fluid, the density has a flat profile at the core as suggested to solve the core-cusp problem with $p_r<0$ and $p_\theta>0$. \subref{fig:model1_MEC_pve} The matter energy conditions are fulfilled at $r>r_0$. \subref{fig:model1_EEC_pve} The effective energy conditions are fulfilled at $r>r_0$.  We set $r_0=1$ pc, for the galaxy NGC 2366, where $\sigma_c=15 \times 10^{-3}~M_\odot$/pc$^{3}$ and $r_c=3$ kpc \cite{Banares-Hernandez:2023axy}.}
\label{Fig:Model1_EC_pve}
\end{figure}

For large positive $\beta\geq 1.840\times 10^{13}$, the matter density and pressures are given in Fig. \ref{Fig:Model1_EC_pve}\subref{fig:model1_fluid_pve}. It can be noted that all energy conditions are satisfied for both the matter and the effective (including the non-minimal coupling effect) sectors at $r>r_0$ as seen in Figs. \ref{Fig:Model1_EC_pve}\subref{fig:model1_MEC_pve} and \ref{Fig:Model1_EC_pve}\subref{fig:model1_EEC_pve}. However, the flaring-out condition and the Lorentzian signature conditions are clearly broken as shown in Fig. \ref{Fig:Model1_EC_pve}\subref{fig:model1_shape_pve}. This confirms the exclusion the interval $\beta\geq 1.840\times 10^{13}$ as it cannot represent a traversable WH.

\begin{figure}
\centering
\subfigure[~The shape function]{\label{fig:model1_shape_nve}\includegraphics[scale=.21]{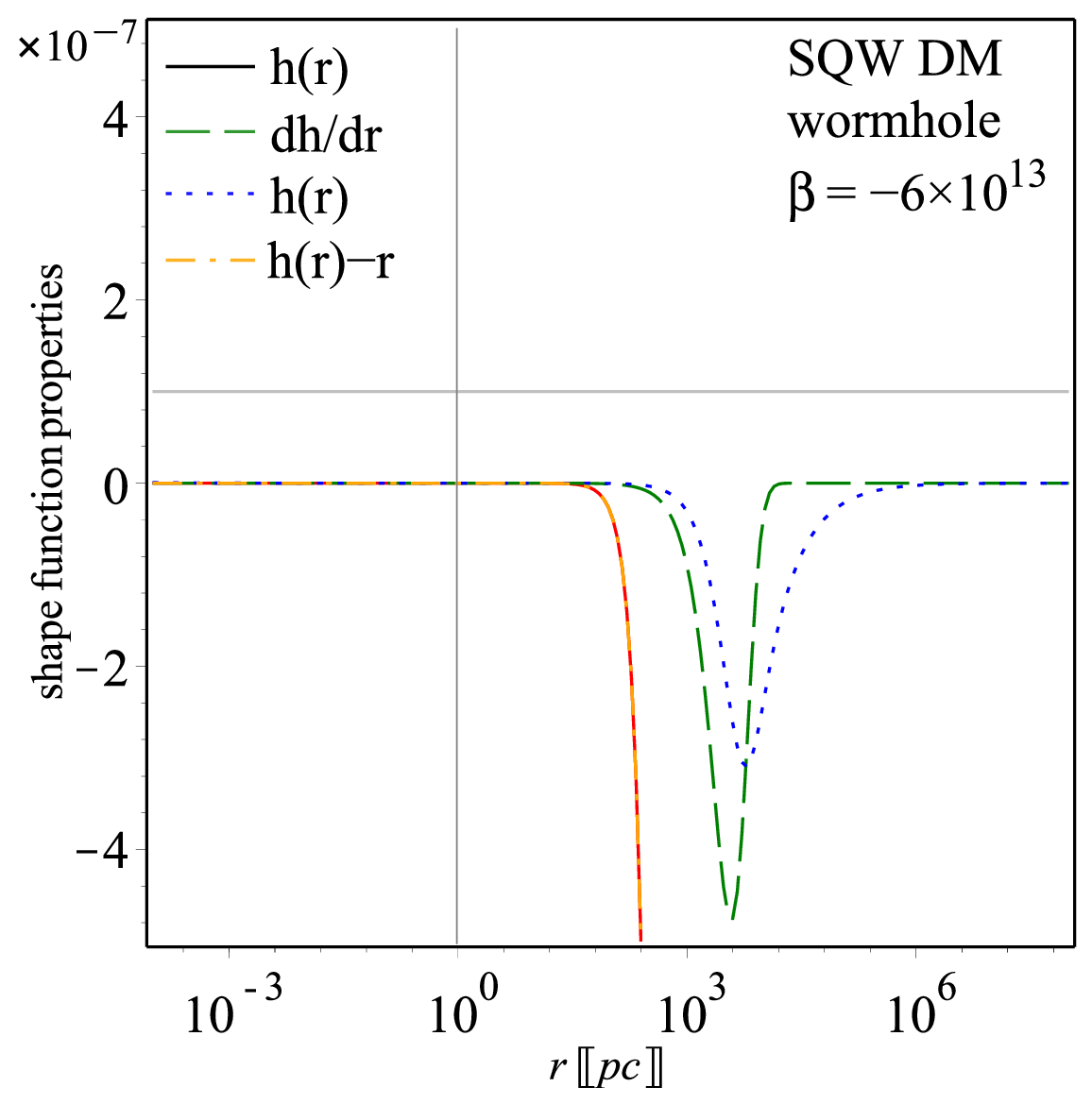}}
\subfigure[~The WH fluid]{\label{fig:model1_fluid_nve}\includegraphics[scale=.22]{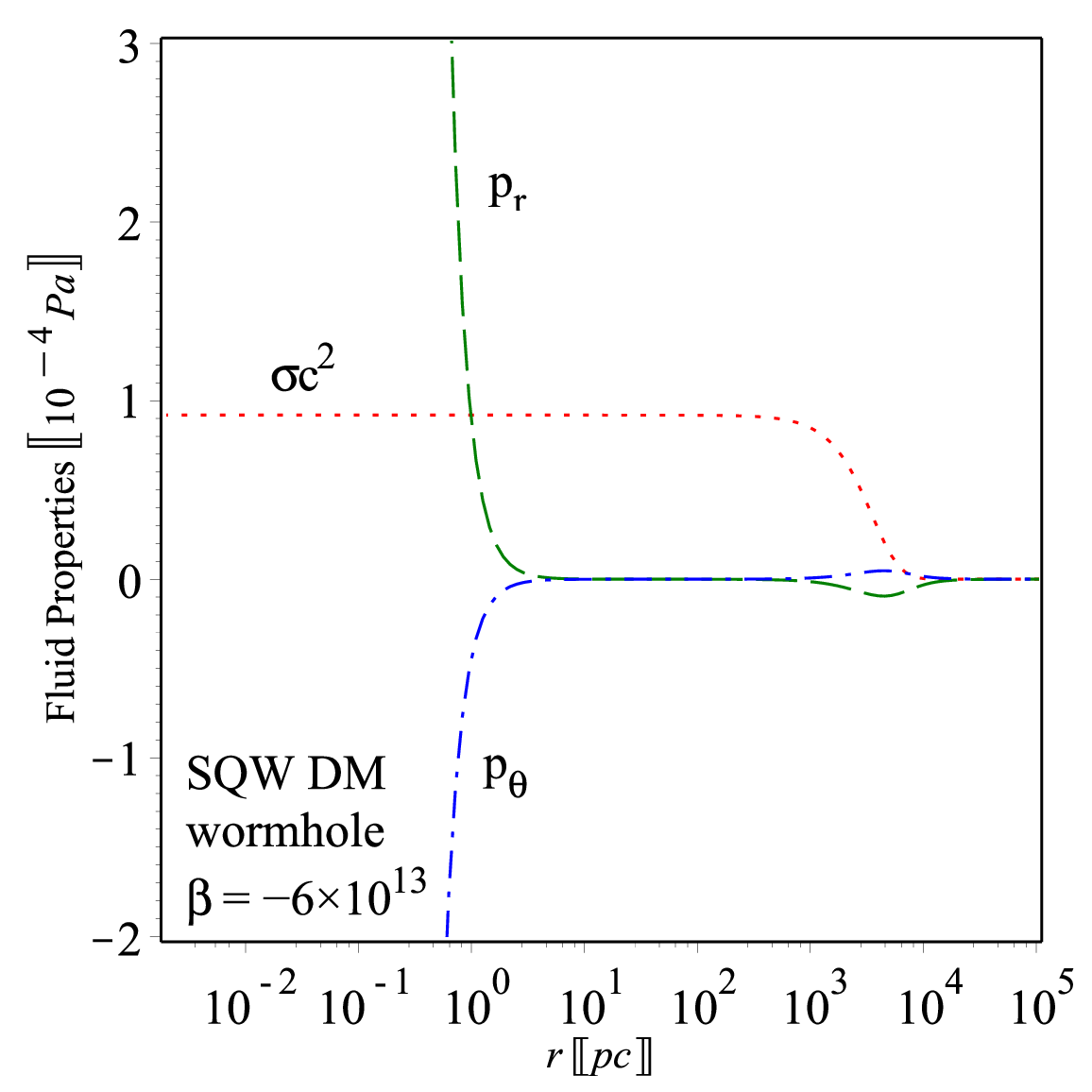}}
\subfigure[~Matter energy conditions]{\label{fig:model1_MEC_nve}\includegraphics[scale=.22]{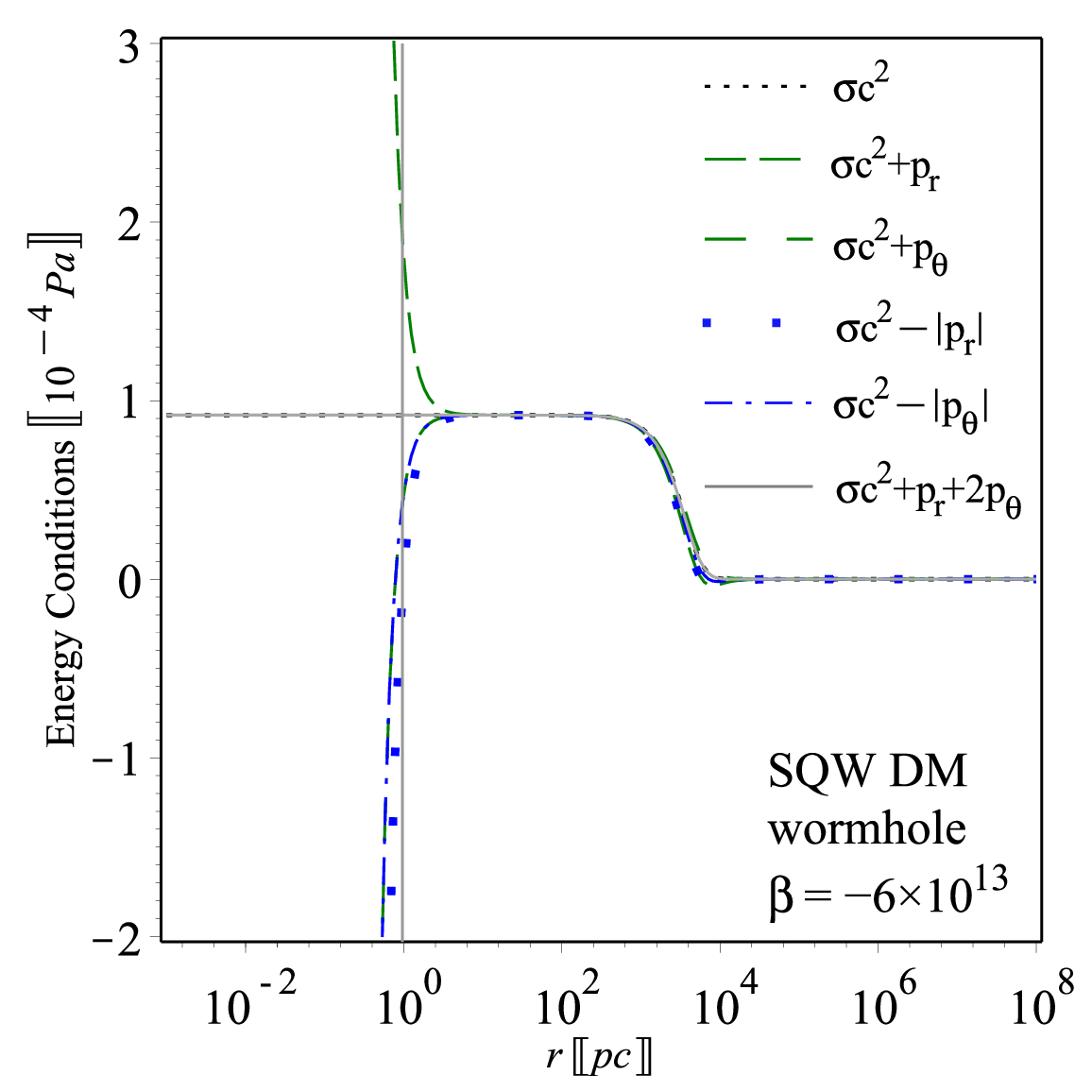}}
\subfigure[~Effective energy conditions]{\label{fig:model1_EEC_nve}\includegraphics[scale=.22]{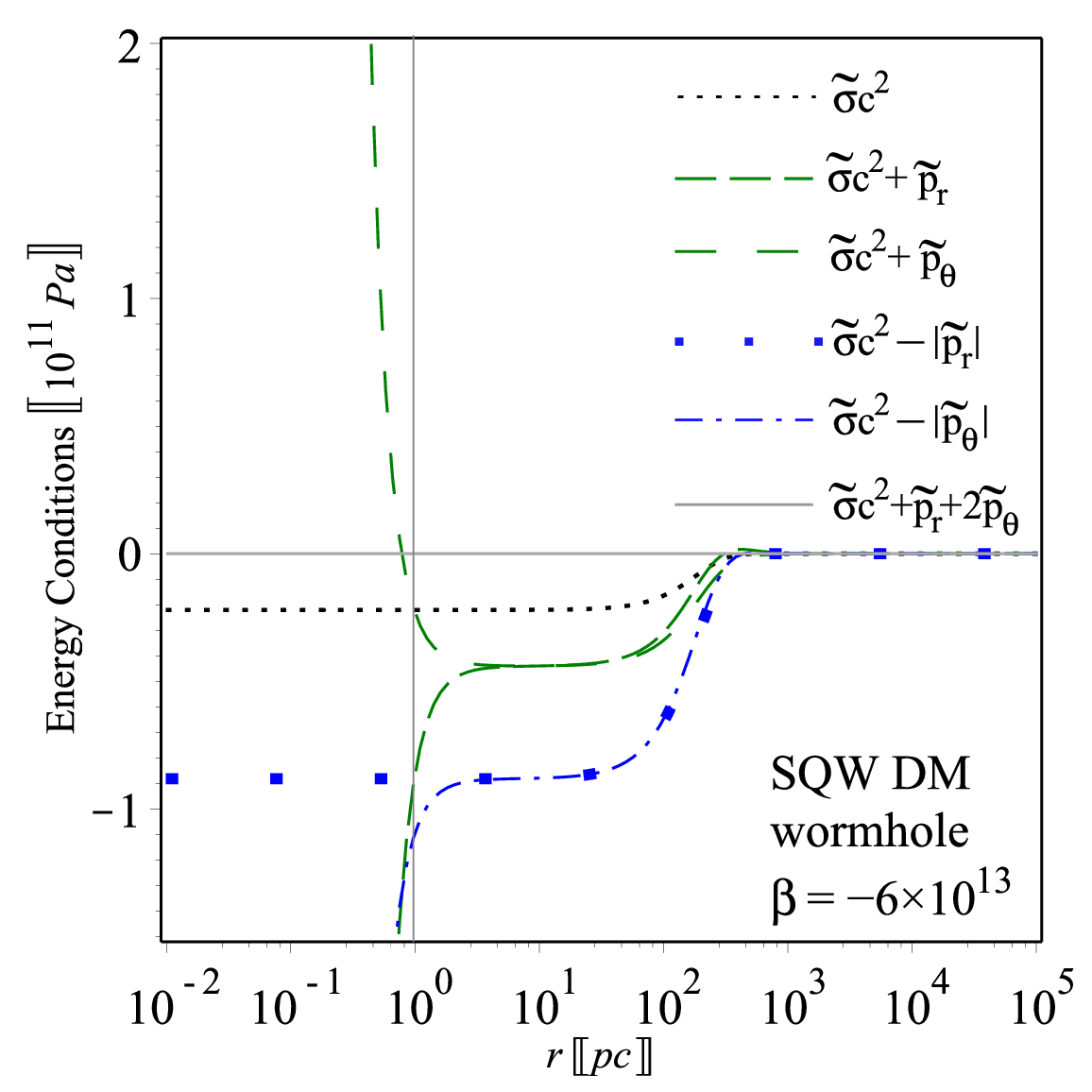}}
\caption[figtopcap]{Model I with large negative coupling $\beta\leq -5.519\times 10^{13}$: \subref{fig:model1_shape_nve} The WH is Lorentzian as $h(r)<r$ at $r>r_0$, additionally the flaring-out condition is fulfilled. \subref{fig:model1_fluid_nve}  The matter fluid, the density has a flat profile at the core as suggested to solve the core-cusp problem with $p_r>0$ and $p_\theta<0$. \subref{fig:model1_MEC_nve} The matter energy conditions are fulfilled at $r>r_0$. \subref{fig:model1_EEC_nve} The effective energy conditions are broken at $r>r_0$ which represents the GR case.  We set $r_0=1$ pc, for the galaxy NGC 2366, where $\sigma_c=15 \times 10^{-3}~M_\odot$/pc$^{3}$ and $r_c=3$ kpc \cite{Banares-Hernandez:2023axy}.}
\label{Fig:Model1_EC_nve}
\end{figure}

For large negative $\beta\leq -5.519\times 10^{13}$, the matter density and pressures are given in Fig. \ref{Fig:Model1_EC_nve}\subref{fig:model1_fluid_nve}. It can be noted that all energy conditions are satisfied for the matter fluid at $r>r_0$ as seen in Figs. \ref{Fig:Model1_EC_nve}\subref{fig:model1_MEC_nve}. On the contrary, the energy conditions are broken on the effective sector as seen by Fig. \ref{Fig:Model1_EC_nve}\subref{fig:model1_EEC_nve}, which is must occur if the flaring-out of the WH is hold. In Fig. \ref{Fig:Model1_EC_nve}\subref{fig:model1_shape_nve}, we show that the WH solution, in this case, satisfies all conditions related to the shape function and its derivative.

In conclusion, the above calculations confirm the possibility to find a healthy WH solution where the energy conditions are broken only on the effective sector but satisfied on the matter sector. Consequently, no exotic matter is needed to form a physical WH as required in the GR framework, if non-minimal coupling between matter and geometry has been considered and stays within certain bounds.
%%%%%%%%%%%%%%%%%%%%%%%%%%%%%%%%%%%%%%%%%%%%%%%%%%%%%%%%%%
\subsection{Model II}\label{Sec:ECWHII}
\begin{figure}
\centering
\subfigure[~The WH fluid]{\label{fig:model2_fluid}\includegraphics[scale=.22]{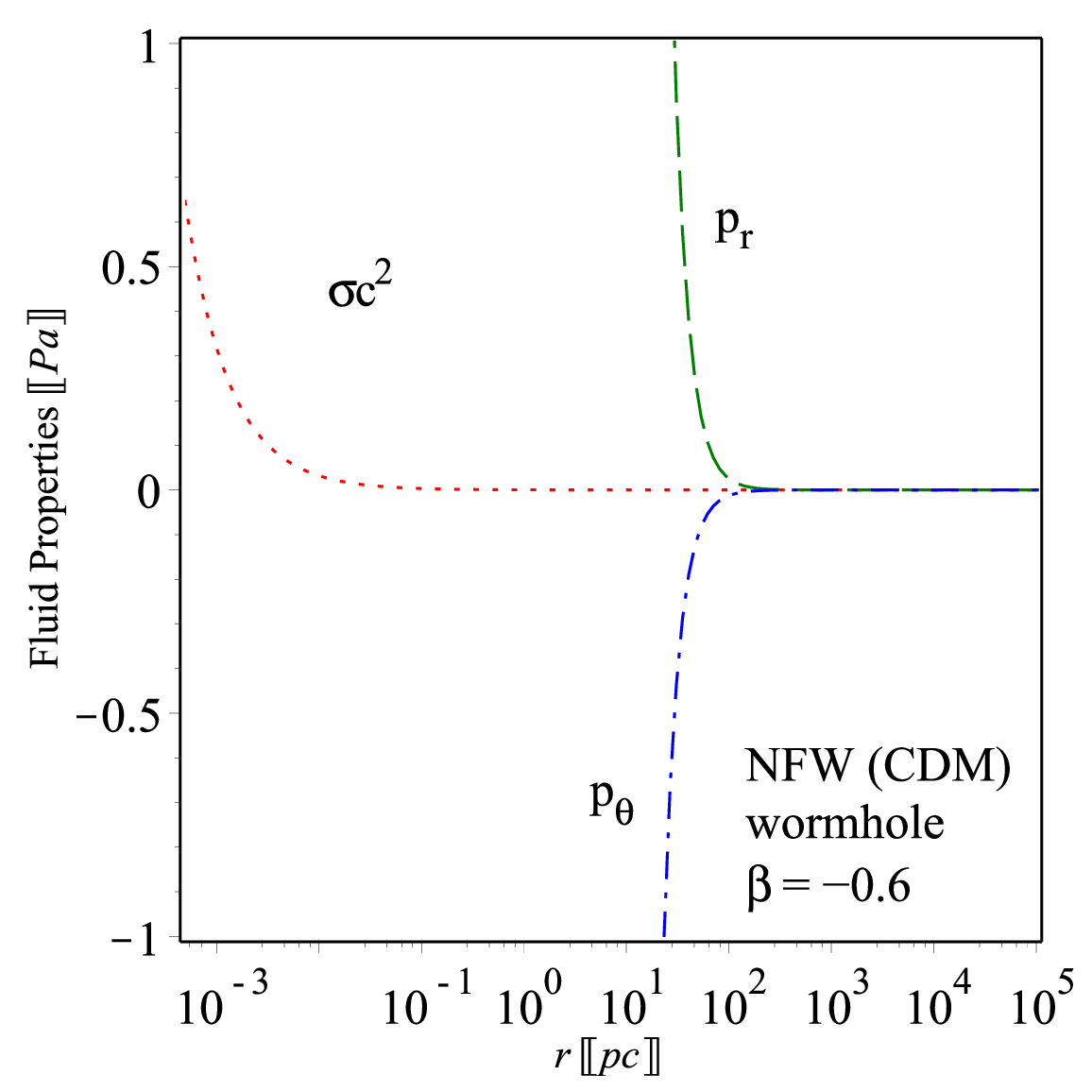}}\hspace{0.5cm}
\subfigure[~Matter energy conditions]{\label{fig:model2_MEC}\includegraphics[scale=.22]{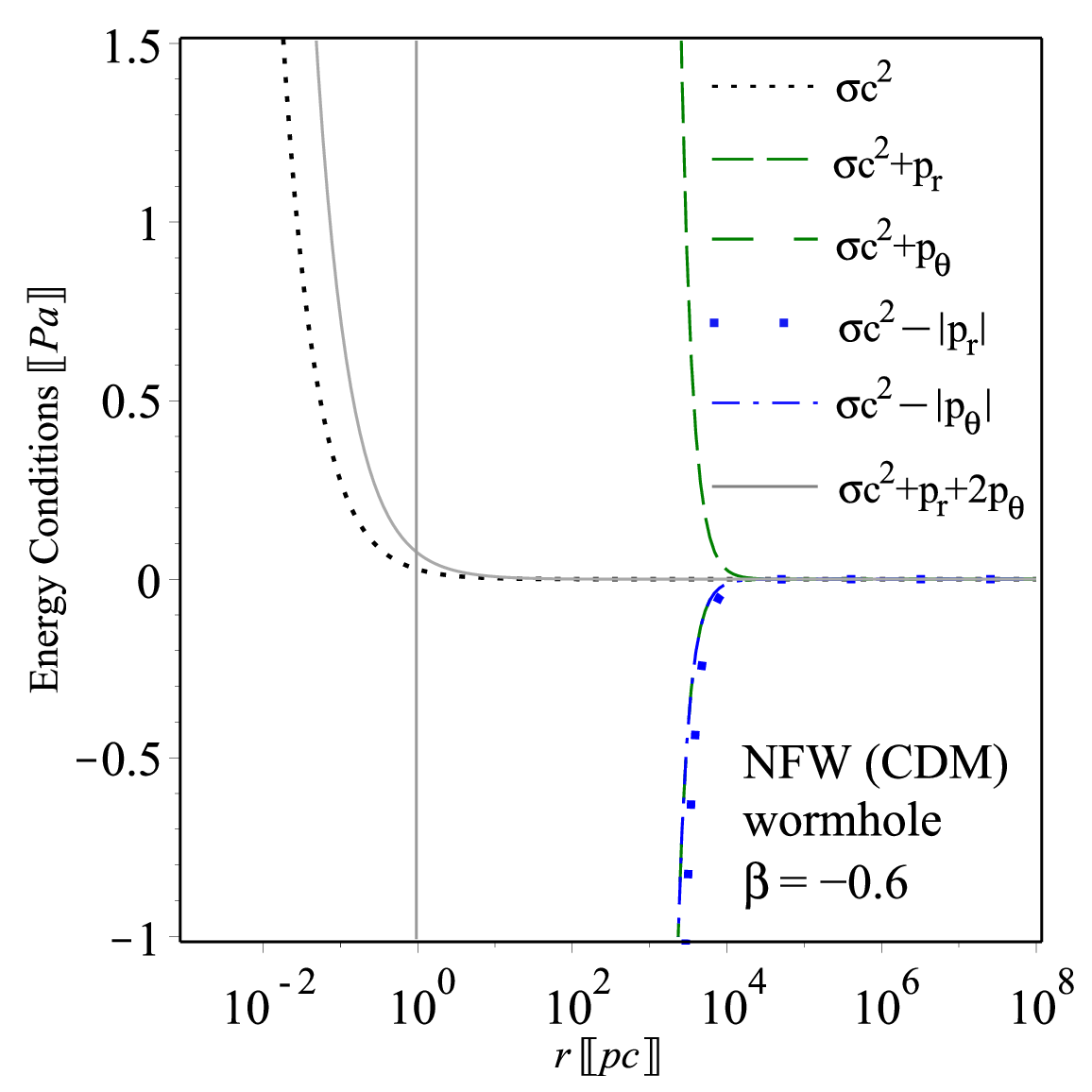}}\hspace{0.5cm}
\subfigure[~Effective energy conditions]{\label{fig:model2_EEC}\includegraphics[scale=.22]{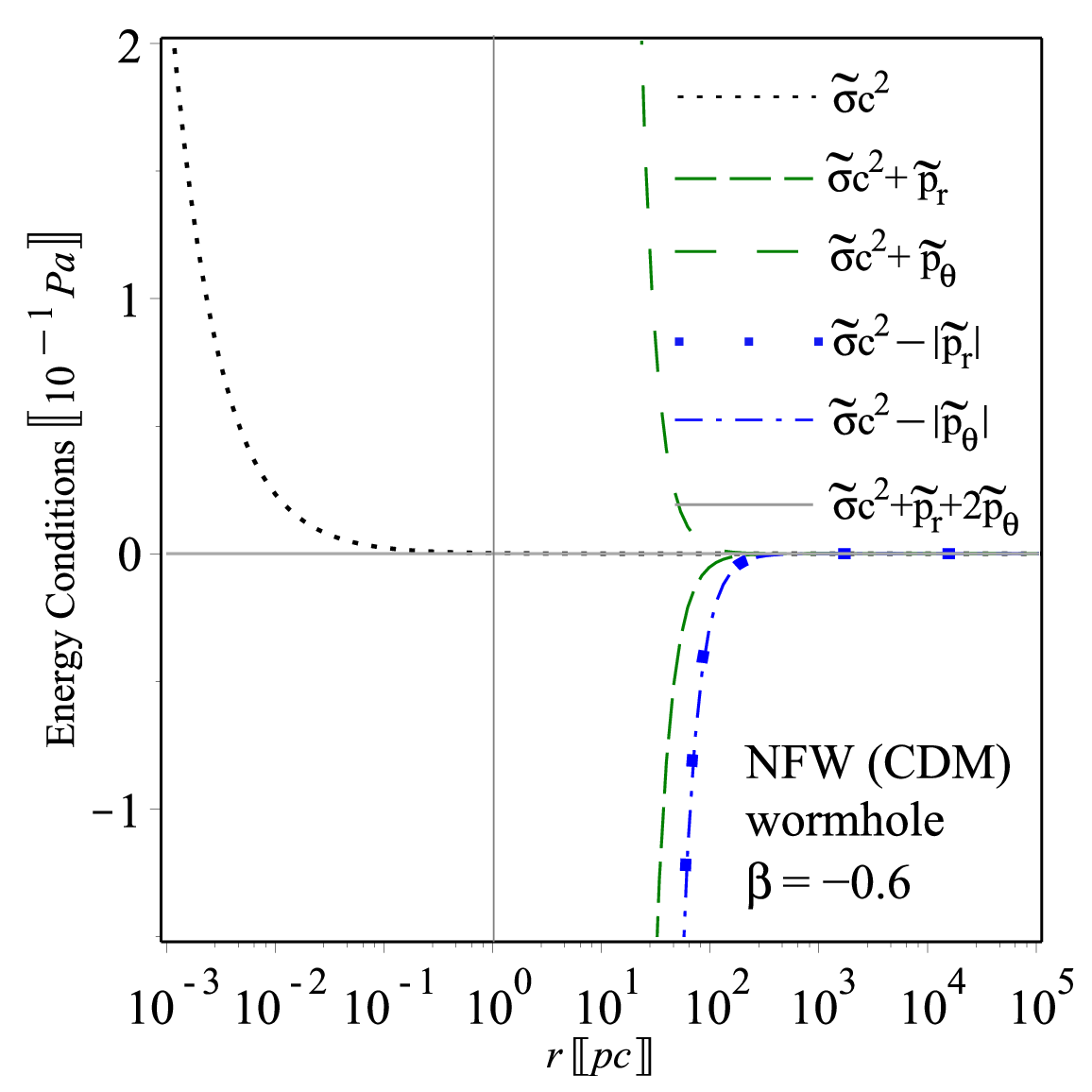}}
\caption[figtopcap]{Model II with $\beta=-0.6$: \subref{fig:model2_fluid} The matter fluid, the density has a flat profile at the core as suggested to solve the core-cusp problem with $p_r>0$ and $p_\theta<0$. \subref{fig:model2_MEC} The matter energy conditions, where the NEC $\sigma c^2+p_r>0$ is satisfied and $\sigma c^2+p_\theta<0$ at $r>r_0$. \subref{fig:model2_EEC} The effective energy conditions, where the NEC $\tilde{\sigma} c^2+\tilde{p}_r>0$ is broken, but $\tilde{\sigma} c^2+\tilde{p}_\theta>0$ is satisfied at $r>r_0$. The alternative behavior of the NEC in the matter and the effective sector is understood since $\beta=-0.6<-1/2$, see Eqs. \eqref{eq:linearEC1} and \eqref{eq:linearEC2}. Other energy conditions, namely SEC and DEC, are broken in both sectors. We set $r_0=1$ pc, for the galaxy NGC 2366, $\sigma_s=3.11 \times 10^{-3}~\text{M}_{\odot}/\text{pc}^3$ and ${r_s} = 1.447$ kpc \cite{Banares-Hernandez:2023axy}.}
\label{Fig:Model2_EC}
\end{figure}

Substituting the NFW shape function \eqref{eq:NFW_h} into Eqs. \eqref{eq:linearrpress} and \eqref{eq:lineartpress}, we obtain
\begin{eqnarray}
p_{r}&=&-\frac{\left[r_0+\tilde{A}\frac{(r-r_0)}{(r+r_s)}+\tilde{B} \ln\left(\frac{r_s+r}{r_s+r_0}\right)\right]}{(1+2\beta)\kappa^2 r^3}+\frac{4\beta\left[(\tilde{A}+\tilde{B})r_s+\tilde{A} r_0+\tilde{B} r\right]}{3(r_s+r)^2(1+2\beta)(1+4\beta)\kappa^2 r^2}\, ,\label{s2pr}\\
p_{\theta}&=&\frac{\left[r_0+\tilde{A}\frac{(r-r_0)}{(r+r_s)}+\tilde{B} \ln\left(\frac{r_s+r}{r_s+r_0}\right)\right]}{2(1+2\beta)\kappa^2 r^3}+\frac{(3+4\beta)\left[(\tilde{A}+\tilde{B})r_s+\tilde{A} r_0+\tilde{B} r\right]}{6(r_s+r)^2(1+2\beta)(1+4\beta)\kappa^2 r^2}\,.\label{s2pt}
\end{eqnarray}
We select $\beta=-0.6$ which is consistent with the flaring-out condition and the NEC $\sigma c^2+p_r>0$ (i.e. $\beta<-1/2$). We plot the density profile of the CDM model of galaxy NGC 2366, the radial and the tangential pressures as shown in Fig. \ref{Fig:Model2_EC}\subref{fig:model2_fluid}. The figure shows the cuspy behavior of the density profile at the core region, while $p_r>0$ and $p_\theta<0$ for $r>r_0$ where $r_0=1$ pc. The corresponding energy conditions on the matter fluid are presented in Fig. \ref{Fig:Model2_EC}\subref{fig:model2_MEC}, which clearly show that the NEC $\sigma c^2+p_r>0$ is fulfilled at $r>r_0$. Since $\beta<-1/2$, the NEC is broken effectively as $\tilde{\sigma} c^2+\tilde{p}_r<0$ as shown in Fig. \ref{Fig:Model2_EC}\subref{fig:model2_EEC}. On the other hand, one can find that $\sigma c^2+p_\theta<0$, where $\tilde{\sigma} c^2+\tilde{p}_\theta>0$. Other energy constraints on the matter and the effective sectors are also shown in Figs. \ref{Fig:Model2_EC}\subref{fig:model2_MEC} and \ref{Fig:Model2_EC}\subref{fig:model2_EEC}.

Substituting \eqref{eq:NFWdensprof}, \eqref{s2pr} and \eqref{s2pt} into the energy conditions, one can accordingly set some constraints on the non-minimal coupling parameter $\beta$. By solving the energy conditions on the matter sector, we find the following constraints $\beta\leq -1.842\times 10^{11}$ or $\beta\geq 6.140\times 10^{10}$. Remarkably, for the same $r_0$, we have shown that the faring-out condition gives $\beta<-3/8-2.545\times 10^{-13}$ or $-3/8<\beta<6.140\times 10^{10}$. Therefore, the flaring-out condition excludes the interval $\beta\geq 6.140\times 10^{10}$. In the following we discuss both cases in more detail.

\begin{figure}
\centering
\subfigure[~The shape function]{\label{fig:model2_shape_pve}\includegraphics[scale=.22]{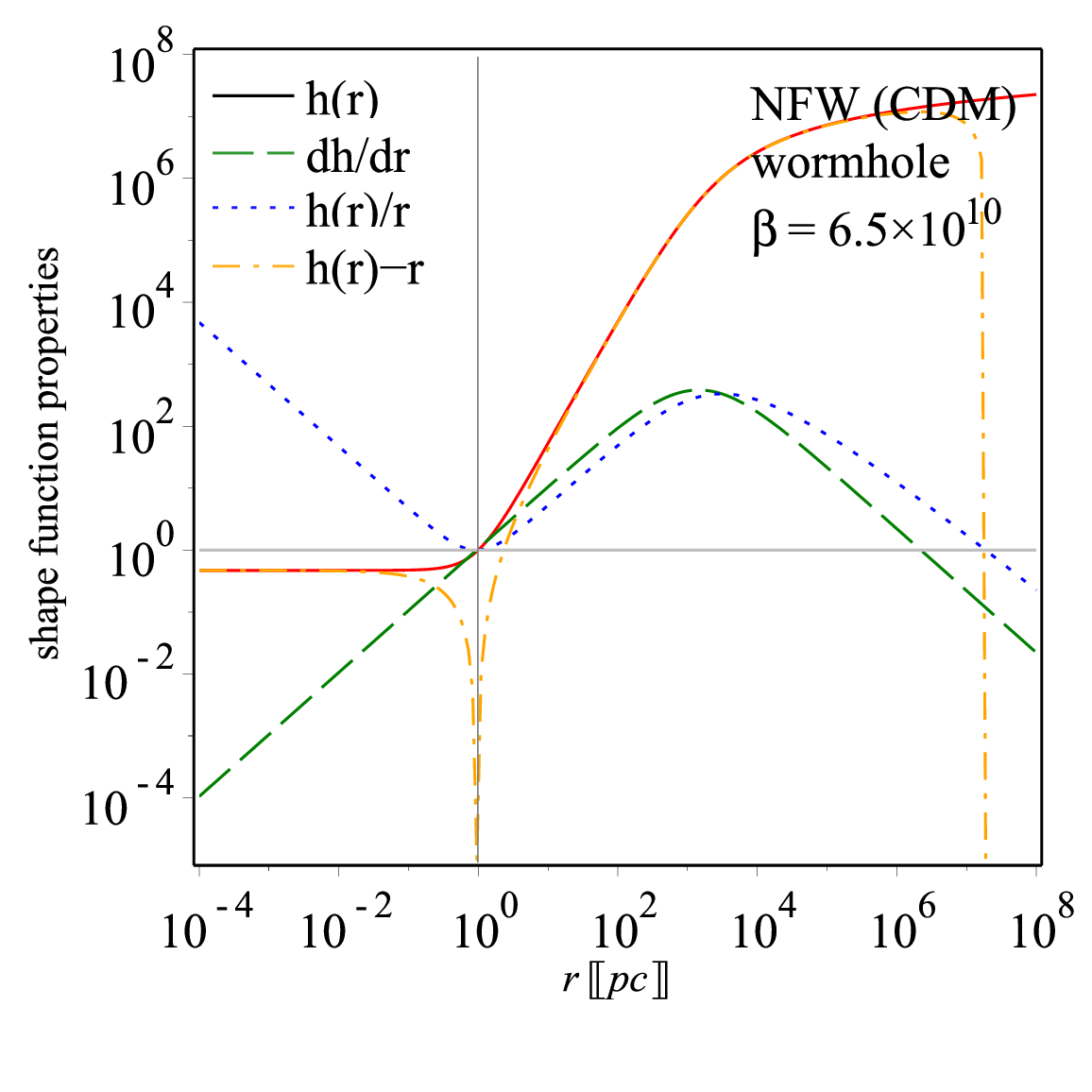}}
\subfigure[~The WH fluid]{\label{fig:model2_fluid_pve}\includegraphics[scale=.22]{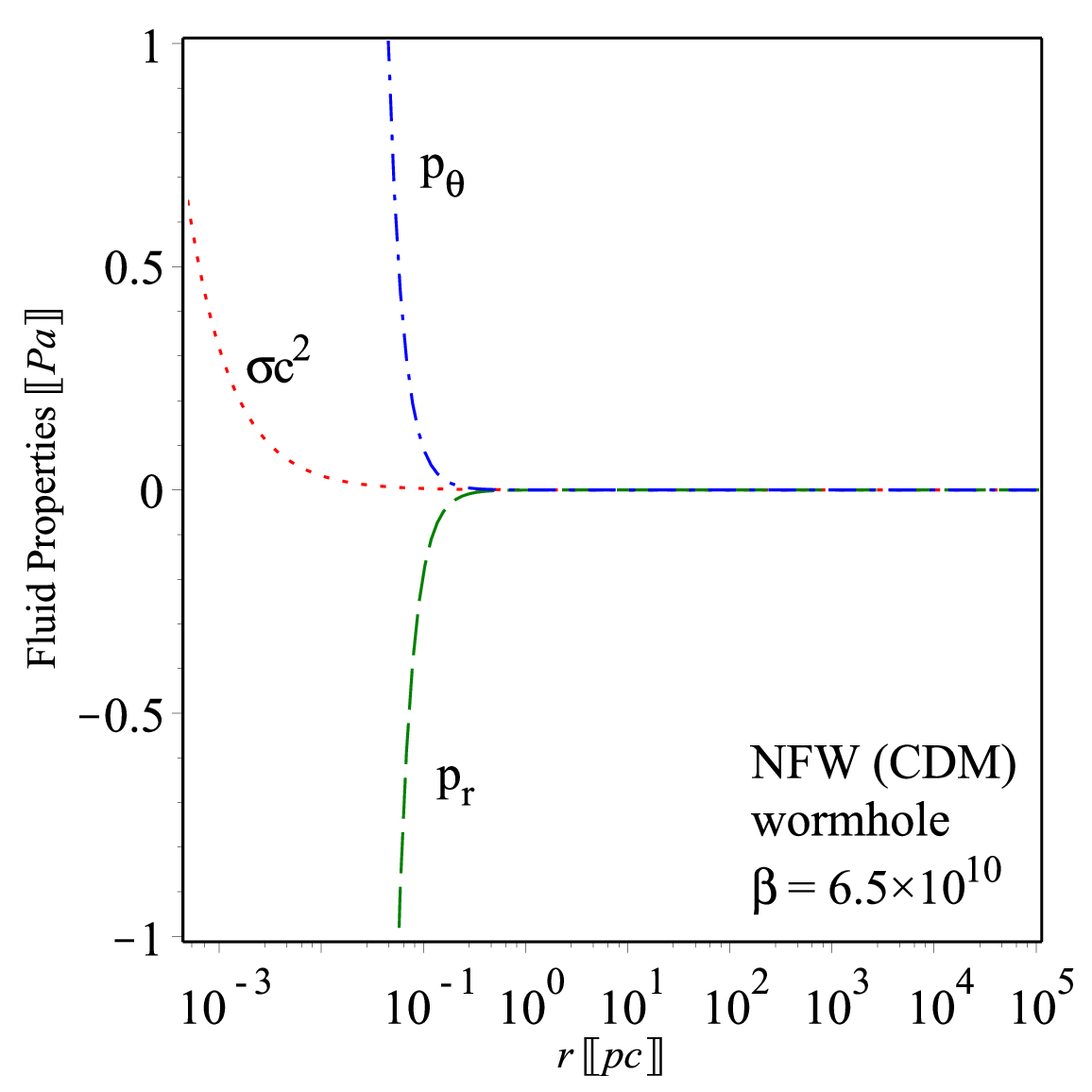}}
\subfigure[~Matter energy conditions]{\label{fig:model2_MEC_pve}\includegraphics[scale=.22]{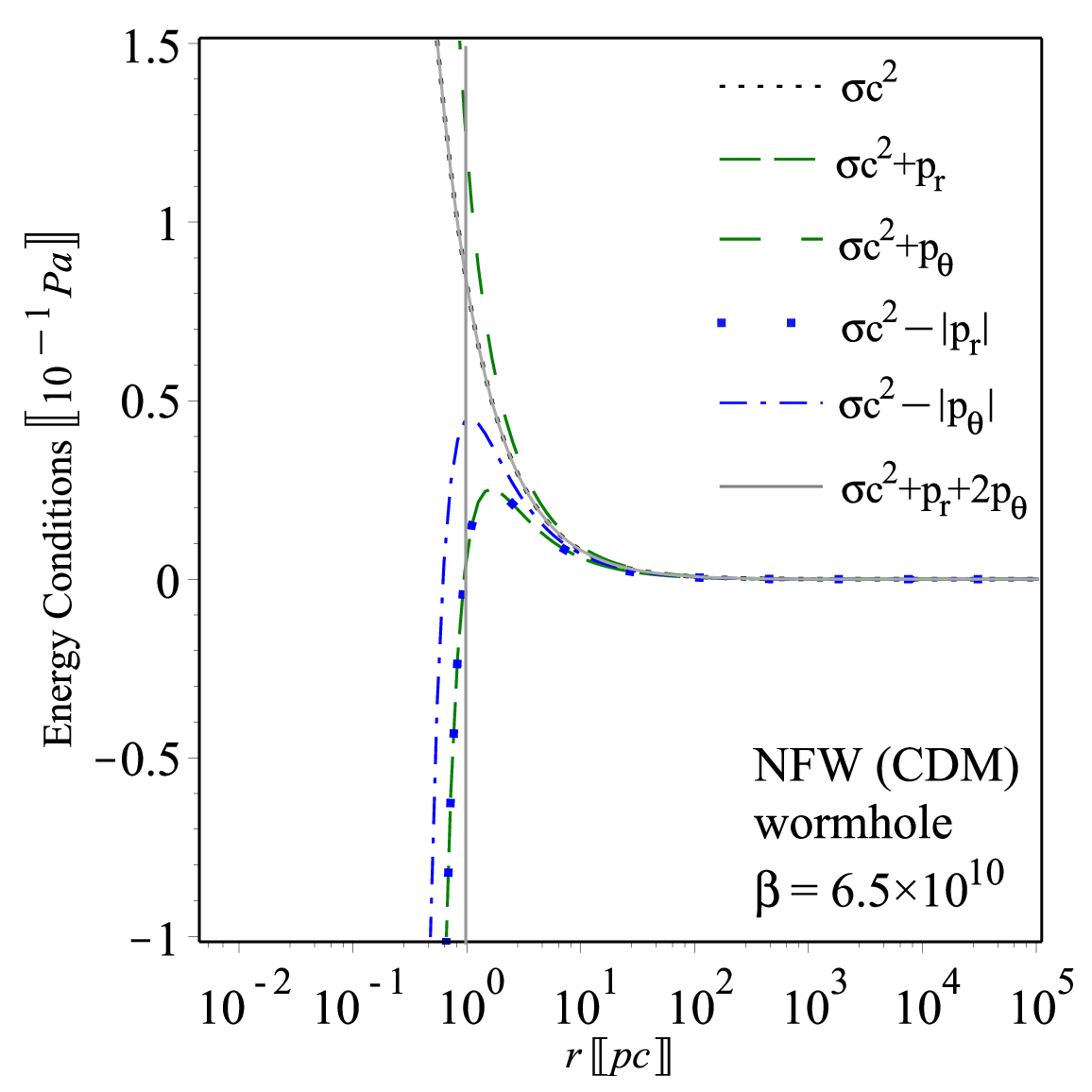}}
\subfigure[~Effective energy conditions]{\label{fig:model2_EEC_pve}\includegraphics[scale=.22]{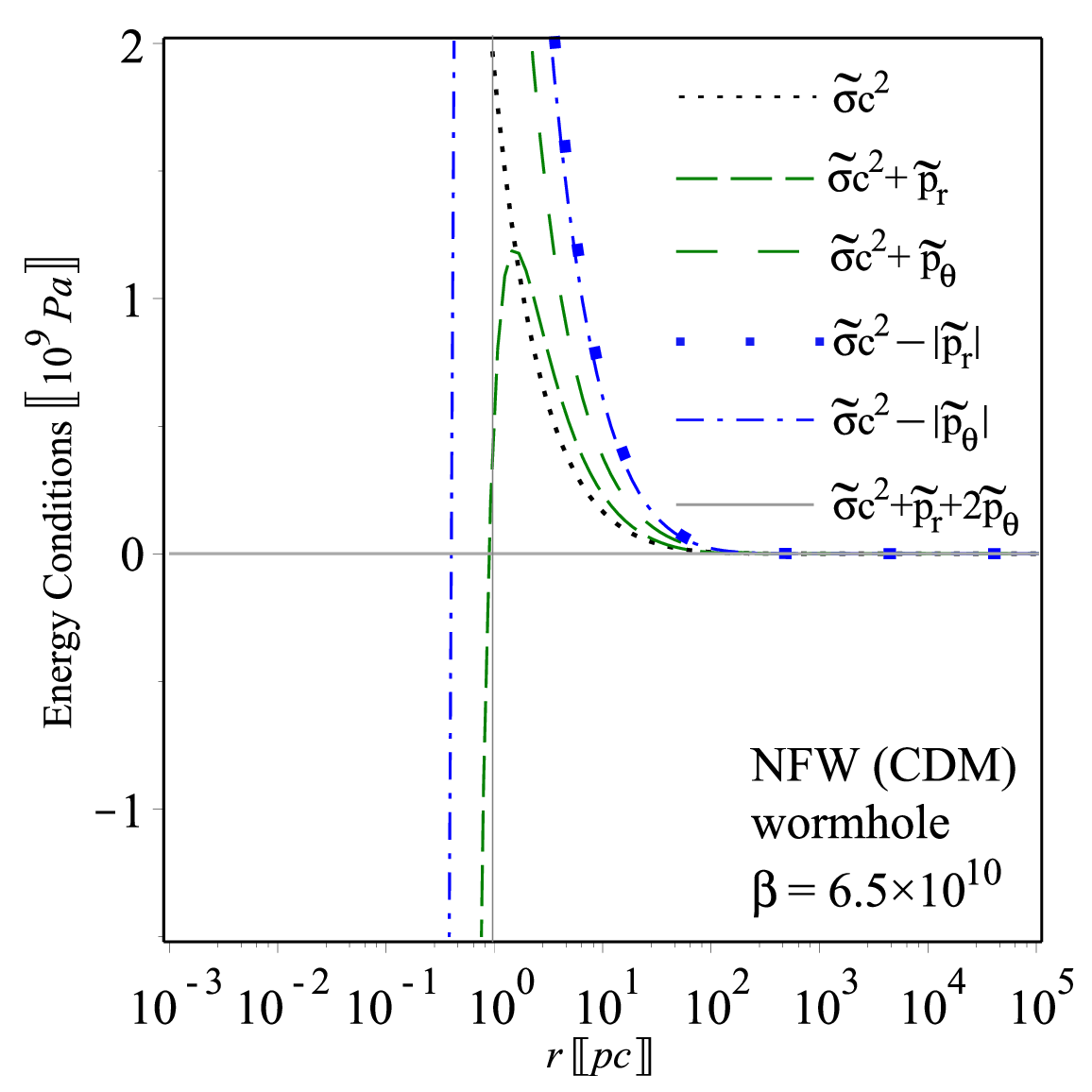}}
\caption[figtopcap]{Model II with large positive coupling $\beta\geq 6.140 \times 10^{10}$:
\subref{fig:model2_shape_pve} The WH is not Lorentzian as $h(r)>r$ at $r>r_0$, additionally the flaring-out condition is broken.
\subref{fig:model2_fluid_pve} The matter fluid, the density has a cuspy profile at the core region, whereas $p_r<0$ and $p_\theta>0$.
\subref{fig:model2_MEC_pve} The matter energy conditions are fulfilled at $r>r_0$.
\subref{fig:model2_EEC_pve}  The effective energy conditions are fulfilled at $r>r_0$.
We set $r_0=1$ pc, for the galaxy NGC 2366, $\sigma_s=3.11 \times 10^{-3}~\text{M}_{\odot}/\text{pc}^3$ and ${r_s} = 1.447$ kpc \cite{Banares-Hernandez:2023axy}.}
\label{Fig:Model2_EC_pve}
\end{figure}

For large positive $\beta\geq 6.140 \times 10^{10}$, the matter density and pressures are given in Fig. \ref{Fig:Model2_EC_pve}\subref{fig:model2_fluid_pve}. It can be noted that all energy conditions are satisfied for both the matter and the effective (including the non-minimal coupling effect) sectors at $r>r_0$ as seen in Figs. \ref{Fig:Model2_EC_pve}\subref{fig:model2_MEC_pve} and \ref{Fig:Model2_EC_pve}\subref{fig:model2_EEC_pve}. However, the flaring-out condition and the Lorentzian signature conditions are clearly broken as shown in Fig. \ref{Fig:Model2_EC_pve}\subref{fig:model2_shape_pve}. This confirms the exclusion the interval $\beta\geq 5.388\times 10^{6}$ as it cannot represent a traversable WH.

\begin{figure}
\centering
\subfigure[~The shape function]{\label{fig:model2_shape_nve}\includegraphics[scale=.22]{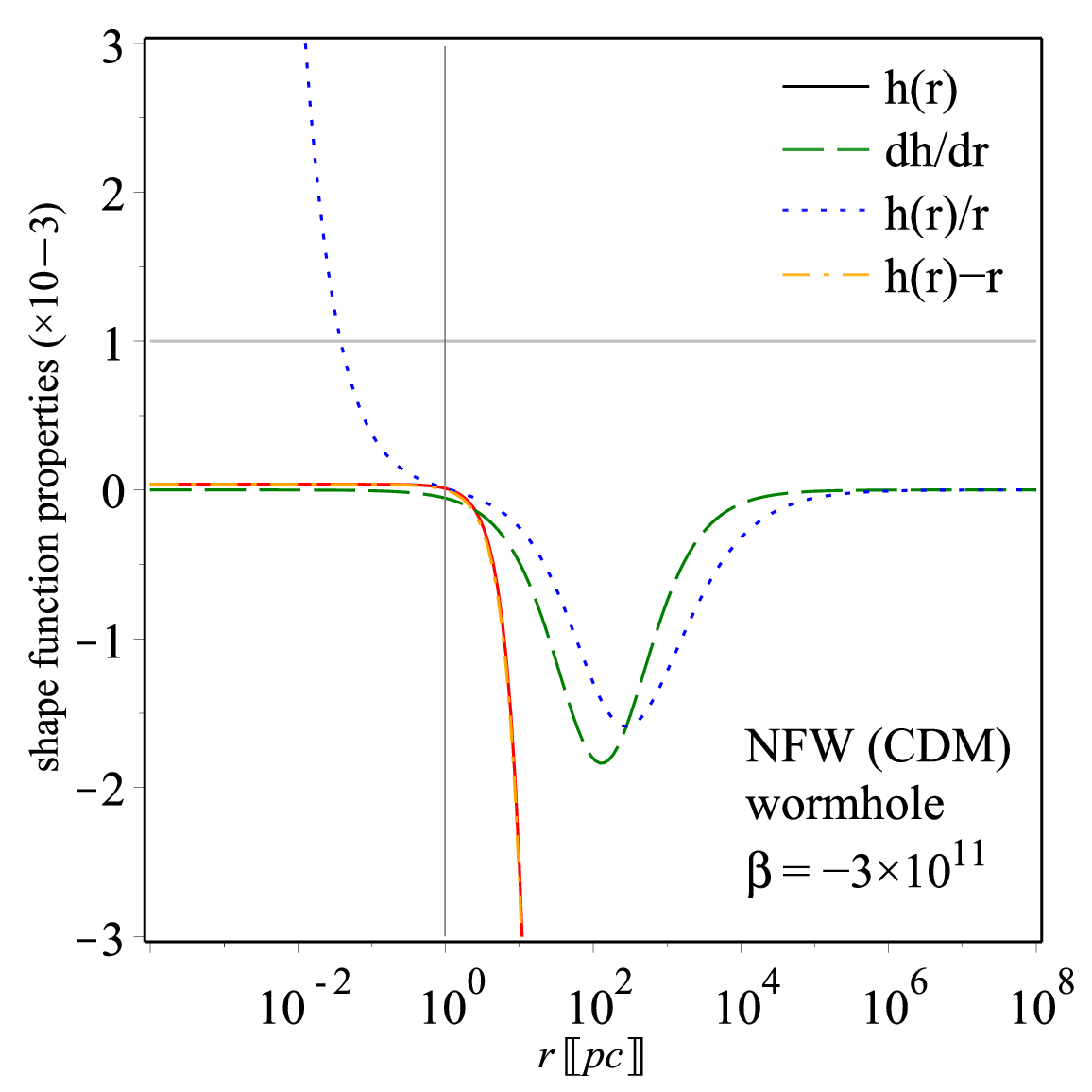}}
\subfigure[~The WH fluid]{\label{fig:model2_fluid_nve}\includegraphics[scale=.22]{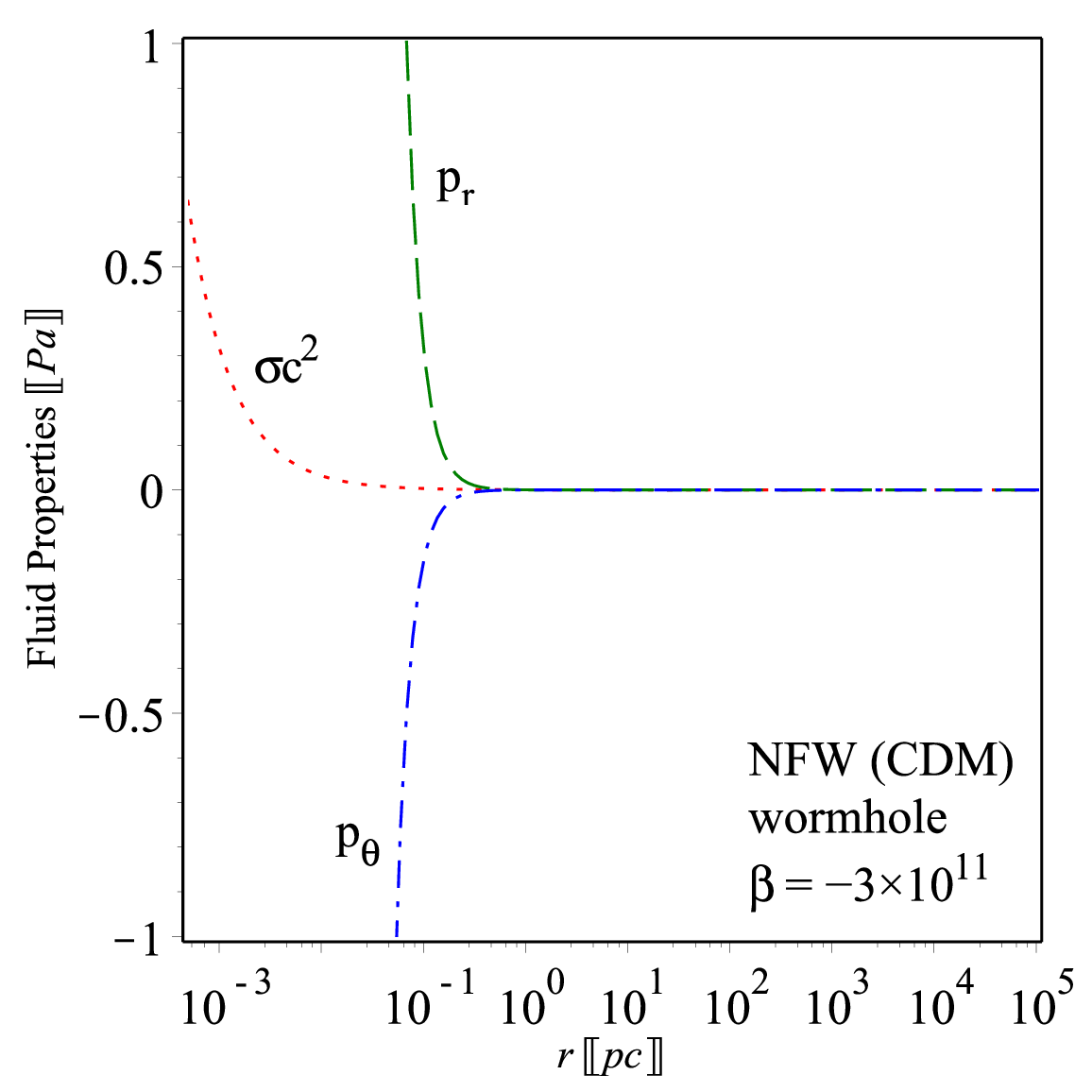}}
\subfigure[~Matter energy conditions]{\label{fig:model2_MEC_nve}\includegraphics[scale=.22]{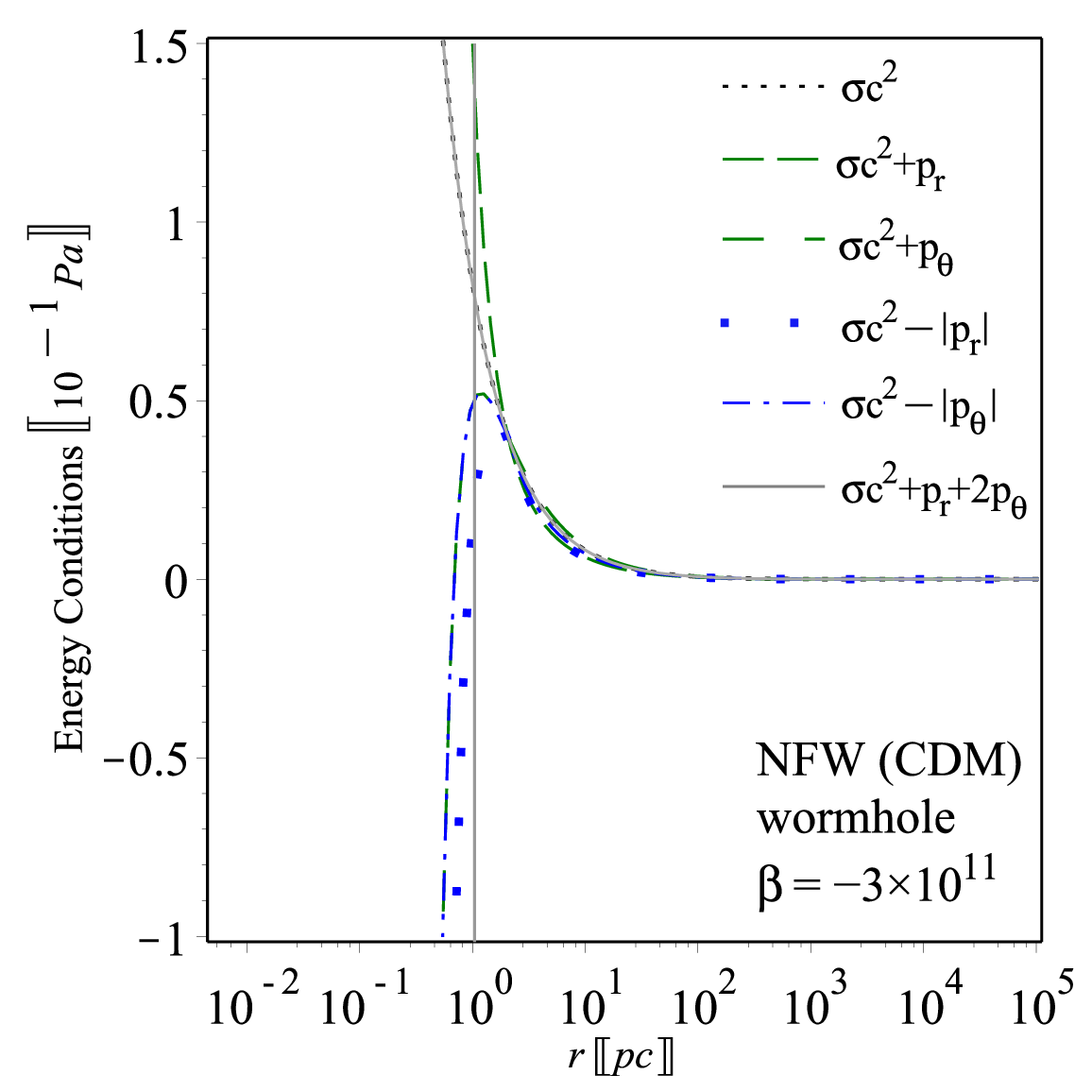}}
\subfigure[~Effective energy conditions]{\label{fig:model2_EEC_nve}\includegraphics[scale=.22]{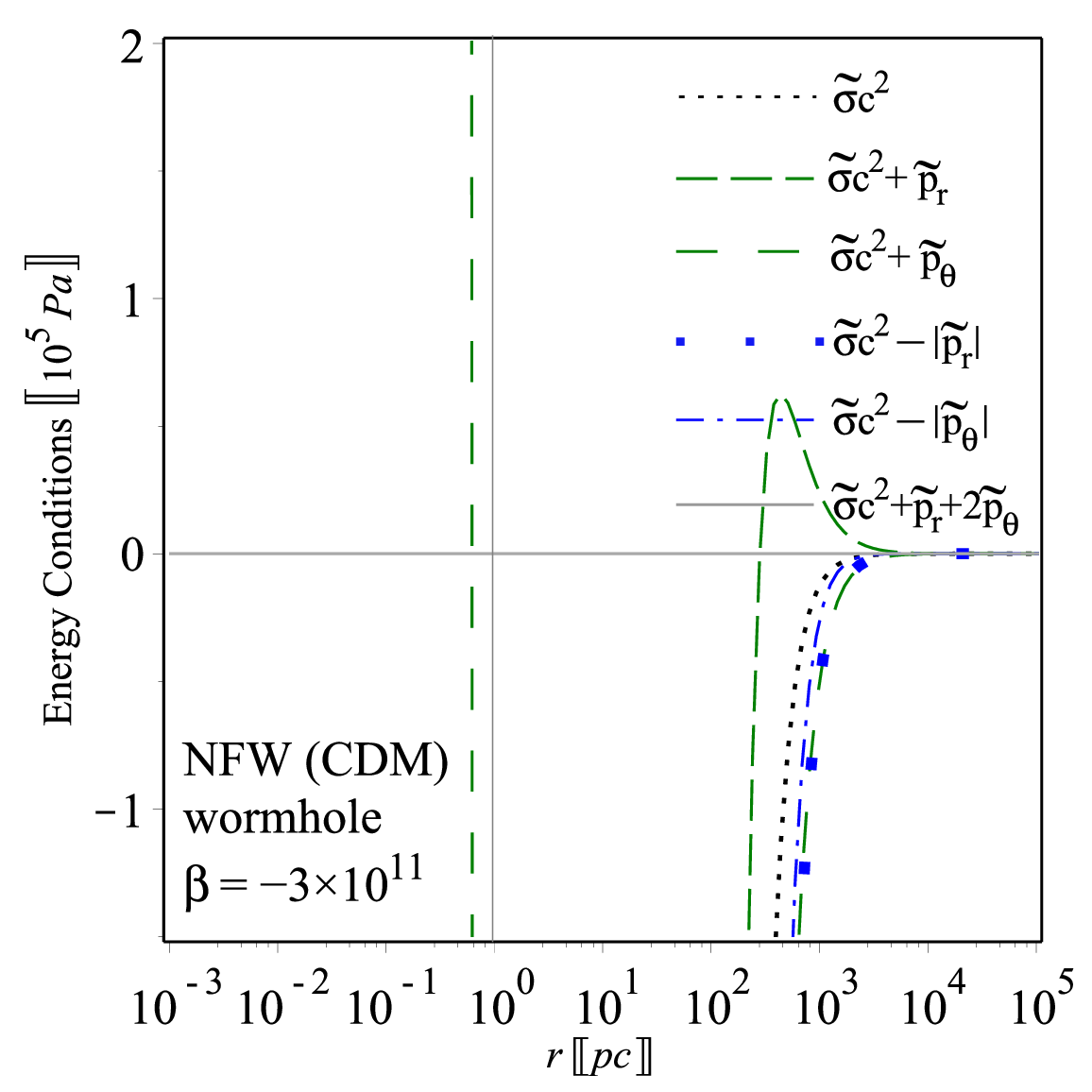}}
\caption[figtopcap]{Model II with large negative coupling $\beta\leq -1.842\times 10^{11}$: \subref{fig:model2_shape_nve} The WH is Lorentzian as $h(r)<r$ at $r>r_0$, additionally the flaring-out condition is fulfilled. \subref{fig:model2_fluid_nve} The matter fluid, the density has a flat profile at the core as suggested to solve the core-cusp problem with $p_r>0$ and $p_\theta<0$. \subref{fig:model2_MEC_nve} The matter energy conditions are fulfilled at $r>r_0$. \subref{fig:model2_EEC_nve} The effective energy conditions are broken at $r>r_0$ which represents the GR case. We set $r_0=1$ pc, for the galaxy NGC 2366, $\sigma_s=3.11 \times 10^{-3}~\text{M}_{\odot}/\text{pc}^3$ and ${r_s} = 1.447$ kpc \cite{Banares-Hernandez:2023axy}.}
\label{Fig:Model2_EC_nve}
\end{figure}

For large negative $\beta\leq -1.842\times 10^{11}$, the matter density and pressures are given in Fig. \ref{Fig:Model2_EC_nve}\subref{fig:model2_fluid_nve}. It can be noted that all energy conditions are satisfied for the matter fluid at $r>r_0$ as seen in Figs. \ref{Fig:Model2_EC_nve}\subref{fig:model2_MEC_nve}. On the contrary, the energy conditions are broken on the effective sector as seen by Fig. \ref{Fig:Model2_EC_nve}\subref{fig:model2_EEC_nve}, which is must occur if the flaring-out of the WH is hold. In Fig. \ref{Fig:Model2_EC_nve}\subref{fig:model2_shape_nve}, we show that the WH solution, in this case, satisfies all conditions related to the shape function and its derivative.

In conclusion, the above calculations confirm the possibility to find a healthy WH solution where the energy conditions are broken only on the effective sector but satisfied on the matter sector. Consequently, no exotic matter is needed to form a physical WH as required in the GR framework, if non-minimal coupling between matter and geometry has been considered.
%%%%%%%%%%%%%%%%%%%%%%%%%%%%%%%%%%%%%%%%%%%%%%%%%%%%%%%%%%%%%%%%%%%%%%%%%%%%%%%%%%%%%%%%%%%%%%%%
\section{Stability of the wormhole solutions}\label{SEC:VII}
%%%%%%%%%%%%%%%%%%%%%%%%%%%%%%%%%%%%%%%%%%%%%%%%%%%%%%%%%%%%%%%%%%%%%%%%%%%%%%%%%%%%%%%%%%%%%%%%
In this section, we investigate the stability of the obtained WH solutions. Stability is often verified using hydrostatic equilibrium constraints, known as the Tolman-Oppenheimer-Volkoff (TOV) equation \cite{Oppenheimer:1939ne,Gorini:2008zj,Kuhfittig:2020fue}, showing that these WHs can be dynamically stable. Recalling the field equations \eqref{eq:Einstein_frame}, which show that the conservation is no longer hold on the matter sector $\nabla_\mu \cal{T}{^\mu}{_\nu} \neq 0$, since $\nabla_\mu \cal{T}{^\mu}{_\nu^{(\text{NMG})}} \neq 0$. However, the total (effective) stress-energy tensor satisfies $\nabla_\mu \tilde{\cal T}{^\mu}{_\nu} = 0$ as required by applying Bianchi identity to the left hand side of the field equations. This modifies TOV equation introducing a new force due to non-minimal coupling between matter and geometry. For the present theory, $f({\cal Q},{\cal T})={\cal Q}+\beta {\cal T}$, we write the following  modified TOV equation:
\begin{eqnarray}\label{eq:TOV}
p'_r=-\frac{3(1+2\beta)(\sigma c^2+{p_r})}{3+7\beta}\xi'+\beta\frac{3\sigma' c^2-2 p'_\theta}{3+7\beta}+\frac{6(1+2\beta)({p_\theta}-{p_r})}{(3+7\beta)r}.
\end{eqnarray}
Then, we define the contributing forces in the above equation, where the hydrostatic, gravitational, coupling and anisotropic forces are respectively given as
\begin{eqnarray}\label{eq:TOV_forces}
F_h=-p'_r,~F_g=-\frac{3(1+2\beta)(\sigma c^2+{p_r})}{3+7\beta}\xi',~ F_c=\beta\frac{3\sigma' c^2-2 p'_\theta}{3+7\beta},~ F_a=\frac{6(1+2\beta)({p_\theta}-{p_r})}{(3+7\beta)r}\,.
\end{eqnarray}
Clearly, by setting the non-minimal coupling coefficient $\beta=0$, the coupling force $F_c$ vanishes, and the above hydrostatic equilibrium equation reduces to the GR version
\begin{equation*}\label{eq:GR_TOV}
p'_r=-(\sigma c^2+{p_r})\xi'+\frac{2}{r}({p_\theta}-{p_r}).
\end{equation*}
In order for the WH solutions to be stable, the sum of the forces $F_h$, $F_g$, $F_c$ and $F_a$ should be zero. Given the redshift function $\xi(r)=\xi_0= \text{constant}$ as is assumed in this study, there is no tidal force ${\mathcal F_g}=0$, resulting in the stability constraint being reformulated as:
\begin{equation}
\label{eq:TOV_forces2}
F_h+ F_c+ F_a=0.
\end{equation}
We note that the coefficient of the anisotropic force in Eq. \eqref{eq:TOV_forces}, i.e. $\frac{1+2\beta}{3+7\beta}=\frac{2}{7}+\frac{1}{7(3+7\beta)}\to \frac{2}{7}$ for large $|\beta|$, in addition it is always positive as $\beta<-1/2$ as required by the NEC. In this case, the anisotropic force could be repulsive (attractive) when $p_r<p_\theta$ ($p_r>p_\theta$). For the WH solution, assuming $f(\cal{Q}, \cal{T})=\cal{Q}+\beta \cal{T}$ with $\xi(r)=\xi_0$, the forces can be written in terms of the shape function as
\begin{eqnarray}
    F_h&=&-\frac{3h}{(1+2\beta)\kappa^2 r^4}+\frac{(3+20 \beta) h'}{3(1+2\beta)(1+4\beta)\kappa^2 r^3}-\frac{4\beta h''}{3(1+2\beta)(1+4\beta)\kappa^2 r^2},\label{eq:hforce}\\
    F_c&=&\frac{3\beta h}{(1+2\beta)(3+7\beta)\kappa^2 r^4}-\frac{\beta(27+68 \beta) h'}{3(1+2\beta)(1+4\beta)(3+7\beta)\kappa^2 r^3}+\frac{4\beta h''}{3(1+2\beta)(1+4\beta)\kappa^2 r^2},\label{eq:cforce}\\
    F_a&=&\frac{9h}{(3+7\beta)\kappa^2 r^4}-\frac{3 h'}{(3+7\beta)\kappa^2 r^3}.\label{eq:aforce}
\end{eqnarray}
One can easily fulfill the hydrostatic equilibrium constraint, namely \eqref{eq:TOV_forces2}, by virtue of the above equations.
%%%%%%%%%%%%%%%%%%%%%%%%%%%%%%%%%%%%%%%%%%%%%%%%%%%%%%%%%%%%%%%%%%%%%
\subsection{Model I}\label{Sec:TOVI}
\begin{figure}
\centering
\subfigure[~TOV forces, $\beta=-0.6$]{\label{fig:Model1_Forces}\includegraphics[scale=.22]{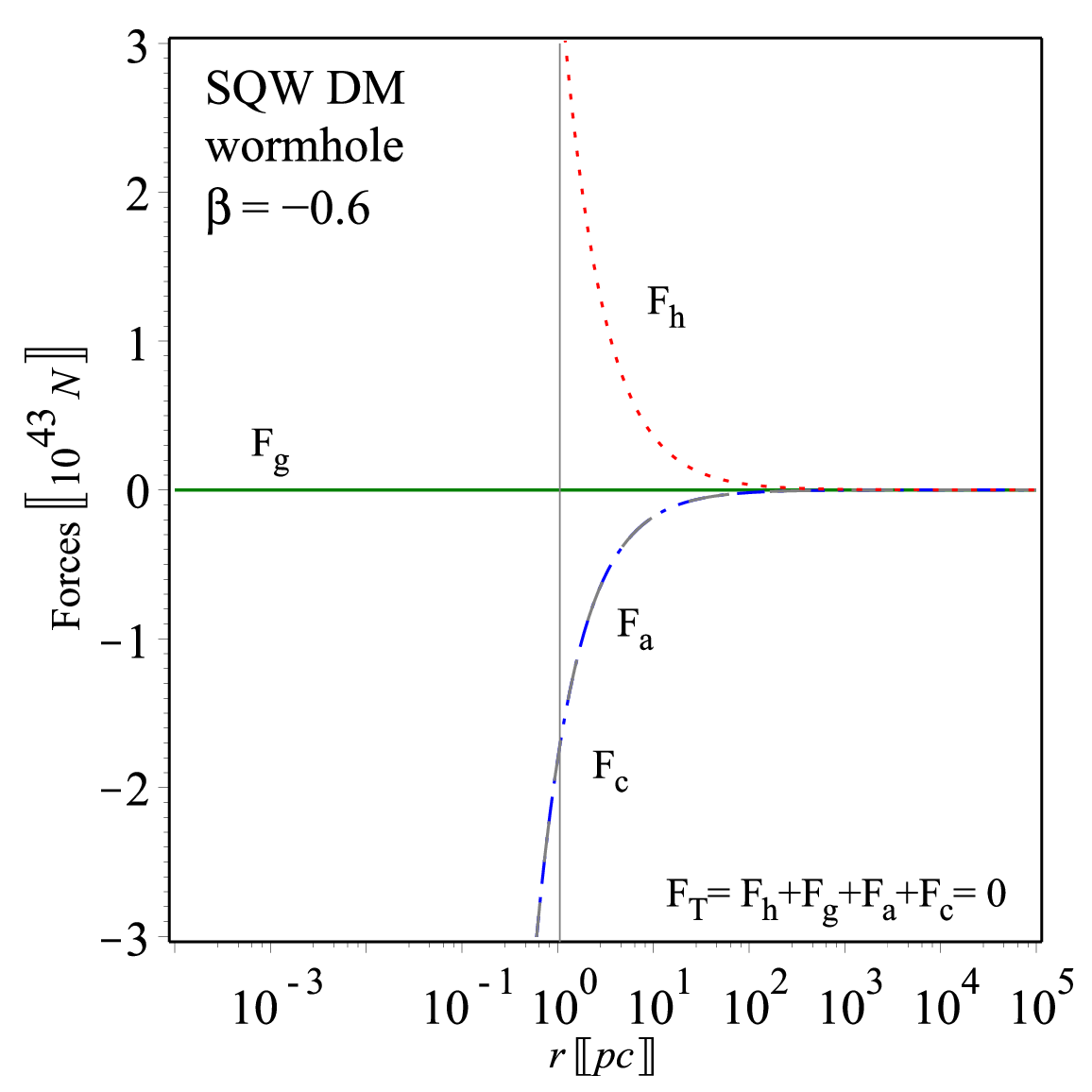}}\hspace{0.5cm}
\subfigure[~TOV forces, large $\beta>0$]{\label{fig:Model1_Forces_pve}\includegraphics[scale=.22]{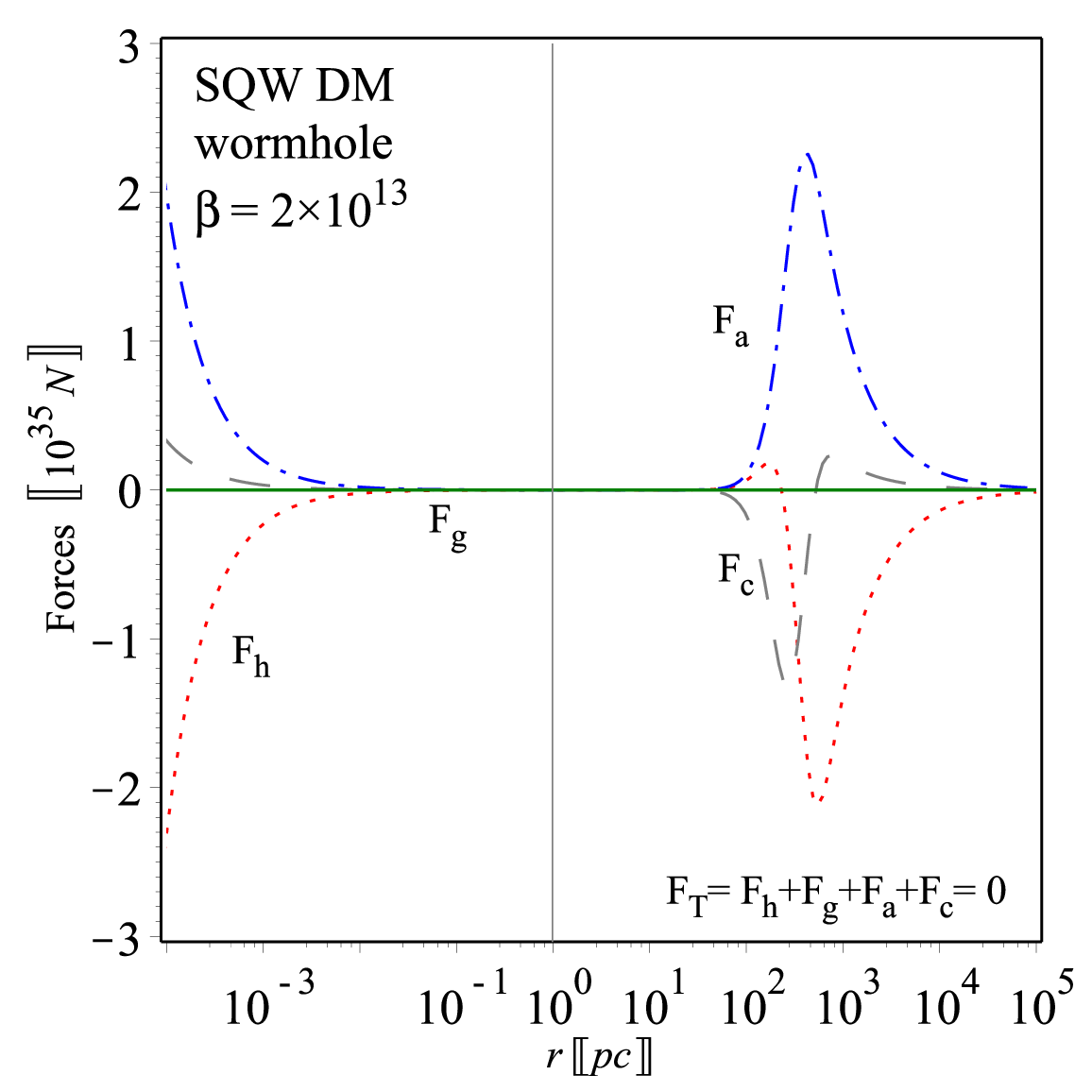}}\hspace{0.5cm}
\subfigure[~TOV forces, large $\beta<0$]{\label{fig:Model1_Forces_nve}\includegraphics[scale=.22]{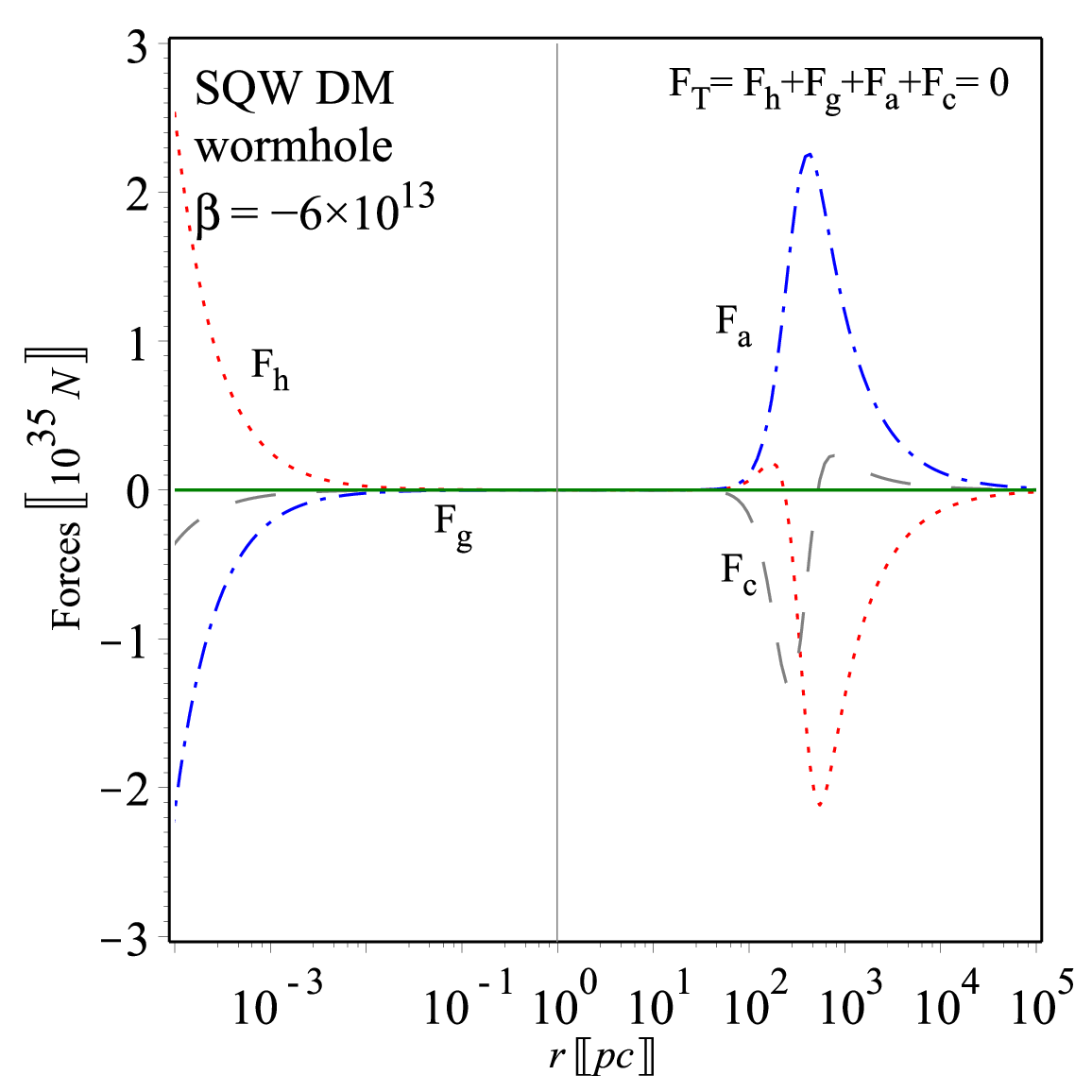}}
\caption[figtopcap]{Model I constituent forces of the modified TOV equation \eqref{eq:TOV}. The absence of the gravitational force is due to our choice of the redshift function $\xi(r)=\xi_0$, while other constituent forces $F_h$, $F_c$ and $F_a$ as given by \eqref{eq:Model1_Fh}--\eqref{eq:Model1_Fa}.
\subref{fig:Model1_Forces} For $\beta=-0.6<-1/2$, which is consistent with the flaring-out in addition to the NEC, the repulsive hydrostatic force $F_h>0$ is compensated by other two attractive forces, those are the anisotropic force $F_a<0$ (i.e. $p_r>p_\theta$) in addition to the coupling force $F_c$ which is attractive in this case.
\subref{fig:Model1_Forces_pve} For $\beta\geq 1.840\times 10^{13}$, which satisfies all the energy conditions but breaks the flaring-out condition, the anisotropic force is repulsive (i.e. $p_r>p_\theta$) and balanced by the other two forces, the hydrostatic force and the force due to the non-minimal coupling between matter and geometry, those are interpolating between attractive and repulsive behavior at different radial distance $r>r_0$. However, the resultant force is null in all distances.
\subref{fig:Model1_Forces_nve}  For $\beta\leq -5.519\times 10^{13}$, similar to large positive $\beta$ case, but it satisfies all the energy conditions in addition to the flaring-out condition.
 We set $r_0=1$ pc, for the galaxy NGC 2366, where $\sigma_c=15 \times 10^{-3}~M_\odot$/pc$^{3}$ and $r_c=3$ kpc \cite{Banares-Hernandez:2023axy}.}
\label{Fig:Model1_TOV_forces}
\end{figure}

For soliton model of DM, we insert the shape function \eqref{eq:soliton_h} into Eqs \eqref{eq:hforce}--\eqref{eq:aforce}, { which derives the corresponding forces $F_h$, $F_c$, and $F_a$ which we list them in Appendix \ref{appEI}.} We plot the above mentioned forces for the galaxy NGC 2366, setting $r_0=1$ pc, $\sigma_c=15 \times 10^{-3}~M_\odot$/pc$^{3}$ and $r_c=3$ kpc, for three different values of the non-minimal coupling parameter $\beta$ as seen in Fig. \ref{Fig:Model1_TOV_forces}.
(i) For $\beta=-0.6<-1/2$, see Fig. \ref{Fig:Model1_TOV_forces}\subref{fig:Model1_Forces}, which is consistent with the flaring-out in addition to the NEC, the repulsive hydrostatic force $F_h>0$ is compensated by other two attractive forces, those are the anisotropic force $F_a<0$ (i.e. $p_r>p_\theta$) in addition to the coupling force $F_c$ which is attractive in this case.
(ii) For $\beta\geq 1.840\times 10^{13}$, see Fig. \ref{Fig:Model1_TOV_forces}\subref{fig:Model1_Forces_pve}, which satisfies all the energy conditions but breaks the flaring-out condition, the anisotropic force is repulsive (i.e. $p_r>p_\theta$) and balanced by the other two forces, the hydrostatic force and the force due to the non-minimal coupling between matter and geometry, those are interpolating between attractive and repulsive behavior at different radial distance $r>r_0$. However, the resultant force is null in all distances.
(iii)  For $\beta\leq -5.519\times 10^{13}$, see Fig. \ref{Fig:Model1_TOV_forces}\subref{fig:Model1_Forces_nve}, similar to large positive $\beta$ case, but it satisfies all the energy conditions in addition to the flaring-out condition.
%%%%%%%%%%%%%%%%%%%%%%%%%%%%%%%%%%%%%%%%%%%%%%%%%%%%%%%%%%%%%%%%%%%%%%%%%%%%%%%%%%%%%%%
\subsection{Model II}\label{Sec:TOVII}
\begin{figure}
\centering
\subfigure[~TOV forces, $\beta=-0.6$]{\label{fig:Model2_Forces}\includegraphics[scale=.22]{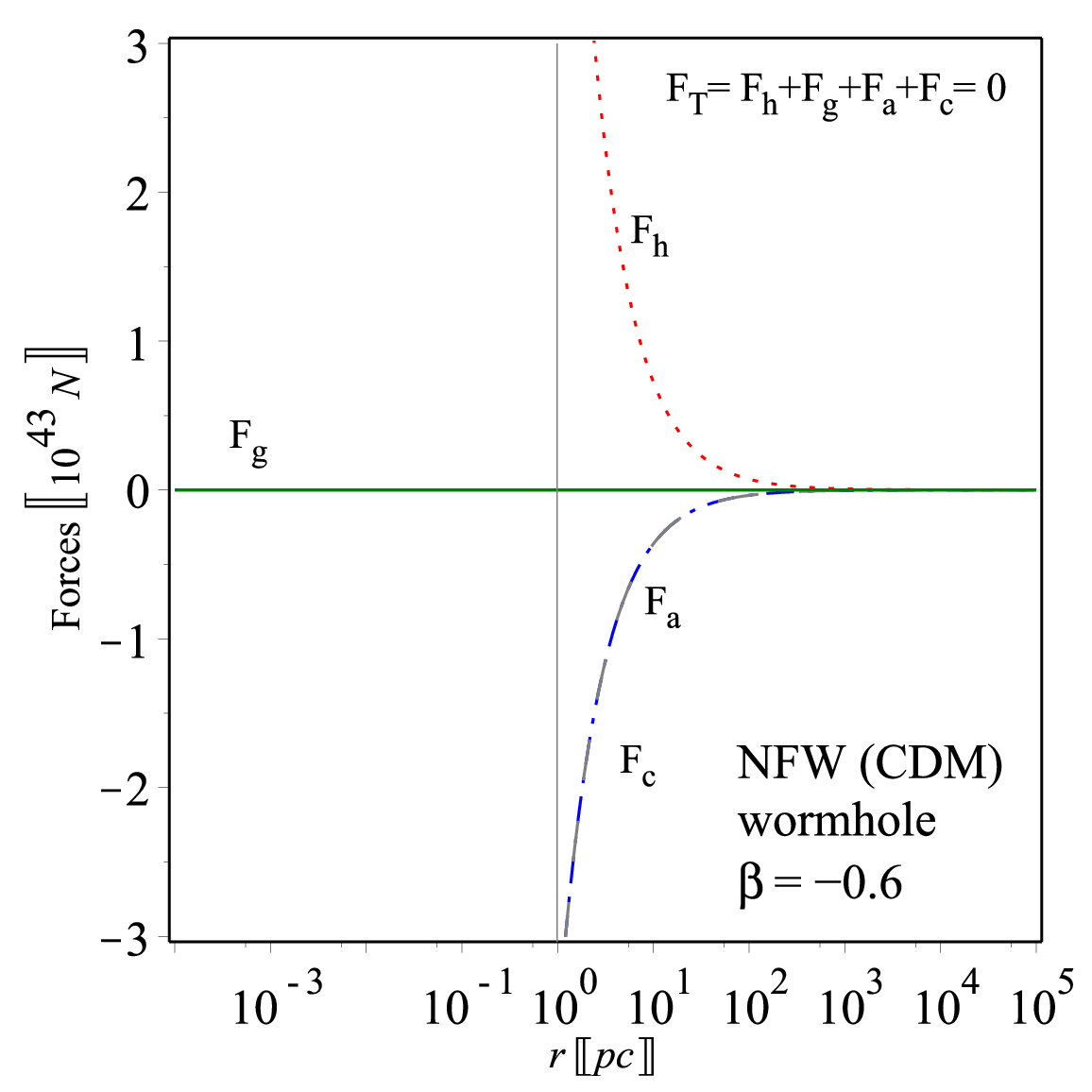}}\hspace{0.5cm}
\subfigure[~TOV forces, large $\beta>0$]{\label{fig:Model2_Forces_pve}\includegraphics[scale=.22]{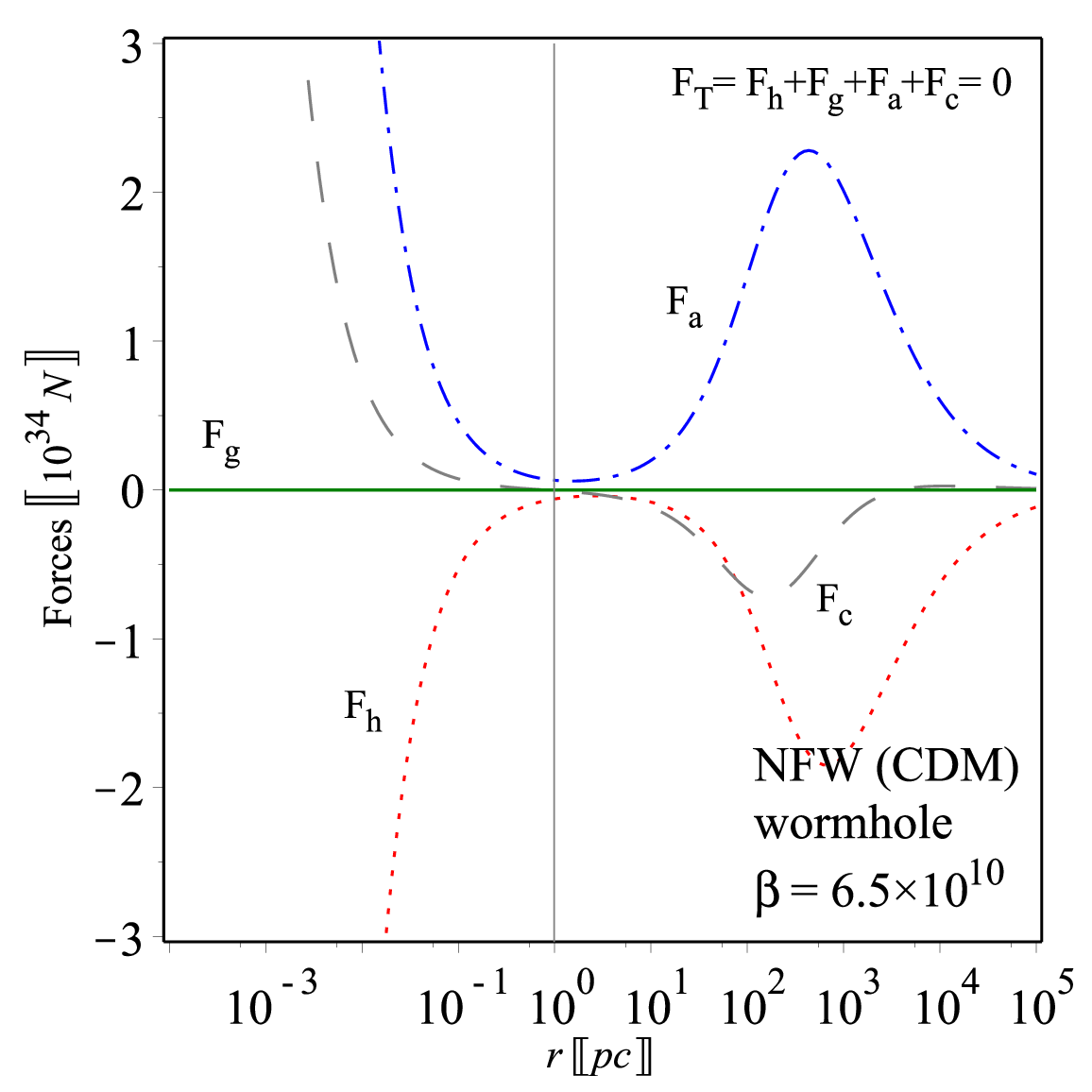}}\hspace{0.5cm}
\subfigure[~TOV forces, large $\beta<0$]{\label{fig:Model2_Forces_nve}\includegraphics[scale=.22]{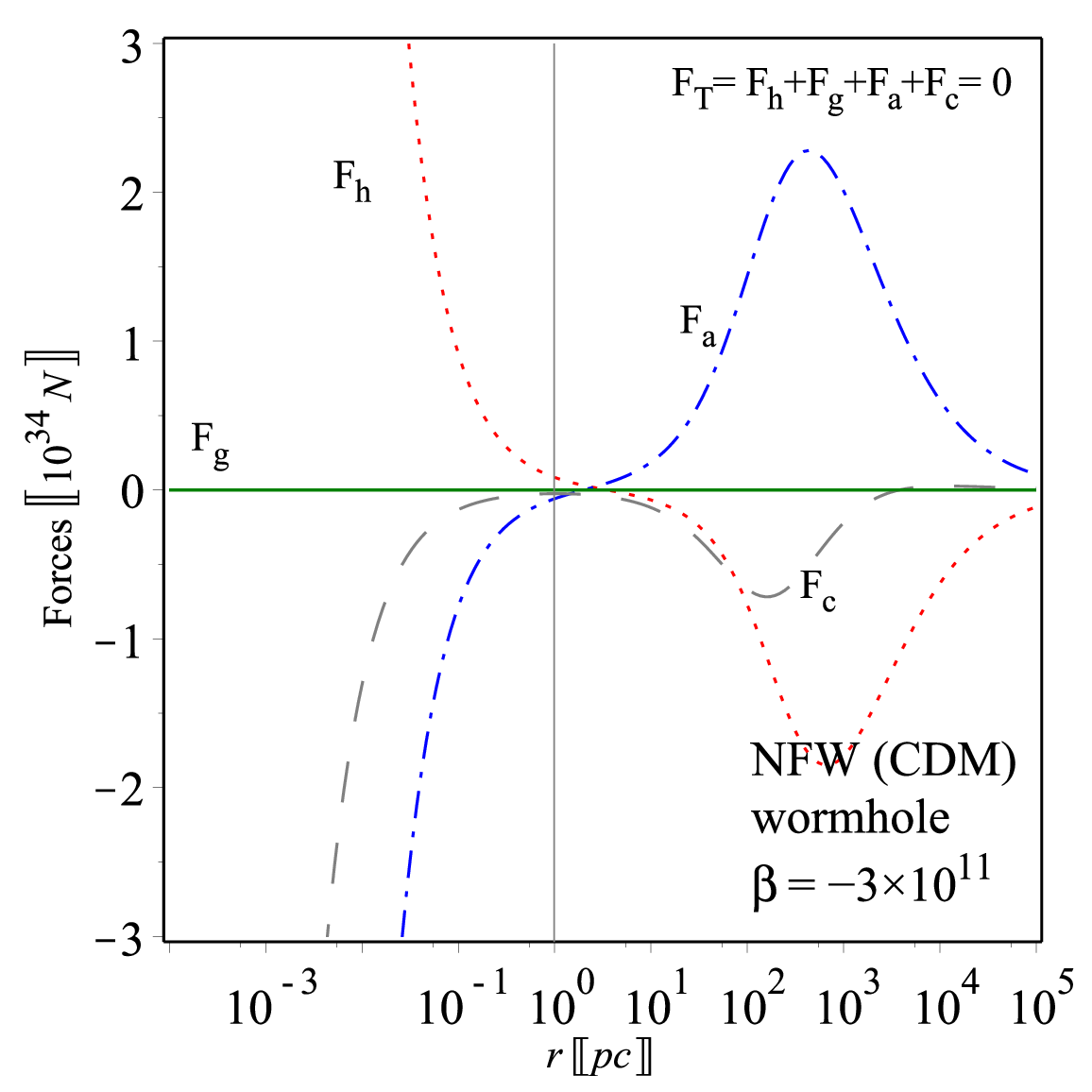}}
\caption[figtopcap]{Model II constituent forces of the modified TOV equation \eqref{eq:TOV}. The absence of the gravitational force is due to our choice of the redshift function $\xi(r)=\xi_0$, while constituent forces $F_h$, $F_c$ and $F_a$ as given by \eqref{eq:Model2_Fh}--\eqref{eq:Model2_Fa}.
\subref{fig:Model2_Forces} For $\beta=-0.6<-1/2$, which is consistent with the flaring-out in addition to the NEC, the repulsive hydrostatic force $F_h>0$ is compensated by the attractive anisotropic force $F_a<0$ (i.e. $p_r>p_\theta$) in addition to the coupling force $F_c$ which is attractive in this case.
\subref{fig:Model2_Forces_pve} For $\beta\geq 6.140 \times 10^{10}$, which satisfies all the energy conditions but breaks the flaring-out condition, the anisotropic force is repulsive (i.e. $p_r>p_\theta$) and balanced by the other two forces, the hydrostatic force and the force due to the non-minimal coupling between matter and geometry, those are interpolating between attractive and repulsive behavior at different radial distance $r>r_0$. However, the resultant force is null in all distances.
\subref{fig:Model2_Forces_nve} For $\beta\leq -1.842\times 10^{11}$, similar to large positive $\beta$ case, but it satisfies all the energy conditions in addition to the flaring-out condition.
We set $r_0=1$ pc, for the galaxy NGC 2366, $\sigma_s=3.11 \times 10^{-3}~\text{M}_{\odot}/\text{pc}^3$ and ${r_s} = 1.447$ kpc \cite{Banares-Hernandez:2023axy}.}
\label{Fig:Model2_TOV_forces}
\end{figure}

For NFW model of DM, we insert the shape function \eqref{eq:NFW_h} into Eqs \eqref{eq:hforce}--\eqref{eq:aforce},  { which derives the corresponding forces $F_h$, $F_c$, and $F_a$ which we list them in Appendix \ref{appEII}.} We plot the above mentioned forces for the galaxy NGC 2366, setting $r_0=1$ pc, $\sigma_s=3.11 \times 10^{-3}~\text{M}_{\odot}/\text{pc}^3$ and ${r_s} = 1.447$ kpc, for three different values of the non-minimal coupling parameter $\beta$ as seen in Fig. \ref{Fig:Model2_TOV_forces}.
(i) For $\beta=-0.6<-1/2$, see Fig. \ref{Fig:Model2_TOV_forces}\subref{fig:Model2_Forces}, which is consistent with the flaring-out in addition to the NEC, the repulsive hydrostatic force $F_h>0$ is compensated by other two attractive forces, those are the anisotropic force $F_a<0$ (i.e. $p_r>p_\theta$) in addition to the coupling force $F_c$ which is attractive in this case.
(ii) For $\beta\geq 6.140 \times 10^{10}$, see Fig. \ref{Fig:Model2_TOV_forces}\subref{fig:Model2_Forces_pve}, which satisfies all the energy conditions but breaks the flaring-out condition, the anisotropic force is repulsive (i.e. $p_r>p_\theta$) and balanced by the other two forces, the hydrostatic force and the force due to the non-minimal coupling between matter and geometry, those are interpolating between attractive and repulsive behavior at different radial distance $r>r_0$. However, the resultant force is null in all distances.
(iii)  For $\beta\leq -1.842\times 10^{11}$, see Fig. \ref{Fig:Model2_TOV_forces}\subref{fig:Model2_Forces_nve}, similar to large positive $\beta$ case, but it satisfies all the energy conditions in addition to the flaring-out condition.

{ The lack of a dynamical or perturbative stability investigation for the WH solutions in $f({\cal Q},{\cal T})$ gravity is a significant drawback of our current approach. Despite our thorough analysis of the static stability using a modified Tolman-Oppenheimer-Volkoff (TOV) equilibrium condition, the evolution of minor perturbations around the static backdrop is not taken into consideration by this method. Since the non-metricity ${\cal Q}$ and the trace of the stress-energy tensor ${\cal T}$ support the structure of the WHs in $f({\cal Q},{\cal T})$ gravity, it is essential to examine their dynamical behavior under perturbations in order to completely evaluate their stability.
The linearized Einstein field equations for the perturbed quantities would be derived by perturbing the matter variables and WH metric in order to perform a dynamical stability analysis. This would enable us to ascertain if the solutions are stable or unstable under perturbations by examining whether any minor fluctuations increase or decrease over time. Furthermore, given the assertion that exotic matter is avoided because of the coupling structure, it may be crucial to comprehend how these solutions react to perturbations in order to evaluate their resilience \cite{DelAguila:2021awa,Kang:2019ost}.

In subsequent work, we intend to expand this study by using a linear perturbation theory to investigate the stability of the WH configurations. In order to get the evolution equations for both scalar and tensor perturbations, we would need to perturb the background geometry using a time-dependent metric ansatz. For assessing the feasibility of these WH solutions in more dynamic situations, the derived equations would allow us to ascertain how the system would behave under minor departures from equilibrium. In order to examine the stability of the WH solutions in the presence of perturbations, it would also be intriguing to evaluate their quasinormal modes. These modes offer crucial information on the propagation of perturbations and whether they result in decay, which denotes stability, or exponential development, which indicates instability. A more thorough understanding of the physical characteristics of the solutions would be provided by such a study, which would supplement our present findings, c.f. \cite{Hassan:2025vko} for more details.}
%%%%%%%%%%%%%%%%%%%%%%%%%%%%%%%%%%%%%%%%%%%%%%%%%%%%%%%%%%%%%%%%%%%%%%%%%%%%%%%%%%%%%%%%%%%%%%%%
\section{Potential observational signatures}\label{SEC:VIII}
%%%%%%%%%%%%%%%%%%%%%%%%%%%%%%%%%%%%%%%%%%%%%%%%%%%%%%%%%%%%%%%%%%%%%%%%%%%%%%%%%%%%%%%%%%%%%%%%
{ Current evidence strongly supports that the centers of galaxies are supermassive BHs, not WHs. However, there are several observational signatures which distinguish WHs from BHs could be revealed by future observations with higher precision. One such feature is gravitational lensing, where WH geometries may give rise to additional relativistic images or unusual magnification patterns that differ from those expected in BH spacetimes \cite{Perlick:2004tq,Tsukamoto:2016zdu}.{Another observable differences between BHs and WHs, other than gravitational lensing, like accretion disk behavior has been discussed \cite{Paul:2019trt}, where WH accretion disks appear on both sides of the WH throat. This unique feature of accretion disk images could play an important role to verify the existence of WHs.} Recent analyses have shown that wormholes can produce shadows with distinct sizes and shapes compared to those of standard BHs, which could be investigated using current and future very-long-baseline interferometry observations, such as those conducted by the Event Horizon Telescope \cite{Nedkova:2013msa,Shaikh:2018kfv}. Moreover, in the context of gravitational wave astronomy, WHs may leave imprints in the form of echoes during the post-merger ringdown phase, offering an additional tool to discriminate them from BHs \cite{Cardoso:2016rao,Bueno:2017hyj}. { It has been proposed that the GW scattering within time-independent scattering theory would provide a tool to confirm the existence of WHs \cite{Bao:2022iaz}. The GW echo signatures in the two universes on both sides of the WH present a unique feature. For a certain range of the WH mass, the transmitted GW that passes through the WH exhibits a unique isolated chirp without an inspiral waveform which is typically seen in binary mergers, whereas the reflected wave has the anti-chirp behavior}. Although still speculative, these observational prospects highlight the relevance of WH solutions as potential candidates for exotic compact objects. { Several methods have been proposed to trace the light rays in the vicinity of WHs to study their shadows. We follow \citep{Nedkova:2013msa} to derive the basic calculations needed to find the impact parameters which characterize the photon's trajectories in the present study}.

%%%%%%%%%%%%%%%%%%%%%%%%%%%%%%%%%%%%%%%%%%%%%%%%%%%%%%%%%%%%%%%%
\subsection{Null geodesics in a wormhole spacetime}
\label{sec:null_geodesics}
%%%%%%%%%%%%%%%%%%%%%%%%%%%%%%%%%%%%%%%%%%%%%%%%%%%%%%%%%%%%%%%%%
Null geodesics, $ds^2 = 0$, in the spacetime of the WHs (\ref{eq:MTWHmetric}) have already been studied in \citep{Nedkova:2013msa}. However, we summarize them here. The Lagrangian describing the motion of a photon in the spacetime of the  WH (\ref{eq:MTWHmetric}), where  we set  redshift function $\xi(r) = 0$,  is given by
\begin{equation}
2\mathcal{L}=- \dot{t}^2+\frac{\dot{r}^2}{1-\frac{h}{r}}+r^2
\left[\dot{\theta}^2+\sin^2\theta\,\dot{\Phi}^2
\right],
\end{equation}
where an overdot represents a differentiation with respect to the affine parameter $\lambda$. Since the Lagrangian is independent of $t$ and $\Phi$, we have two constants of motion, namely, the energy $E$ and the angular momentum $L$ (about the axis of symmetry) of the photon:
\begin{equation}
\mathfrak{p}_t=\frac{\partial \mathcal{L}}{\partial \dot{t}}=\dot{t}=E, \qquad
\mathfrak{p}_\Phi=\frac{\partial \mathcal{L}}{\partial \dot{\Phi}}=r^2 \sin^2\theta \,\dot{\Phi}=L.
\end{equation}
Solving the last two equations, we obtain
\begin{equation}
\dot{t}=E, \quad \dot{\Phi}=\frac{L}{r^2\,\sin^2\theta}.
\end{equation}
The $r$- and $\theta$-component of the momentum are, respectively, given by
\begin{equation}
\mathfrak{p}_r=\frac{\partial \mathcal{L}}{\partial \dot{r}}=\frac{\dot{r}}{1-\frac{h}{r}}, \quad \mathfrak{p}_\theta=\frac{\partial \mathcal{L}}{\partial \dot{\theta}}=r^2
\dot{\theta}.
\label{eq:r_momentum}
\end{equation}
The $r$- and $\theta$-part of the geodesic equations can be obtained by solving the Hamilton-Jacobi equation for photon as,
\begin{equation}
\frac{\partial S}{\partial \lambda}=-\frac{1}{2}g^{\mu\nu}\frac{\partial S}{\partial x^{\mu}}\frac{\partial S}{\partial x^{\nu}},
\label{eq:HJ}
\end{equation}
where $S$ is the Jacobi action. If there is a separable solution, then in terms of the already known constants of the motion, it must take the form
\begin{equation}
S=- E t + L \Phi + S_{r}(r)+S_{\theta}(\theta).
\label{eq:action_ansatz}
\end{equation}
Since the shape function $h(r)$ of the WH (\ref{eq:MTWHmetric}) is function of the radial coordinates $r$ then the Hamilton-Jacobi equation is separable. Inserting Eq. (\ref{eq:action_ansatz}) into Eq. (\ref{eq:HJ}) and separating out the $r$- and $\theta$-part, we obtain \citep{Nedkova:2013msa}
\begin{equation}\label{eq:HJ_theta}
\left(\frac{dS_\theta}{d\theta}\right)^2=Q- \frac{L^2}{\sin^2\theta}, \qquad \left(1-\frac{h}{r}\right)\left(\frac{dS_r}{dr}\right)^2=E^2- \frac{Q }{r^2},
\end{equation}
where $Q$ is the Carter constant. Since $\mathfrak{p}_r=\frac{\partial S}{\partial r}=\frac{dS_r}{dr}$ and $\mathfrak{p}_\theta=\frac{\partial S}{\partial \theta}=\frac{dS_\theta}{d\theta}$, using Eqs. (\ref{eq:r_momentum}), and (\ref{eq:HJ_theta}), we obtain \citep{Nedkova:2013msa}
\begin{equation}
\frac{1}{\left(1-\frac{h}{r}\right)^{1/2}}\frac{dr}{d\lambda} = \pm\sqrt{R(r)}, \quad
r^2\frac{d\theta}{d\lambda} = \pm\sqrt{T(\theta)},
\end{equation}
where
\begin{equation}
R(r)=E^2 - \frac{Q }{r^2}, \qquad \qquad
T(\theta) = Q - \frac{L^2}{\sin^2\theta}.
\end{equation}
Although there are three constants of motion $E$, $L$, and $Q$, the geodesic motion of a photon is characterized by two independent parameters defined by \citep{Nedkova:2013msa}
\begin{equation}
\zeta=\frac{L}{E},\quad \eta=\frac{Q}{E^2}.
\end{equation}
These parameters are intrinsic to the photon's motion and do not depend on the affine parameter or energy scale.  By introducing a new affine parameter $\tilde{\lambda}= E\lambda$, we can redefine the functions $R(r)$ and $T(\theta)$ as
\begin{equation}
R(r)= 1 -  \frac{\eta}{r^2}, \quad T(\theta)= \eta -\frac{\zeta^2}{\sin^2\theta}.
\end{equation}
Using the above data, we write down the radial equation of motion in the following form:
\begin{equation}
\left(\frac{dr}{d\tilde{\lambda}}\right)^2 + V_{eff} = 0, \quad V_{eff} =- \left(1-\frac{h}{r}\right)R(r)=- \left(1-\frac{h}{r}\right)\left( 1 -  \frac{\eta}{r^2}\right),
\label{eq:effective_pot}
\end{equation}
where $V_{eff}$ is the effective potential describing the geodesic motion of a photon which has the form,
\begin{eqnarray}
V_{eff} = 0,\quad \frac{dV_{eff}}{dr} = 0,\quad \frac{d^2 V_{eff}}{dr^2} \leq 0.
\label{eq:pot_maxima}
\end{eqnarray}}
{ The above calculations are the basic equations to simulate multiple photon trajectories around a static spherically symmetric WH using the Morris-Thorne metric \eqref{eq:MTWHmetric} where the shape functions \eqref{eq:soliton_h} and \eqref{eq:NFW_h} are applied. Then, by solving the null geodesic equations for different impact parameters the resulting paths can be plotted to visualize how photons interact with the WH geometry in presence of non-minimally coupled gravity.}

%%%%%%%%%%%%%%%%%%%%%%%%%%%%%%%%%%%%%%%%%%%%%%%%%%%%%%%%%%%%%%%%%%%%%%%%%%%%%%%%%%%%%%%%%%%%%%%%
\section{Summary and Conclusion}\label{SEC:IX}
%%%%%%%%%%%%%%%%%%%%%%%%%%%%%%%%%%%%%%%%%%%%%%%%%%%%%%%%%%%%%%%%%%%%%%%%%%%%%%%%%%%%%%%%%%%%%%%%
In this study, we have explored WH models within the framework of the linear form of  $f({\cal Q},{\cal T})$ gravity. By adopting a spherically symmetric and static spacetime in the presence of anisotropic matter, we derived the field equations that govern the structure of WHs, with a particular focus on linear $f({\cal Q},{\cal T})={\cal Q}+\beta {\cal T}$ gravity, where GR is recovered by setting $\beta=0$. In the GR context, the flaring-out condition, which plays a fundamental role in WH physics, necessarily violates the NEC. Therefore, the price to have a WH in GR is the existence of exotic matter. However, in a different context, as in modified gravity, which accounts for non-minimal coupling between matter and geometry, we may obtain different results. This was the aim of the present study.

It has been shown that DM could be consistent with WH structure at galactic halos \cite{Rahaman:2013xoa, Sarkar:2024klp} and even at the central regions \cite{Rahaman:2014pba}. In the present study, we used Solition+NFW DM density profile to derive the corresponding shape functions, where soliton model is applied at the galactic core region and the NFW model is applied at the outer regions of the galactic halo of the dwarf galaxy NGC 2366. We use the model parameters for the two models as appear in \cite{Banares-Hernandez:2023axy}: Model I (soliton), the dwarf galaxy NGC 2366 core radius is $r_c=3$ kpc and the central density is $\sigma_c = 15\times 10^{-3}~\text{M}_{\odot}/\text{pc}^3$. Model II (NFW), the dwarf galaxy NGC 2366 scale radius is ${r_s} = 1.447$ kpc and the corresponding scale density is $\sigma_s=3.11 \times 10^{-3}~\text{M}_{\odot}/\text{pc}^3$.

Recalling the flaring-out condition at the WH throat, we found the following constraints on the coupling strength: In model I, we obtained the viable parameter space $\{\beta, r_0\}$ for an arbitrary throat size $r_0$ as given by Eqs. \eqref{eq:model1_beta1} and \eqref{eq:model1_beta2}, and by setting $r_0=1$ pc, these give $\beta<-3/8-8.494\times 10^{-16}$ or $-3/8<\beta<1.840\times 10^{13}$. In model II, similarly, we obtained the viable parameter space $\{\beta, r_0\}$ as given by Eqs. \eqref{eq:model2_beta1} and \eqref{eq:model2_beta2}, and by setting $r_0=1$ pc, these give $\beta<-3/8-2.545\times 10^{-13}$ or $-3/8<\beta<6.140\times 10^{10}$. In general, for the corresponding valid regions of the coupling parameter $\beta$, the WH models fulfill the following constraints:
\begin{itemize}
    \item [(i)] At the throat $r=r_0$; the shape function $h(r_0)=r_0$, $h'(r_0)<1$, $z(r_0)=0$ and $z'(r_0)\to \infty$.
    \item [(ii)] At finite radius $r>r_0$; the shape function $h(r)<r$, $h'(r)<1$, $z(r)$ is finite and $z'(r)$ is finite.
    \item [(iii)]  At infinite distance from the WH $r\to \infty$; the shape function satisfies $h(r)/r\to 0$, $h'(r)\to 0$, $z(r)\to \pm \infty$ and $z'(r)\to 0$.
\end{itemize}

Interestingly, we have shown that the flaring-out condition can be reconciled with the NEC in the context of $f(\cal{Q}, \cal{T})$ modified gravity. More precise, one must violate the NEC only effectively when the flaring-out condition holds. However, for the matter sector, one could have $\sigma c^2+p_r>0$ where $\beta<-1/2$, which is not recovered in the GR regime, and therefore the NEC is fulfilled. In this case, the NEC condition is broken only effectively, whereas the matter sector preserves the NEC with no need for exotic matter. We remark that $\beta<-1/2$ is consistent with our findings by applying the flaring-out condition.

In addition, we have shown that all energy conditions can be satisfied for certain bounds on the strength of the matter-geometry non-minimal coupling: For model I, with $r_0=1$ pc, we found the following constraints $|\beta| \gtrapprox 10^{13}$, while for the same $r_0$, the faring-out condition excludes exactly the positive coupling. Similarly, for model II, with $r_0=1$ pc, the energy conditions set the following constraints $|\beta| \gtrapprox 10^{11}$, while the faring-out condition excludes exactly the positive solution. In general, for large positive coupling parameter, the null energy condition (NEC) among other energy conditions, can be satisfied at the WH throat, meaning exotic matter is not needed, while the WH is no longer Lorentzian and the flaring-out condition is broken. However, for large negative coupling parameters, the NEC among other conditions, can be satisfied, allowing for healthy WHs without exotic matter, provided the coupling strength stays within certain bounds. In the latter case, the NEC is broken only effectively.

The $f(\cal{Q}, \cal{T})$ theory modifies gravity considering possible non-minimal between matter and geometry, this requires breaking of matter conservation. Nevertheless, the matter-geometry coupling compensates for it by additional non-conservative sector holds the conservation law effectively. In this sense, the new coupling term contributes by adding extra force to the hydrostatic equilibrium equation, known as TOV equation. We investigated the stability of both WH models by virtue of a modified version of TOV equation, which includes a new force due to matter-geometry non-minimal, showing that these WHs are dynamically stable. { Although our study does not include dynamical simulations nor waveforms explicitly, we acknowledge this as a promising direction for future work.}

In conclusion, the present study confirms the possibility to find a healthy WH solution where the energy conditions are broken only on the effective sector but satisfied on the matter sector. Consequently, no exotic matter is needed to form a physical WH as required in the GR framework, if non-minimal coupling between matter and geometry has been considered.

%%%%%%%%%%%%%%%%%%%%%%%%%%%%%%%%%%%%%%%%%%%%%%%%%%%%%%%%%%%%%%%%%%%%%%%%%%%%%
\appendix
%%%%%%%%%%%%%%%%%%%%%%%%%%%%%%%%%%%%%%%%%%%%%%%%%%%%%%%%%%%%%%%%%%%%%%%%%%%%%

\section{The matter fluid and the effective fluid in $f(\mathcal{Q}, \mathcal{T})$ gravity}\label{appA}
%%%%%%%%%%%%%%%%%%%%%%%%%%%%%%%%%%%%%%%%%%%%%%%%%%%%%%%%%%%%%%%%%%%%%%%%%%%%%
In this appendix, we rewrite the field equations \eqref{eq:fieldeq1}, \eqref{eq:fieldeq2} and \eqref{eq:fieldeq3} in the form
\begin{equation}\label{eq:density1}
\kappa^2  \sigma c^2=\frac{(r-h)}{2 r^3} \left[f_{\cal Q} \left\{\frac{(2 r-h) \left(r h'-h\right)}{(r-h)^2}+\frac{h \left(2 r \xi '+2\right)}{r-h}\right\}+\frac{2 h r f_{\cal{QQ}} {\cal Q}'}{r-h}+\frac{f r^3}{r-h}-\frac{2r^3 f_{\cal T} (P+\sigma c^2)}{(r-h)}\right],
\end{equation}
\begin{equation}\label{eq:rpress1}
\kappa^2  p_r=-\frac{(r-h)}{2 r^3} \left[f_{\cal Q} \left\{\frac{h }{r-h}\left(\frac{r h'-h}{r-b}+2 r \xi '+2\right)-4 r \xi '\right\}+\frac{2 h r f_{\cal{QQ}} {\cal Q}'}{r-h}+\frac{f r^3}{r-h}-\frac{2r^3 f_{\cal T} \left(P-p_r\right)}{(r-h)}\right],
\end{equation}
\begin{equation}\label{eq:tpress1}
\kappa^2 p_\theta=-\frac{(r-h)}{4 r^2} \left[f_{\cal Q} \left\{\frac{\left(r h'-h\right) \left(\frac{2 r}{r-h}+2 r \xi '\right)}{r (r-h)}+\frac{4 (2h-r) \xi '}{r-h}-4 r \xi'^2-4 r \xi ''\right\}-4 r f_{\cal{QQ}} {\cal Q}' \xi '+\frac{2 f r^2}{r-h}-\frac{4r^2 f_{\cal T} \left(P-p_\theta\right)}{(r-h)}\right].
\end{equation}
To adequately ascertain the limits of $\sigma$, $p_r$ and $p_\theta$, one could write the field equations for traversable WHs in GR like-frame, which yields
\begin{equation}\label{eq:Einstein_frame}
    G_{\mu \nu}= \kappa^2 \left(\mathcal{T}_{\mu \nu}+\mathcal{T}_{\mu\nu}^\text{NMG}\right)=\kappa^2 \tilde{\mathcal{T}}_{\mu\nu},
\end{equation}
where $\mathcal{T}_{\mu\nu}^\text{NMG}$ denotes the non-minimal coupling between matter and gravity, while $\tilde{\mathcal{T}}_{\mu\nu}$ denotes the effective (total) stress-energy tensor, i.e. $\tilde{\mathcal{T}}{^\mu}{_\nu}=diag(\tilde{\sigma} c^2,~\tilde{p}_r,~\tilde{p}_\theta,~\tilde{p}_\theta)$. Therefore, we write
\begin{equation}\label{eq:eff_dens1}
\tilde{\sigma} c^2=\frac{h'}{\kappa^2  r^2},
\end{equation}
\begin{equation}\label{eq:eff_rpress1}
\tilde{p}_r=\frac{1}{\kappa^2}\left[2\left(1-\frac{h}{r}\right)\frac{\xi '}{r}-\frac{h}{r^3}\right],
\end{equation}
\begin{equation}\label{eq:eff_tpress1}
\tilde{p}_\theta=\frac{1}{\kappa^2}\left(1-\frac{h}{r}\right)\left[\xi '' +{\xi '}^2-\frac{(rh'-h)\xi '}{2r(r-h)}-\frac{rh'-h}{2r^2 (r-h)}+\frac{\xi '}{r}\right].
\end{equation}
Here $\tilde{\sigma}$, $\tilde{p}_r$ and $\tilde{p}_\theta$ represent the effective density, radial and tangential pressures. The above equations show that the solution for the effective sector is nothing but the GR one. It proves convenient to write the effective fluid in terms of the matter fluid including the effects due to non-minimal coupling between gravity and matter. This shows how the matter sector in modified gravity is different from the GR framework. The comparison between the set of equations \eqref{eq:fieldeq1}-\eqref{eq:fieldeq3} and Eqs. \eqref{eq:eff_dens1}-\eqref{eq:eff_tpress1} enables us to obtain the following important relations \cite{Tayde:2022vbn}
\begin{equation}\label{eq:eff_dens2}
\tilde{\sigma} c^2=\frac{2 (r-h)}{(2 r-h) f_{\cal Q}}\left[\sigma c^2-\frac{1}{\kappa^2  r^2}\left(1-\frac{h}{r}\right) \left(\frac{h r f_{\cal{QQ}} {\cal Q}'}{r-b}+h f_{\cal Q} \left(\frac{ r \xi '+1}{r-h}-\frac{2 r-h}{2 (r-h)^2}\right)+\frac{f r^3}{2 (r-h)}\right)+\frac{f_{\cal T} (P+\sigma c^2)}{\kappa^2 }\right],
\end{equation}
\begin{equation}\label{eq:eff_rpress2}
\tilde{p}_r=\frac{2 h}{f r^3}\left[p_r +\frac{1}{2\kappa^2 r^2}\left(1-\frac{h}{r}\right) \left(f_{\cal Q} \left(\frac{h \left(\frac{r h'-h}{r-h}+2 r \xi '+2\right)}{r-h}-4 r \xi '\right)+\frac{2 h r f_{\cal{QQ}} {\cal Q}h'}{r-g}\right)+\frac{fr^3 (r-h)\xi '}{\kappa^2 h  r^2}-\frac{f_{\cal T} \left(P-p_r\right)}{\kappa^2 }\right],
\end{equation}
\begin{multline}\label{eq:eff_tpress2}
\tilde{p}_\theta=\frac{1}{f_{\cal Q} \left(\frac{r}{r-h}+r \xi '\right)}\left[p_\theta +\frac{1}{4\kappa^2 r}\left(1-\frac{h}{r}\right) \left(f_{\cal Q} \left(\frac{4 (2 h-r) \xi '}{r-h}-4 r \left(\xi '\right)^2-4 r \xi ''\right)+\frac{2 f r^2}{r-h}-4 r f_{\cal{QQ}} Q' \xi '\right) \right.\\ \left.
+\frac{1}{\kappa^2}\left(1-\frac{h}{r}\right)\left(\xi '' +{\xi '}^2-\frac{(rh'-h)\xi '}{2r(r-h)}+\frac{\xi '}{r}\right)f_{\cal Q} \left(\frac{r}{r-h}+r \xi '\right)-\frac{f_{\cal T} \left(P-p_\theta\right)}{\kappa^2 }\right].
\end{multline}

%%%%%%%%%%%%%%%%%%%%%%%%%%%%%%%%%%%%%%%%%%%%%%%%%%%%%%%%%%%%%%%%%%%%%%%%%%%%%
\section{The energy conditions of the matter and the effective fluids in $f(\mathcal{Q}, \mathcal{T})$ gravity with a constant redshift function}\label{appB}
%%%%%%%%%%%%%%%%%%%%%%%%%%%%%%%%%%%%%%%%%%%%%%%%%%%%%%%%%%%%%%%%%%%%%%%%%%%%%
In this appendix, we show how the system of differential equations, Eqs.~\eqref{eq:eff_dens3}--\eqref{eq:eff_tpress3}, satisfies the energy conditions.
By considering the effective density and pressures as indicated in Eqs.~\eqref{eq:eff_dens3}--\eqref{eq:eff_tpress3}, we obtain \cite{Tayde:2022vbn}:
{ For a constant redshift function, i.e. $\xi=\xi_0$, the effective fluid density and pressures, namely Eqs. \eqref{eq:eff_dens2}, \eqref{eq:eff_rpress2} and \eqref{eq:eff_tpress2}, allow the following energy conditions on the effective fluid \cite{Tayde:2022vbn}}
\begin{align}
\tilde{\sigma} c^2+ \tilde{p_r}=& \frac{h (h-2 r) f_{\cal Q}^2 \left[h (3 h-2 r)-h r h'\right]-f r^4 (h-r)^2 \left[2 h f_{\cal{QQ}} {\cal Q}'+r^2 \left\{f-2 f_{\cal T} (P+\sigma c^2)+2\kappa^2  p_r\right\}\right]}{\kappa^2  f r^6 (h-2 r) (h-r) f_{\cal Q}}\nonumber\\
&+\frac{2 (r-h) \left(\sigma c^2+p_r \right)}{(2 r-h) f_{\cal Q}}+\frac{h r (h-r) f_{\cal Q} \left[2 (h-2 r) \left\{{\cal Q} f_{\cal{QQ}} {\cal Q}'+r^2 \left(f_T \left(p_r-P\right)+\kappa^2  p_r\right)\right\}-h f r^2\right]}{\kappa^2  f r^6 (h-2 r) (h-r) f_{\cal Q}}\,,
\end{align}
\begin{equation}
\tilde{\sigma} c^2+ \tilde{p_\theta}=\frac{2 (r-h) \left(\sigma+p_\theta \right)}{(2 r-h) f_{\cal Q}}-\frac{2 h^2 f_{\cal Q}+r (r-h) \left[2 r f_{\cal T} \left(+2 \sigma  r-h p_\theta+h P+2 r p_\theta\right)-h \left(4 f_{\cal{QQ}} {\cal Q}'+f r+2\kappa^2  r p_\theta\right)\right]}{2\kappa^2  r^3 (h-2 r) f_{\cal Q}}\,,
\end{equation}
\begin{multline}
\tilde{\sigma} c^2- \tilde{p_r}=-\frac{h (h-2 r) f_{\cal Q}^2 \left(h (3 h-2 r)-h r h'\right)+f r^4 (h-r)^2 \left[2 h f_{\cal{QQ}} {\cal Q}'+r^2 \left\{f-2 f_{\cal T} (P+\sigma )-2\kappa^2  p_r\right\}\right]}{\kappa^2  f r^6 (h-2 r) (h-r) f_{\cal Q}}\\
+\frac{2 (r-h) \left(\sigma -p_r\right)}{(2 r-h) f_Q}-\frac{h r (h-r) f_{\cal Q} \left[2 (h-2 r) \left\{h f_{\cal{QQ}} {\cal Q}'+r^2 \left(f_{\cal T} \left(p_r-P\right)+\kappa^2  p_r\right)\right\}+h f r^2\right]}{\kappa^2  f r^6 (h-2 r) (h-r) f_{\cal Q}}\,,
\end{multline}
\begin{equation}
\tilde{\sigma} c^2- \tilde{p_\theta}=\frac{2 (r-h) \left(\sigma -p_\theta\right)}{(2 r-h) f_{\cal Q}}-\frac{2 h^2 f_{\cal Q}+r (r-h) \left[2 \left\{r f_{\cal T} \left((h-2 r) p_\theta-h P+4 P r+2 \sigma  r\right)-2 b f_{\cal{QQ}} {\cal Q}'+\kappa^2  h r p_\theta\right\}+f r (h-4 r)\right]}{2\kappa^2  r^3 (h-2 r) f_{\cal Q}}\,,
\end{equation}
\begin{multline}
\hspace{-0.5cm}\tilde{\sigma} c^2+ \tilde{p_r} + 2\tilde{p_\theta}=
+\frac{2 (r-h) \left(p_r+2 p_\theta+\sigma \right)}{(2 r-h) f_{\cal Q}}+\frac{h\, r }{\kappa^2  r^6}\left[\frac{2 \left(h f_{\cal{QQ}} {\cal Q}'+r^2 \left\{f_{\cal T} \left(p_r-P\right)+\kappa^2  p_r\right\}\right)}{f}-\frac{h r^2}{h-2 r}\right]\\
+\frac{h^2 f_{\cal Q} \left(3 h-r h'-2 r\right)}{\kappa^2 r^6\,f\, (h-r)}+\frac{r^4 (h-r) \left[2 r f_{\cal T} \left(h P-(h-2 r) p_\theta-P r+\sigma  r\right)-2 h f_{\cal{QQ}} {\cal Q}'+f r (r-h)-2\kappa^2  r \left(h p_\theta+r p_r\right)\right]}{\kappa^2 r^6\,(h-2 r) f_{\cal Q}}\,.
\end{multline}
Hence, the energy condition requirements for a WH model within $f({\cal Q},{\cal T})$ theory, using the previously mentioned equations, lead to the following important results which relate the energy conditions on the matter sector and the effective sector \cite{Tayde:2022vbn}.

$\bullet$ $\tilde{\sigma}\geq 0  \Rightarrow \sigma \geq 0$, where
$$\frac{(2 r-h) f_{\cal Q}}{r-h} > 0 \text{~and~} \frac{2 (r-h) \left[\frac{\left(1-\frac{h}{r}\right) \left(\frac{h r f_{\cal{QQ}} {\cal Q}'}{r-h}+\frac{h f_{\cal Q}}{r-h}-\frac{h (2 r-h) f_{\cal Q}}{2 (r-h)^2}+\frac{f r^3}{2 (r-h)}\right)}{\kappa^2  r^2}-\frac{f_{\cal T} (P+\sigma )}{\kappa^2 }\right]}{(2 r-h) f_{\cal Q}} \leq 0.$$

$\bullet$ $\tilde{\sigma}+\tilde{p_r}\geq 0 \Rightarrow \sigma + p_r \geq 0$, where
\begin{gather*}
\frac{(2 r-h) f_{\cal Q}}{r-h} > 0 \text{~and~} \frac{h r (h-r) f_{\cal Q} \left[2 (h-2 r) \left(h f_{\cal{QQ}} {\cal Q}'+r^2 \left(f_{\cal T} \left(p_r-P\right)+\kappa^2  p_r\right)\right)-b f r^2\right]}{\kappa^2  f r^6 (b-2 r) (b-r) f_Q}\\
+\frac{h (h-2 r) f_{\cal Q}^2 \left(h (3 h-2 r)-h r h'\right)-f r^4 (h-r)^2 \left[2 h f_{\cal{QQ}} {\cal Q}'+r^2 \left(-2 f_{\cal T} (P+\sigma )+f+2\kappa^2  p_r\right)\right]}{\kappa^2  f r^6 (h-2 r) (h-r) f_{\cal Q}} \geq 0.
\end{gather*}

$\bullet$ $\tilde{\sigma}+\tilde{p_\theta}\geq 0 \Rightarrow \sigma + p_\theta \geq 0$, where
$$\frac{(2 r-h) f_{\cal Q}}{r-h} > 0 \text{~and~} \frac{2 h^2 f_{\cal Q}+r (r-h) \left[2 r f_{\cal Q} \left(-h p_\theta+h P+2 r p_\theta+2 \sigma  r\right)-h \left(4 f_{\cal{QQ}} {\cal Q}'+f r+2\kappa^2  r p_\theta\right)\right]}{2\kappa^2  r^3 (h-2 r) f_{\cal Q}} \leq 0.$$

$\bullet$ $\tilde{\sigma}-\tilde{p_r}\geq 0 \Rightarrow \sigma - p_r \geq0$, where
\begin{gather*}
\frac{(2 r-h) f_{\cal Q}}{r-h} > 0 \text{~and~} \frac{h r (h-r) f_{\cal Q} \left[2 (h-2 r) \left\{h f_{\cal{QQ}} {\cal Q}'+r^2 \left(f_{\cal T} \left(p_r-P\right)+\kappa^2  p_r\right)\right\}+b f r^2\right]}{\kappa^2  f r^6 (b-2 r) (b-r) f_Q}\\
+\frac{h (h-2 r) f_{\cal Q}^2 \left[h (3 h-2 r)-h r h'\right]+f r^4 (h-r)^2 \left[2 h f_{\cal{QQ}} {\cal Q}'+r^2 \left(-2 f_{\cal T} (P+\sigma )+f-2\kappa^2  p_r\right)\right]}{\kappa^2  f r^6 (b-2 r) (h-r) f_{\cal Q}} \leq 0.
\end{gather*}

$\bullet$ $\tilde{\sigma}-\tilde{p_\theta}\geq 0 \Rightarrow \sigma - p_\theta \geq 0$, where
$$\frac{(2 r-h) f_{\cal Q}}{r-h} > 0 \text{~and~} \frac{2 h^2 f_{\cal Q}+r (r-h) \left[2 \left(r f_{\cal T} \left((h-2 r) p_\theta-h P+4 P r+2 \sigma  r\right)-2 b f_{\cal{QQ}} {\cal Q}'+\kappa^2  h r p_\theta\right)+f r (h-4 r)\right]}{2\kappa^2  r^3 (h-2 r) f_{\cal Q}} \leq 0.$$

$\bullet$ $\tilde{\sigma}+\tilde{p_r} + 2\tilde{p_\theta} \geq 0 \Rightarrow \sigma + p_r + 2 p_\theta \geq 0$, where
\begin{gather*}
\frac{(2 r-h) f_{\cal Q}}{r-h} > 0 \text{~and~} \frac{h r \left[\frac{2 \left(h f_{\cal{QQ}} {\cal Q}'+r^2 \left(f_{\cal T} \left(p_r-P\right)+\kappa^2  p_r\right)\right)}{f}-\frac{h r^2}{h-2 r}\right]}{\kappa^2  r^6}\\
+\frac{\frac{h^2 f_{\cal Q} \left(3 h-r h'-2 r\right)}{f (h-r)}+\frac{r^4 (h-r) \left[2 r f_{\cal T} \left(h P-(h-2 r) p_\theta-P r+\sigma  r\right)-2 h f_{\cal{QQ}} {\cal Q}'+f r (r-h)-2\kappa^2  r \left(h p_\theta+r p_r\right)\right]}{(h-2 r) f_{\cal Q}}}{\kappa^2  r^6} \geq 0.
\end{gather*}
The above mentioned energy conditions are not satisfied in GR theory for traversable WH solutions. In particular, for positive energy density, the NEC must be violated as a consequence of the flaring-out condition at the WH throat. This requires presence of an exotic matter at the throat of the WH. Therefore, it is straightforward to extend this conclusion to modified gravity where the NEC must be violated effectively as verified by Eq. \eqref{eq:Einstein_frame}. However, the coupling between matter and geometry introduces additional degrees of freedom, which can mimic the effects of exotic matter or even replace it entirely keeping the physical matter healthy under specific conditions. For more details see the discussion in Sec. \ref{SEC:VI}.

%%%%%%%%%%%%%%%%%%%%%%%%%%%%%%%%%%%%%%%%%%%%%%%%%%%%%%%%%%%%%%%%%%%%%%%%%%%%%
\section{The complementary part of the shape function of model I}\label{appC}
%%%%%%%%%%%%%%%%%%%%%%%%%%%%%%%%%%%%%%%%%%%%%%%%%%%%%%%%%%%%%%%%%%%%%%%%%%%%%
The explicit form of the function $\mathcal{F}(r)$ in Eq.~\eqref{eq:soliton_h} is given by:
\begin{align}
&\mathcal{F}(r)=
680680r_c^{8}{r_0}^{11}{r}^{7}{\alpha}^{9}+ 680680r_c^{8}{r_0}^{7}{r}^{11}{\alpha}^{9}+3465{r}^{13 }{\alpha}^{13}{r_0}^{13}+3465r_c^{26} -48580r_c^ {24}\alpha{r_0}^{2}-48580r_c^{24}\alpha{r}^{2}-92323 r_c^{22}{\alpha}^{2}{r}^{4}\nonumber\\
&-92323r_c^{22}{\alpha}^{2 }{r_0}^{4}-101376r_c^{20}{\alpha}^{3}{r_0}^{6}- 101376r_c^{20}{\alpha}^{3}{r}^{6}-65373r_c^{18}{ \alpha}^{4}{r_0}^{8}-23100r_c^{16}{\alpha}^{5}{r_0} ^{10}-3465r_c^{14}{\alpha}^{6}{r_0}^{12}-65373r_c^{18}{\alpha}^{4}{r}^{8}\nonumber\\
&-23100r_c^{16}{\alpha}^{5}{r}^{10}- 3465r_c^{14}{\alpha}^{6}{r}^{12}+65373r_c^{4}{r_0}^{9}{r}^{13}{\alpha}^{11}+23100r_c^{2}{r_0}^{13}{r}^ {11}{\alpha}^{12}-1155r_c^{2}{r_0}^{12}{r}^{12}{\alpha}^ {12}+23100r_c^{2}{r_0}^{11}{r}^{13}{\alpha}^{12}\nonumber\\
&+65373 r_c^{4}{r_0}^{13}{r}^{9}{\alpha}^{11}-7392r_c^{4} {r_0}^{12}{r}^{10}{\alpha}^{11}+154308r_c^{4}{r_0}^ {11}{r}^{11}{\alpha}^{11}-7392r_c^{4}{r_0}^{10}{r}^{12}{ \alpha}^{11}-127820r_c^{8}{r_0}^{10}{r}^{8}{\alpha}^{9}+ 92323r_c^{8}{r_0}^{5}{r}^{13}{\alpha}^{9}\nonumber\\
&-28952r_c^{8}{r_0}^{6}{r}^{12}{\alpha}^{9}+1245013r_c^{8}{r_0}^{9}{r}^{9}{\alpha}^{9}+101376r_c^{6}r_0^7{r }^{13}{\alpha}^{10}+101376r_c^6r_0^{13}{r}^7 \alpha^10-19899r_c^{6}r_0^12{r}^{8}{\alpha}^{10}+ 437712r_c^6r_0^{11}{r}^{9}{\alpha}^{10}\nonumber\\
&-47388r_c^{6}{r_0}^{10}{r}^{10}{\alpha}^{10}+437712r_c^{6 }{r_0}^{9}{r}^{11}{\alpha}^{10}-19899r_c^{6}{r_0}^{ 8}{r}^{12}{\alpha}^{10}-72835r_c^{24}\alpha r r_0-165088 r_c^{22}{\alpha}^{2}r_0{r}^{3}-165088r_c^{22} {\alpha}^{2}{r_0}^{3}r\nonumber\\
&-505148r_c^{22}{\alpha}^{2}{r_0}^{2}{r}^{2}-868912r_c^{20}{\alpha}^{3}{r_0}^{4}{r}^{ 2}+151268r_c^{20}{\alpha}^{3}{r_0}^{3}{r}^{3}-222651r_c^{20}{\alpha}^{3}{r_0}^{5}r-222651r_c^{20}{ \alpha}^{3}r_0{r}^{5}-868912r_c^{20}{\alpha}^{3}{r_0}^{2}{r}^{4}\nonumber\\
&-1134763r_c^{18}{\alpha}^{4}{r_0}^{4}{r}^ {4}+804020r_c^{18}{\alpha}^{4}{r_0}^{5}{r}^{3}-896280{ r_c}^{18}{\alpha}^{4}{r_0}^{2}{r}^{6}-186648r_c^{18 }{\alpha}^{4}r_0{r}^{7}+804020r_c^{18}{\alpha}^{4}{r_0}^{3}{r}^{5}-896280r_c^18\alpha^4r_0^6r^2\nonumber\\
&-186648r_c^{18}{\alpha}^{4}{r_0}^{7}r-95865r_c^{16}{\alpha}^{5}r_0{r}^{9}-982072r_c^{16}{ \alpha}^{5}{r_0}^{6}{r}^{4}+1146824r_c^{16}{\alpha}^{5}{ r_0}^{3}{r}^{7}-982072r_c^{16}{\alpha}^{5}{r_0}^{4} {r}^{6}-553476r_c^{16}{\alpha}^{5}{r_0}^{8}{r}^{2}\nonumber\\
&-95865 r_c^{16}{\alpha}^{5}{r_0}^{9}r+2249233r_c^{16}{ \alpha}^{5}{r}^{5}{r_0}^{5}-553476r_c^{16}{\alpha}^{5}{r_0}^{2}{r}^{8}+1146824r_c^{16}{\alpha}^{5}{r_0}^{7} {r}^{3}-27720r_c^{14}{\alpha}^{6}{r_0}^{11}r-542073r_c^{14}{\alpha}^{6}{r_0}^{4}{r}^{8}\nonumber\\
&+2689232r_c^{14 }{\alpha}^{6}{r_0}^{5}{r}^{7}-27720r_c^{14}{\alpha}^{6}r_0{r}^{11}+2689232r_c^{14}{\alpha}^{6}{r_0}^{7}{r }^{5}+830760r_c^{14}{\alpha}^{6}{r_0}^{9}{r}^{3}-189420 r_c^{14}\alpha^6r_0^2r^{10}-542073r_c^{14}\alpha^6r_0^8r^4\nonumber\\
&+830760r_c^{14}{\alpha} ^{6}{r_0}^{3}{r}^{9}-858928r_c^{14}{\alpha}^{6}{r_0 }^{6}{r}^{6}-189420r_c^{14}{\alpha}^{6}{r_0}^{10}{r}^{2} +1766023r_c^{12}{r_0}^{5}{r}^{9}{\alpha}^{7}+1942472r_c^{10}{r_0}^{7}{r}^{9}\alpha^8-28952r_c^8r_0^{12}r^6\alpha^9.\nonumber\\
&+92323r_c^{8}{r_0}^{13}{ r}^{5}{\alpha}^{9}-522032r_c^{12}{r_0}^{6}{r}^{8}{\alpha }^{7}-27720r_c^{12}{r_0}^{2}{r}^{12}{\alpha}^{7}-172760 r_c^{12}{r_0}^{10}{r}^{4}{\alpha}^{7}+622076r_c ^{10}{r_0}^{11}{r}^{5}\alpha^8-24185r_c^{10}r_0^4r^{12}\alpha^8\nonumber\\
&-186424r_c^{10}{r_0}^{6}{r}^{ 10}{\alpha}^{8}-345583r_c^{10}{r_0}^{8}{r}^{8}{\alpha}^{ 8}+622076r_c^{10}{r_0}^{5}{r}^{11}\alpha^8+1942472 r_c^{10}r_0^9r^7\alpha^8+48580r_c^{10 }r_0^3r^{13}\alpha^8+48580r_c^{10}r_0^{13}r^3\alpha^8\nonumber\\
&-186424r_c^{10}{r_0}^{10}{r}^{6}{ \alpha}^{8}-24185r_c^{10}{r_0}^{12}{r}^{4}{\alpha}^{8}- 3465r_c^{12}r_0{r}^{13}{\alpha}^{7}+312340r_c ^{12}{r_0}^{3}{r}^{11}{\alpha}^{7}+3026128r_c^{12}r_0^7r^7\alpha^7-27720r_c^{12}r_0^{12}r^2\alpha^7\nonumber\\
&-3465r_c^{12}{r_0}^{13}r{\alpha}^{7}+ 1766023r_c^{12}{r_0}^{9}{r}^{5}{\alpha}^{7}+312340r_c^{12}{r_0}^{11}{r}^{3}{\alpha}^{7}-522032r_c^{12 }{r_0}^{8}{r}^{6}{\alpha}^{7}-172760r_c^{12}{r_0}^{ 4}{r}^{10}{\alpha}^{7}-127820r_c^8r_0^8r^{10}\alpha^9\,.
\end{align}

%%%%%%%%%%%%%%%%%%%%%%%%%%%%%%%%%%%%%%%%%%%%%%%%%%%%%%%%%%%%%%%%%%%%%%%%%%%%%
\section{The radial and tangential pressures of model I}\label{appD}
%%%%%%%%%%%%%%%%%%%%%%%%%%%%%%%%%%%%%%%%%%%%%%%%%%%%%%%%%%%%%%%%%%%%%%%%%%%%%
The explicit forms of the radial $p_{r}$ and the tangential $p_{\theta}$ pressures of the matter fluid of the soliton WH model in Sec. \ref{Sec:ECWHI}:
\begin{align}
&p_{r}=-\frac{A  \left[\arctan \left( {\frac {\sqrt {\alpha}r}{r_c}} \right)-\arctan \left( {\frac {\sqrt {\alpha}r_0}{r_c}} \right)\right]}{ {\kappa}^2{r}^{3} \left( 1+2\beta\right)}+\frac {1}{3 \left( {r_c}^ {2}+\alpha{r}^{2} \right) ^{8}{\kappa}^{2} \left( 1+2\beta \right)  \left( 8\beta+1 \right) {r}^{3}}\left[136\beta{r}^{2}\mathcal{F}\left( r \right) \alpha r_0+8 \beta{r}^{2}  \mathcal{F}'(r){r_c}^{2}\right.\nonumber\\
&\left.+8\beta{r}^{4} \mathcal{F}'(r) \alpha-16\beta r\mathcal{F}(r) r_c^2 -128\beta r^3\mathcal{F}(r) \alpha+3\mathcal{F}(r) r_0\alpha r^2+24\beta \mathcal{F}(r) r_0r_c^2+56\beta{r}^{3}A\alpha^{\frac{3}2}r_c^{13}+168 \beta r^5A\alpha^{\frac{5}2}r_c^{11}\right.\nonumber\\
&\left.+280\beta{r}^{7}A{ \alpha}^{\frac{7}2}{r_c}^{9}+280\beta{r}^{9}A{\alpha}^{\frac{9}2}{r_c}^{7}+168\beta{r}^{11}A{\alpha}^{\frac{11}2}{r_c}^{5}+56\beta{ r}^{13}A{\alpha}^{\frac{13}2}{r_c}^{3}+8\beta{r}^{15}A{\alpha}^{\frac{15} 2}r_c-192\beta r_0{r_c}^{14}\alpha{r}^{2}-672 \beta r_0{r_c}^{12}{\alpha}^{2}{r}^{4}\right.\nonumber\\
&\left.-1344\beta r_0{r_c}^{10}{\alpha}^{3}{r}^{6}-1680\beta r_0{r_c}^{8}{\alpha}^{4}{r}^{8}-1344\beta r_0{r_c}^{6}{ \alpha}^{5}{r}^{10}-672\beta r_0{r_c}^{4}{\alpha}^{6}{r }^{12}-192\beta r_0{r_c}^{2}{\alpha}^{7}{r}^{14}+8 \beta r A\sqrt {\alpha}{r_c}^{15} \right.\nonumber\\
&\left.-8\beta r \mathcal{F}'(r) r_0{r_c}^{2}-8\beta{r}^ {3} \mathcal{F}'(r) r_0\alpha -24r_0{r_c}^{14}\alpha{r}^{2}-84r_0{r_c} ^{12}{\alpha}^{2}{r}^{4}-168r_0{r_c}^{10}{\alpha}^{3}{r} ^{6}-210r_0{r_c}^{8}{\alpha}^{4}{r}^{8}-168r_0{ r_c}^{6}{\alpha}^{5}{r}^{10}\right.\nonumber\\
&\left.-84r_0{r_c}^{4}{\alpha} ^{6}{r}^{12}-24r_0{r_c}^{2}{\alpha}^{7}{r}^{14}-24 \beta r_0{\alpha}^{8}{r}^{16}-3 \mathcal{F}(r) r{r_c }^{2}-3\mathcal{F}(r){r}^{3}\alpha+3\mathcal{F}(r) r_0{r_c}^{2}-3r_0{r_c}^{16}-3r_0{\alpha} ^{8}{r}^{16}-24\beta r_0{r_c}^{16}\right]\,,
\nonumber\\ \label{s1pr}\\
&p_{\theta}=\frac{A  \left[\arctan \left( {\frac {\sqrt {\alpha}r}{r_c}} \right)-\arctan \left( {\frac {\sqrt {\alpha}r_0}{r_c}} \right)\right]}{ 2{\kappa}^2{r}^{3} \left( 1+2\beta\right)}+\frac{1}{6\left( {r_c}^{2 }+\alpha\,{r}^{2} \right) ^{8}{\kappa}^{2}{r}^{3} \left( 1+10\, \beta+16\,{\beta}^{2} \right)}\left[ -\left( 8\beta+3 \right) r_c\,A{r}^{15}{\alpha} ^{15/2}\right.\nonumber\\
&\left.-21\left( 8\beta+3\right) r_c^5A r^{11}\alpha^{\frac{11}2}-7\left(8 \beta+3 \right)r_c^3A{r}^{13}\alpha^{\frac{13}2}-35 \left( 8\beta+3 \right) r_c^7A r^9\alpha^{ 9/2}-7\left(8 \beta+3\right) r_c^{13}A r^3\alpha^{3/ 2}-21 \left(8\beta+3 \right) {r_c}^{11}A{r}^{5}{\alpha}^{5/2 }\right.\nonumber\\
&\left.-35 \left( 8\beta+3 \right) {r_c}^{9}A{r}^{7}{\alpha}^{7/2}+ 3r_0\, \left(8 \beta+1 \right)  \left( {r_c}^{2}+\alpha \,{r}^{2} \right) ^{8}- \left( 8\beta+3\right)  \left( r-r_0 \right) r \left( {r_c}^{2}+\alpha\,{r}^{2} \right)\mathcal{F}'(r) -\left(8\beta+3 \right) {r_c}^{15}Ar \sqrt {\alpha}\right.\nonumber\\
&\left.+ \left(  \left( 128\,\beta+42 \right) \alpha\,{r}^{3}- r_0\, \left( 45+136\beta \right) \alpha\,{r}^{2} +16\,\beta\,r{r_c}^{2}-3r_0\,{r_c}^{2} \left( 8\beta+ 1 \right)  \right) \mathcal{F}(r) \right]\,.\label{s1pt}
\end{align}

%%%%%%%%%%%%%%%%%%%%%%%%%%%%%%%%%%%%%%%%%%%%%%%%%%%%%%%%%%%%%%%%%%%%%%%%%%%%%
\section{The contributing forces of the hydrostatic equilibrium equation}\label{appE}
%%%%%%%%%%%%%%%%%%%%%%%%%%%%%%%%%%%%%%%%%%%%%%%%%%%%%%%%%%%%%%%%%%%%%%%%%%%%%
\subsection{Model I}\label{appEI}
%%%%%%%%%%%%%%%%%%%%%%%%%%%%%%%%%%%%%%%%%%%
The radial dependence of the contributing forces of the hydrostatic equilibrium equation of the soliton WH model in Sec. \ref{Sec:TOVI}:
\begin{align}
F_h&=\frac{3}{(1+2\beta)\kappa^2 r^4}\left[r_0+A\, \left[\arctan(\sqrt{\alpha}r/r_c)-\arctan(\sqrt{\alpha}r_0/r_c) \right]+B\, \frac{r-r_0}{(r_c^2+\alpha r^2)^7}\mathcal{F}(r)\right]\nonumber\\
    &+\frac{(3+20 \beta)}{3(1+2\beta)(1+4\beta)\kappa^2 r^3}\left[\frac{A\sqrt {\alpha}}{{r_c} \left( 1+{\frac {{r}^{2}\alpha}{{r_c}^{2}}} \right)}+\frac{  {\mathcal{F}' \left( r \right)   \left( r-r_0 \right) }}{ \left( {r_c}^{2} +{r}^{2}\alpha \right) ^{7}}+{\frac {\mathcal{F} \left( r \right) }{ \left( {r_c}^{2}+{r}^{2}\alpha \right) ^{7}}}-14\,{\frac {\mathcal{F} \left( r \right)  \left(r -r_0 \right) r\alpha}{ \left( {r_c}^{2}+{ r}^{2}\alpha \right) ^{8}}}\right]\nonumber\\
    &-\frac{4\beta}{3(1+2\beta)(1+4\beta)\kappa^2 r^2}\left[ -\frac{2\,A{\alpha}^{3/2}r}{{r_c}^{3} \left( 1+{\frac {{r}^{2}\alpha}{{r_c}^{2}}} \right) ^2}+\frac {{\mathcal{F}''\left( r \right)   \left(r -r_0 \right) }}{ \left( {r_c}^{2}+{r}^{2}\alpha \right) ^{7}}+\frac{2\,\mathcal{F}' \left( r \right) }{ \left( {r_c}^{2}+{r}^{2}\alpha \right) ^{7}} -\frac { 28\mathcal{F}' \left( r \right) \left( r-r_0 \right) r\alpha}{ \left( {r_c}^{2}+{r}^{2} \alpha \right) ^{8}}\right.\nonumber\\
    &\left.-\frac{28\mathcal{F} \left( r \right) r\alpha}{ \left( {r_c}^{2}+{r}^{2}\alpha \right) ^{8}}+\frac{224\mathcal{F} \left( r \right)  \left(r -r_0 \right) {r}^{2}{\alpha}^{2}}{ \left( {r_c}^{2}+{r}^{2}\alpha \right) ^{9}}-\frac{ 14\mathcal{F}\left( r \right)  \left( r-r_0 \right) \alpha}{ \left( {r_c}^{2}+{r }^{2}\alpha \right) ^{8}}\right]\,,\label{eq:Model1_Fh}
\end{align}
\begin{align}
    F_c&=\frac{3\beta }{(1+2\beta)(3+7\beta)\kappa^2 r^4}\left[r_0+A\, \left[\arctan(\sqrt{\alpha}r/r_c)-\arctan(\sqrt{\alpha}r_0/r_c) \right]+B\, \frac{r-r_0}{(r_c^2+\alpha r^2)^7}\mathcal{F}(r)\right]\nonumber\\
    &-\frac{\beta(27+68 \beta)}{3(1+2\beta)(1+4\beta)(3+7\beta)\kappa^2 r^3}\left[\frac{A\sqrt {\alpha}}{{r_c} \left( 1+{\frac {{r}^{2}\alpha}{{r_c}^{2}}} \right)}+\frac{  {\mathcal{F}' \left( r \right)   \left( r-r_0 \right) }}{ \left( {r_c}^{2} +{r}^{2}\alpha \right) ^{7}}+{\frac {\mathcal{F} \left( r \right) }{ \left( {r_c}^{2}+{r}^{2}\alpha \right) ^{7}}}-14\,{\frac {\mathcal{F} \left( r \right)  \left(r -r_0 \right) r\alpha}{ \left( {r_c}^{2}+{ r}^{2}\alpha \right) ^{8}}}\right]\nonumber\\
    &+\frac{4\beta }{3(1+2\beta)(1+4\beta)\kappa^2 r^2}\left[ -\frac{2\,A{\alpha}^{3/2}r}{{r_c}^{3} \left( 1+{\frac {{r}^{2}\alpha}{{r_c}^{2}}} \right) ^2}+\frac {{\mathcal{F}''\left( r \right)   \left(r -r_0 \right) }}{ \left( {r_c}^{2}+{r}^{2}\alpha \right) ^{7}}+\frac{2\,\mathcal{F}' \left( r \right) }{ \left( {r_c}^{2}+{r}^{2}\alpha \right) ^{7}} -\frac { 28\mathcal{F}' \left( r \right) \left( r-r_0 \right) r\alpha}{ \left( {r_c}^{2}+{r}^{2} \alpha \right) ^{8}}\right.\nonumber\\
    &\left.-\frac{28\mathcal{F} \left( r \right) r\alpha}{ \left( {r_c}^{2}+{r}^{2}\alpha \right) ^{8}}+\frac{224\mathcal{F} \left( r \right)  \left(r -r_0 \right) {r}^{2}{\alpha}^{2}}{ \left( {r_c}^{2}+{r}^{2}\alpha \right) ^{9}}-\frac{ 14\mathcal{F}\left( r \right)  \left( r-r_0 \right) \alpha}{ \left( {r_c}^{2}+{r }^{2}\alpha \right) ^{8}}\right]\,,\label{eq:Model1_Fc}\\
\end{align}
\begin{align}
    F_a&=\frac{9}{(3+7\beta)\kappa^2 r^4}\left[r_0+A\, \left[\arctan(\sqrt{\alpha}r/r_c)-\arctan(\sqrt{\alpha}r_0/r_c) \right]+B\, \frac{r-r_0}{(r_c^2+\alpha r^2)^7}\mathcal{F}(r)\right]\nonumber\\
    &-\frac{3 }{(3+7\beta)\kappa^2 r^3}\left[\frac{A\sqrt {\alpha}}{{r_c} \left( 1+{\frac {{r}^{2}\alpha}{{r_c}^{2}}} \right)}+\frac{  {\mathcal{F}' \left( r \right)   \left( r-r_0 \right) }}{ \left( {r_c}^{2} +{r}^{2}\alpha \right) ^{7}}+{\frac {\mathcal{F} \left( r \right) }{ \left( {r_c}^{2}+{r}^{2}\alpha \right) ^{7}}}-14\,{\frac {\mathcal{F} \left( r \right)  \left(r -r_0 \right) r\alpha}{ \left( {r_c}^{2}+{ r}^{2}\alpha \right) ^{8}}}\right]\,.\hfill\label{eq:Model1_Fa}
\end{align}

%%%%%%%%%%%%%%%%%%%%%%%%%%%%%%%%%%%%%%%%%%%
\subsection{Model II}\label{appEII}
%%%%%%%%%%%%%%%%%%%%%%%%%%%%%%%%%%%%%%%%%%%
The radial dependence of the contributing forces of the hydrostatic equilibrium equation of the NFW WH model in Sec. \ref{Sec:TOVII}:
\begin{align}\label{eq:Model2_Fh}
F_h&=\frac{3}{\left( 1+2\,\beta \right){\kappa}^{2}{r}^{4}} \left[ r_0+{\frac {\tilde{A} \left( r-r_0 \right) }{r+r_s}} +\tilde{B}\ln  \left( {\frac {r+r_s}{r_s+r_0}} \right) \right]  +\frac {\left( 3+20\beta \right)  \left[ \tilde{B}r+ \left( \tilde{A}+\tilde{B} \right) r_s+\tilde{A}r_0 \right]}{3\left( r+r_s \right) ^{2} \left( 1+2\,\beta \right)  \left( 1+4\,\beta \right){ \kappa}^{2}{r}^{3}}\nonumber\\
   &+{\frac {4\beta\, \left[\tilde{B}r+ \left( 2\tilde{A}+\tilde{B} \right) r_s+2\tilde{A}r_0 \right] }{ 3\left( r+r_s \right) ^{ 3} \left( 1+2\,\beta \right)  \left( 1+4\,\beta \right)  \left( 3+7\, \eta \right) {\kappa}^{2}{r}^{2}}},
\end{align}
\begin{align}\label{eq:Model2_Fc}
F_c&=\frac{3\beta }{(1+2\beta)(3+7\beta)\kappa^2 r^4}\left[r_0+\tilde{A}\frac{r-r_0}{r_s+r}+
\tilde{B}\ln \left(\frac{r_s+r}{r_s+r_0}\right)\right]-\frac{\beta(27+68 \beta) }{3(1+2\beta)(1+4\beta)(3+7\beta)\kappa^2 r^3}\times\nonumber\\
&\left[\frac{(\tilde{A}+\tilde{B})r_s+\tilde{A} r_0 + \tilde{B} r}{(r+r_s)^2}\right]-\frac{4\beta}{3(1+2\beta)(1+4\beta)\kappa^2 r^2}\left[\frac{(2\tilde{A}+\tilde{B})r_s+2\tilde{A}r_0+\tilde{B}r}{(r+r_s)^3}\right],
\end{align}
\begin{align}\label{eq:Model2_Fa}
F_a&=\frac{9}{\left( 3+7\beta \right){\kappa}^{2}{r}^{4}} \left[ r_0+{\frac {\tilde{A} \left( r-r_0 \right) }{r+r_s}} +\tilde{B}\ln  \left( {\frac {r+r_s}{r_s+r_0}} \right) \right]  -\frac { \left( 3\tilde{A}+3\tilde{B} \right) r_s+3\tilde{A}r_0+3\tilde{B}r}{ \left( r+r_s \right) ^{2} \left( 3+7\beta \right) {\kappa}^{2}{r}^{3}}.
\end{align}
%%%%%%%%%%%%%%%%%%%%%%%%%%%%%%%%%%%%%%%%%%%%%%%%%%%%%%%%%%%%%%%%%%%%%%%%%%%%%%%%%%%%%%
%\bibliographystyle{apsrev}
%\bibliography{JRPHSRef}
%%%%%%%%%%%%%%%%%%%%%%%%%%%%%%%%%%%%%%%%%%%%%%%%%%%%%%%%%%%%%%%%%%%%%%%%%%%%%%%%%%%%%%
%merlin.mbs apsrev4-1.bst 2010-07-25 4.21a (PWD, AO, DPC) hacked
%Control: key (0)
%Control: author (8) initials jnrlst
%Control: editor formatted (1) identically to author
%Control: production of article title (-1) disabled
%Control: page (0) single
%Control: year (1) truncated
%Control: production of eprint (0) enabled
%

\end{document}